\documentclass[11pt]{article}%
\usepackage{eurosym}
\usepackage{amsfonts}
\usepackage{amssymb}
\usepackage{graphicx}
\usepackage{amsmath}
\usepackage{makeidx}
\usepackage{indentfirst}
\usepackage[T1]{fontenc}
\usepackage[utf8]{inputenc}%
\providecommand{\U}[1]{\protect\rule{.1in}{.1in}}
\newcounter{resultnum}[section]
\newcounter{conclusionnum}[section]
\newcounter{conditionnum}[section]
\newcounter{conjecturenum}[section]
\newcounter{examplenum}[section]
\newcounter{exercisenum}[section]
\newcounter{lemmanum}[section]
\newcounter{notationnum}[section]
\newcounter{theoremnum}[section]
\newcounter{definitionnum}[section]
\newcounter{corollarynum}[section]
\newcounter{remarknum}[section]
\newcounter{propositionnum}[section]
\newcounter{acknowledgementnum}[section]
\newcounter{algorithmnum}[section]
\newcounter{axiomnum}[section]
\newcounter{casenum}[section]
\newcounter{claimnum}[section]
\newcounter{summarynum}[section]
\newcounter{problemnum}[section]
\begin{document}
\title{ Nonmetric geometric flows and quasicrystalline topological phases \\ 
for dark energy and dark matter in $f(Q)$ cosmology}
\date{February 29, 2024}
\author{
{\textbf{Lauren\c{t}iu Bubuianu}\thanks{email: laurentiu.bubuianu@tvr.ro and
laurfb@gmail.com}}
\and {\small \textit{SRTV - Studioul TVR Ia\c{s}i} and \textit{University
Appolonia}, 2 Muzicii street, Ia\c{s}i, 700399, Romania} \vspace{.1 in}
\and {\textbf{Erik Nurlan}} \thanks{email: enurlan.edu@gmail.com}
\and {\small \textit{Department of General and Theoretical Physics, 
Eurasian National University, Astana, 010000, Kazakhstan}} \vspace{.1 in} \\
{\textbf{Julia O. Seti }} 
\thanks{email: j.seti@chnu.edu.ua}\\ 
{\small \textit{\ Department of Information Technologies and Computer Physics }}\\
{\small \textit{\ Yu. Fedkovych Chernivtsi National University 2, Chernivtsi, 58012, Ukraine; }}\\ 
{\small \textit{\ Department of Applied Mathematics, Lviv Polytechnic National University, }}\\
{\small \textit{\ Stepan Bandera street, 12, Lviv, 79000, Ukraine}}  \vspace{.1 in} \\
\textbf{Sergiu I. Vacaru} \thanks{%
emails: sergiu.vacaru@fulbrightmail.org ; sergiu.vacaru@gmail.com } \and 
{\small \textit{Department of Physics, California State University at
Fresno, Fresno, CA 93740, USA; }} \and {\small \textit{\ Yu. Fedkovych
Chernivtsi National University 2, Chernivtsi, 58012, Ukraine ;}} \and 
{\small \textit{\ Center for Advanced Studies, Ludwig-Maximilians-Universit\"{a}t, Seestrasse 13, M\"{u}nchen, 80802, Germany}} \vspace{.1 in} 
\and {\textbf{El\c{s}en Veli Veliev }} \thanks{email: elsen@kocaeli.edu.tr and
elsenveli@hotmail.com}\\{\small \textit{\ Department of Physics,\ Kocaeli University, 41380, Izmit,
Turkey }}}
\maketitle

\begin{abstract}
We elaborate on nonmetric geometric flow theory and metric-affine gravity with applications in modern cosmology.  Two main motivations for our research follow from the facts that 1) cosmological models for $f(Q)$ modified gravity theories, MGTs, are efficient for describing recent observational data provided by the James Webb Space Telescope; and 2) the statistical thermodynamic properties of such nonmetric locally anisotropic cosmological models can be studied using generalizations of the concept of G. Perelman entropy. We derive nonmetric distorted R. Hamilton and Ricci soliton equations in such canonical nonholonomic variables when corresponding systems of nonlinear PDEs can be decoupled and integrated in general off-diagonal forms. This is possible if we develop and apply the anholonomic frame and connection deformation method involving corresponding types of generating functions and generating sources encoding nonmetric distortions. Using such generic off-diagonal solutions (when the coefficients of metrics and connections may depend generically on all spacetime coordinates), we model accelerating cosmological scenarios with quasi-periodic gravitational and (effective) matter fields; and study topological and nonlinear geometric properties of respective dark energy and dark matter, DE and DM, models. As explicit examples, we analyze some classes of nonlinear symmetries defining topological
quasicrystal, QC, phases which can modified to generate other types of quasi-periodic and locally anisotropic structures. The conditions when such nonlinear systems possess a behaviour which is similar to that of the Lambda cold dark matter ($\Lambda$CDM) scenario are stated. We conclude that nonmetric geometric and cosmological flows can be considered as an alternative to the $\Lambda$CDM concordance models and speculate on how such theories can be elaborated. This is a partner work with generalizations and applications of the results published by L. Bubuianu, S. Vacaru et all.   EPJC 84 (2024) 211; 80 (2020) 639; 78 (2018) 393; 78 (2018) 969; and CQG 35 (2018) 245009. 

\end{abstract}
\tableofcontents

\section{Introduction and objectives}
\label{sec1}
Modified gravity theories, MGTs, with nonmetricity have been known since 1918 when H. Weyl \cite{weyl1918} considered nonmetric spaces as an attempt to construct a unified geometric theory of gravity and electromagnetism. In the Weyl
geometry, the nonmetricity tensor $Q_{\alpha\beta\gamma}:=D_{\lambda}g_{\beta\gamma}\neq0$ is treated as an additional fundamental geometric object. This is different from the Einstein gravity theory (i.e. general relativity, GR) constructed for symmetric pseudo-Riemannian metrics $g_{\beta\gamma}$ of Lorentz signature $(+++-)$ and when the Levi-Civita, LC,
connection $\nabla_{\lambda}$ is uniquely defined from the conditions of zero torsion, 
$T_{\ \beta\lambda}^{\alpha}[\nabla]=0,$ and metric compatibility, $\nabla_{\lambda}g_{\beta\gamma}=0$.\footnote{In next sections and appendices, we shall provide necessary definitions and explain when our notations will be different from some generally accepted ones.} The monographs \cite{misner,hawking73,wald82,kramer03} contain necessary geometric methods and results on mathematical relativity and exact solutions (they involve generally accepted abstract, frame and coordinate indices, and geometric and physical objects conventions). Nonmetric geometric and gravity theories were
mostly ignored for many decades because of criticisms by A. Einstein and W. Pauli. We cite \cite{hehl95} as the first comprehensive review on metric-affine gravity involving independent metric, $g_{\beta\gamma},$ and linear connection, $D_{\lambda}= \{\Gamma_{\ \beta\lambda}^{\alpha}\}$, structures. Such theories are characterized by nontrivial torsion,
$T_{\ \beta\lambda}^{\alpha}[D],$ and nonmetricity, $Q_{\alpha\beta\gamma}$, tensors and other geometric objects like the non-Riemannian curvature, the Ricci tensors, different scalar curvatures etc. In monograph \cite{vmon05}, the metric-affine geometry and gravity theories were elaborated on (co) tangent super- and/or noncommutative bundle spaces and for various generalized Finsler-Lagrange-Hamilton spaces.

\vskip5pt However, in physical literature, the position of MGTs, including Weyl geometries and various gravitational and matter field theories determined by generalized Lagrangians constructed as nonlinear functionals $f(R,T,Q,T_{m}),$ changed drastically during the last 25 years. That was due to observational pieces of evidence on late-time acceleration cosmology and related dark energy, DE, and dark matter, DM, problems. A plethora of MGTs have been elaborated alternatively to GR and standard particle physics. We recommend \cite{harko21,iosifidis22,khyllep23,koussour23} as some recent reviews of results and applications in modern cosmology.\footnote{In a partner work \cite{kazakh1}, a different system of notations is used. For instance, $F(...)$ is written instead of $f(...)$ because the symbol $f$ was considered for normalizing functions of nonmetric geometric flows. In this paper, we write $f(Q)$ for nonmetric gravity as in \cite{khyllep23,koussour23}, see also references therein. A comprehensive list of references on nonmetricity gravity theories and applications is not provided in this paper. We do not discuss details on increased interest in cosmological phenomenology of $f(Q)$-type and cite only the most important papers that are directly related to the purposes of this work.} In this work, we study MGTs constructed for a gravitational
Lagrangian $f(Q)$ involving a nonmetricity scalar $Q$ (all necessary definitions and notations are provided in the next section and Appendix \ref{appendixa}). Such gravity theories may challenge the standard $\Lambda$CDM scenario, lead to interesting cosmological phenomenology, and can be confronted with various recent observational data provided by James Webb space telescope JWST, \cite{foroconi23,boylan23,biagetti23}.\footnote{We do not provide an exhaustive list of references and do not discuss details on recent observational data provided by James Webb Space Telescope, JWST, which have sparked debates about the validity of many cosmological models. Such results related to the accelerating cosmology and various problems of dark energy, DE, and dark matter, DM, physics motivate the importance for studying in GR and MGTs
 new classes of generic off-diagonal, non-homogeneous and locally anisotropic cosmological solutions.}

\vskip5pt Motivated by a number of new geometric and exciting theoretical properties of nonmetric MGTs, in this and a partner work \cite{kazakh1}, we employ the powerful mathematical tools exploited in theories of relativistic geometric flows and nonholonomic Lorentz manifolds \cite{vacaru20} and consider such constructions for $Q$--deformations on nonholonomic metric-affine spaces. We develop also the anholonomic frame and connection deformation method, AFCDM, for generating exact and parametric cosmological solutions in $f(Q)$ gravity, see \cite{bubuianu17,vacaru18,vreview23} for
recent reviews of methods and results concerning MGTs and GR and applications in modern cosmology.  So, \textsf{the general goal of this work is to show how nonholonomic geometric flow and gravitational equations in $f(Q)$ MGTs and cosmology can be solved in exact and parametric forms using nonholonomic dyadic variables and defining necessary classes of distortions of canonical (non) linear connection structures and nonmetricity scalar $\widehat{\mathbf{Q}}$.}

\vskip5pt The main \textbf{Hypothesis} in this and the partner work \cite{kazakh1} is that: \textit{mathematically self-consistent nonmetric geometric flow and related physically viable nonmetric gravitational and cosmological theories can be constructed considering }$\mathit{Q}$\textit{--deformations of some models on nonholonomic Lorentz manifolds and
certain classes of metric compatible gravity theories. Using nonholonomic dyadic variables and canonically adapted (non) linear connection structures, the corresponding nonmetric geometric evolution flow equations, or dynamical gravitational equations, can be integrated in certain generic off-diagonal forms determined by generating functions and generating sources depending on all spacetime coordinates. Such new classes of nonmetric geometric flow cosmological solutions describe various quasi-periodic spacetime (quasicrystal, QC) and quasicrystalline topological phases \cite{bubuianu17,sv18,else21} being important for determining new features of the DE and DM theories.}

\vskip5pt The formulated hypothesis is supported by an important class of \textsf{nonmetric quasi-stationary solutions which are with Killing symmetry on a time like vector} -- the typical examples include $Q$--deformed black ellipsoid, black holes, wormhole, solitonic etc. configurations \cite{kazakh1}. Using similar geometric methods in dual forms on time and space
coordinates (with corresponding nonholonomic space like Killing symmetry), we can generate various \textsf{nonmetric locally anisotropic cosmological solutions with generic dependence on a time-like coordinate}. Such off-diagonal metrics and generalized connection structures determining nonmetricity fields are constructed and studied in this work. All classes of
quasi-periodic and cosmological solutions are characterized by corresponding nonlinear symmetries and described as $Q$--modified models of Grigori Perelman's thermodynamics introduced in the theory of Ricci flows
\cite{perelman1,hamilton82,friedan2,friedan3}. We cite also \cite{monogrrf1,monogrrf2,monogrrf3}, for reviews of mathematical results and methods, and \cite{vacaru20,bubuianu23,bubuianu23a}, for recent developments and applications in non-standard particle physics, MGTs, and geometric and quantum information flow models.

\vskip5pt The objectives, Objs, of this work are structured for corresponding sections:

The \textbf{Obj 1} stated for section \ref{sec2} is to formulate the $f(Q)$ gravity in nonholonomic (2+2) variables with canonical distortions of the LC-connection. The fundamental geometric and physical objects on such metric--affine spaces are re-defined with respect to nonlinear connection, N-connection, adapted frames. This allows us to prove general decoupling and integration properties of geometric and physically important nonlinear systems of partial differential equations, PDEs, encoding nonmetricity.

\vskip5pt In section \ref{sec3}, we provide a generalization of the $f(Q)$ gravity to a model of nonmetric geometric flow theory. The \textbf{Obj 2} is to define $Q$--distorted relativistic R. Hamilton and D. Friedan equations in canonical nonholonomic variables and to consider respective nonholonomic Ricci soliton equations. Then, the \textbf{Obj 3} is to formulate a statistical thermodynamic model for $f(Q)$ geometric flows derived as a nonholonomic and canonical $Q$--deformation of G. Perelman constructions \cite{perelman1} and non Riemannian generalizations in
\cite{vacaru20,bubuianu23,bubuianu23a,kazakh1}.

\vskip5pt The \textbf{Obj 4} of section \ref{sec4} is to study possible applications of our methods in modern DE and DM physics by constructing new classes of generic off-diagonal cosmological and geometric flow solutions encoding nonmetricity and generating topological QC structures. We prove that quasi-periodic cosmological structures may arise in generic off-diagonal forms from QC like generating functions, or from effective generating sources; and, in a general context, with mixed and different phases of topological QCs. Gravitational polarizations and effective sources are introduced and computed
in such forms when nonmetric QCs are generated as off-diagonal deformations of the $\Lambda$CDM model.

\vskip5pt In section \ref{sec5}, as the \textbf{Obj 5}, we provide explicit examples of how to compute Perelman's thermodynamic variables for nonmetric geometric flows inducing topological QC structures \cite{bubuianu17,sv18,else21} for modeling DE and DM effects.

\vskip5pt Finally, section \ref{sec6} is devoted to conclusions and discussion of the main results of the paper. Appendix \ref{appendixa} contains a reformulation of the AFCDM for constructing exact and parametric solutions in $f(Q)$ gravity and nonmetric geometric flow theories. We provide a brief review of nonholonomic 2+2 spacetime splitting and corresponding topological QC structures in Appendix \ref{appendixb}.

\section{Metric-affine spaces and $f(Q)$ gravity in nonholonomic (2+2) variables}

\label{sec2} In this section, we formulate the four-dimensional, 4-d, $f(Q)$ gravity \cite{khyllep23,koussour23} in nonholonomic dyadic variables with canonical distortion of linear connection structures considered other type nonmetric gravity theories in our partner work \cite{kazakh1}. Such a nonholonomic geometric formalism was developed in \cite{bubuianu17,vacaru18,vacaru20,vreview23} but it will be generalized and applied in section \ref{sec4} for constructing exact and parametric solutions in nonmetric geometric flow and MGTs with generic off-diagonal cosmological metrics and generalized (non) linear connection structures. Appendix \ref{appendixa} contains a summary of necessary notations
and formulas for applications in nonmetric geometric flow and gravity theories of the anholonomic frame and connection deformation method, AFCDM.

\subsection{Nonlinear connections and nonholonomic dyadic splitting}

\label{ss21}Let $V$ be a 4-d Lorentz manifold of necessary smooth class
defined by a metric tensor $g=\{g_{\alpha\beta}(u^{\gamma})\}$ of signature
$(+,+,+,-)$ and corresponding LC-connection $\nabla=\{\breve{\Gamma}%
_{\ \beta\gamma}^{\alpha}(u)\}.$\footnote{Local coordinates are labeled as
$u^{\alpha}=(u^{\acute{\imath}},u^{4}=t)=(x^{i},y^{a}),$ for $\acute{\imath}
=1,2,3$. We follow the conventions for local frames/ coordinates/ indices
used, for instance, in \cite{vmon05,kazakh1} that typical 3-d space indices
run values of type $\acute{\imath}=1,2,3;$ for $u^{4}=y^{4}=ct$; the light
velocity constant $c$ can be always fixed as $c=1$ for corresponding systems
of unities and coordinates; and typical indices of type $i=1,2$ and $a=3,4$
are used for a conventional 2+2 splitting. In brief, we write correspondingly
$u=(x,t)=(x,y).$ Arbitrary local frames $e_{\alpha}=e_{\ \alpha}%
^{\alpha^{\prime}}(u)\partial_{\alpha^{\prime}} $ and (dual) frames, or
co-frames, $e^{\beta}=e_{\beta^{\prime}}^{\ \beta}(u)d^{\beta^{\prime}}$ for
respective coordinate (co) frames $\partial_{\alpha^{\prime}}=\partial
/\partial u^{\alpha^{\prime}}$ and $d^{\beta^{\prime}}=du^{\beta^{\prime}},$
when matrices $e_{\ \alpha}^{\alpha^{\prime}}(u)$ and $e_{\beta^{\prime}%
}^{\ \beta}(u)$ define some tetradic (equivalently, vierbein coefficients).
For primed indices, we may use similar conventions with a necessary 3+1 and/or
2+2 splitting. In our constructions, we can underline necessary indices, or
drop any priming/underlying if that will not result in ambiguities.} We can
endow such a manifold with an independent linear (affine) connection structure
$D=\{\Gamma_{\ \beta\gamma}^{\alpha}(u)\}$ and elaborate on certain physically
important metric-affine geometric flow and gravity models determined by
geometric data $(g,D).$

A nonholonomic structure with 2+2 splitting can be defined by local bases,
$\mathbf{e}_{\nu},$ and co-bases (dual), $\mathbf{e}^{\mu},$
\begin{align}
\mathbf{e}_{\nu}  &  =(\mathbf{e}_{i},e_{a})=(\mathbf{e}_{i}=\partial/\partial
x^{i}-\ N_{i}^{a}(u)\partial/\partial y^{a},\ e_{a}=\partial_{a}%
=\partial/\partial y^{a}),\mbox{ and  }\label{nader}\\
\mathbf{e}^{\mu}  &  =(e^{i},\mathbf{e}^{a})=(e^{i}=dx^{i},\ \mathbf{e}%
^{a}=dy^{a}+\ N_{i}^{a}(u)dx^{i}). \label{nadif}%
\end{align}
We suppose that in local coordinate form a set of coefficients $N_{i}^{a}(u)$
is determined by a nonlinear connection, N-connection structure,
$\mathbf{N}=N_{i}^{a}(x,y)dx^{i}\otimes\partial/\partial y^{a}$, which in
global form can be defined on tangent bundle $TV$ as a Whitney sum:
\begin{equation}
\mathbf{N}:\ TV=hV\oplus vV. \label{ncon}%
\end{equation}
This states a conventional horizontal and vertical splitting ( h- and
v--decomposition) into respective 2-d and 2-d subspaces, $hV$ and
$vV.$\footnote{\label{fnwcoeff}We use the term nonholonomic (equivalently,
anholonomic) because, for instance, a N-elongated basis (\ref{nader})
satisfies certain nonholonomy relations $[\mathbf{e}_{\alpha},\mathbf{e}%
_{\beta}]= \mathbf{e}_{\alpha}\mathbf{e}_{\beta}-\mathbf{e}_{\beta}%
\mathbf{e}_{\alpha}= W_{\alpha\beta}^{\gamma}\mathbf{e}_{\gamma}$, with
nontrivial anholonomy coefficients $W_{ia}^{b}=\partial_{a}N_{i}^{b}%
,W_{ji}^{a}=\Omega_{ij}^{a}=\mathbf{e}_{j}\left(  N_{i}^{a}\right)
-\mathbf{e}_{i}(N_{j}^{a}),$ where $\Omega_{ij}^{a}$ define the coefficients
of N-connection curvature. If all $W_{ia}^{b}$ are zero for a $\mathbf{e}%
_{\alpha},$ such a N-adapted base is holonomic and we can write it as a
partial derivative $\partial_{\alpha}$ with $N_{i}^{a}=0. $ The coefficients
$N_{j}^{a}$ may be nontrivial even all $W_{\alpha\beta}^{\gamma}=0$ (this can
be in general curve coordinate bases but we can always define a diagonal
holonomic base for a corresponding coordinate transforms).} Decomposing
geometric objects (tensors, connections etc.) with respect to N-adapted bases
(\ref{nader}) and (\ref{nadif}), we formulate a nonholonomic dyadic formalism
for metric-affine geometry.

For nonholonomic manifolds enabled with N-connection structure (\ref{nader}),
we shall use boldface symbols and write, for instance, $\mathbf{V},$ and
consider respective nonholonomic tangent, $T\mathbf{V,}$ and cotangent,
$T^{\ast}\mathbf{V}$, bundles; their tensor products, $T\mathbf{V\otimes
}T^{\ast}\mathbf{V,}$ etc. The geometric objects on such geometric spaces
(generalized spacetimes and/or phase spaces) can be N-adapted and written in
boldface form as distinguished geometric objects (in brief, d-objects,
d-vectors, d-tensors etc). For instance, we can write a d--vector as
$\mathbf{X}=(hX,vX)$ and a second rank d-tensor as $\mathbf{F}%
=(hhF,hvF,vhF,vvF).$ Such dyadic decompositions can be written in N-adapted
coefficient forms with respect to N-adapted bases, see details in
\cite{vmon05,bubuianu17,vacaru18,vreview23}.

\subsection{$N$-adapted metric-affine structures}

Any metric tensor $g\in T^{\ast}V\otimes T^{\ast}V$ can be written as a
d-tensor $\mathbf{g}\in T^{\ast}\mathbf{V\otimes}T^{\ast}\mathbf{V}$ and
parameterized in three equivalent forms:
\begin{align}
g  &  =g_{\alpha^{\prime}\beta^{\prime}}(u)e^{\alpha^{\prime}}\otimes
e^{\beta^{\prime}}, \mbox{ with respect to an arbitrary coframe }e^{\alpha
^{\prime}};\nonumber\\
&  = \mathbf{g} =(hg,vg)=\ g_{ij}(x,y)\ e^{i}\otimes e^{j}+\ g_{ab}%
(x,y)\ \mathbf{e}^{a}\otimes\mathbf{e}^{b},\label{dm}\\
&  \qquad\mbox{ in N-adapted form with }hg = \{\ g_{ij}\},vg=\{g_{ab}%
\};\nonumber\\
&  =\underline{g}_{\alpha\beta}(u)du^{\alpha}\otimes du^{\beta},\mbox{
with respect to a coordinate coframe }du^{\beta},\nonumber\\
&  \qquad\mbox{ where } \underline{g}_{\alpha\beta} =\left[
\begin{array}
[c]{cc}%
g_{ij}+N_{i}^{a}N_{j}^{b}g_{ab} & N_{j}^{e}g_{ae}\\
N_{i}^{e}g_{be} & g_{ab}%
\end{array}
\right]  . \label{offd}%
\end{align}
A metric $\mathbf{g}=\{\underline{g}_{\alpha\beta}\}$ is generic off--diagonal
(for 4-d spacetimes, such a matrix can't be diagonalized via coordinate
transforms) if the anholonomy coefficients $W_{\alpha\beta}^{\gamma}$ are not
identical to zero.

Additionally to the metric structure $g,$ we can consider an independent
linear connection structure $D=\{\Gamma_{\ \beta\lambda}^{\alpha}\}$ with
coefficients defined with respect to arbitrary frames and coframes,
$e_{\alpha}$ and $e^{\beta}$ \cite{hehl95}. In general, such a $D$ is not
adapted to a N-connection structure and can be a introduced to be
independent both from the metric and N-connection structures, see details in
\cite{vmon05,vacaru18,vreview23}.

A distinguished connection \textbf{(d--connection),} $\mathbf{D}=(hD,vD),$ is
defined as a linear connection preserving under parallelism the N--connection
splitting (\ref{ncon}). With respect to frames (\ref{nader}) and
(\ref{nadif}), we can write decompositions of $\mathbf{D}$ using h- and
v-indices,
\begin{equation}
\mathbf{D}=\{\mathbf{\Gamma}_{\ \alpha\beta}^{\gamma}=(L_{jk}^{i},\acute
{L}_{bk}^{a};\acute{C}_{jc}^{i},C_{bc}^{a})\},\mbox{ where }hD=(L_{jk}%
^{i},\acute{L}_{bk}^{a})\mbox{ and }vD=(\acute{C}_{jc}^{i},C_{bc}^{a}).
\label{hvdcon}%
\end{equation}
If a general metric-affine space is defined by geometric structures $(V,g,D),$
a nonholonomic metric-affine space can be defined N-adapted form by geometric
data $(\mathbf{V,N,g,D}).$

The fundamental geometric objects of nonholonomic metric-affine space are
defined and computed in standard form as for any linear connection. For a
d-connection $\mathbf{D,}$ we have
\begin{align}
\mathcal{T}(\mathbf{X,Y})  &  := \mathbf{D}_{\mathbf{X}}\mathbf{Y}%
-\mathbf{D}_{\mathbf{Y}}\mathbf{X}-[\mathbf{X,Y}%
],\mbox{ the torsion d-tensor,  d-torsion};\label{fundgeom}\\
\mathcal{R}(\mathbf{X,Y})  &  := \mathbf{D}_{\mathbf{X}}\mathbf{D}%
_{\mathbf{Y}}-\mathbf{D}_{\mathbf{Y}}\mathbf{D}_{\mathbf{X}}-\mathbf{D}%
_{\mathbf{[X,Y]}},\mbox{ the curvature d-tensor, d-curvature};\nonumber\\
\mathcal{Q}(\mathbf{X})  &  := \mathbf{D}_{\mathbf{X}}\mathbf{g,}\mbox{\ the
nonmetricity d-field, d-nonmetricity}.\nonumber
\end{align}
Such geometric d-objects can be written in nonholonomic dyadic form with
respect to N-adapted frames (\ref{nader}) and (\ref{nadif}) with such
parameterizations:
\begin{equation}
\mathcal{R}=\mathbf{\{R}_{\ \beta\gamma\delta}^{\alpha}\},\mathcal{T}%
=\{\mathbf{T}_{\ \alpha\beta}^{\gamma}\},\mathcal{Q}=\mathbf{\{Q}%
_{\gamma\alpha\beta}\}. \label{fundgeomncoef}%
\end{equation}
We note that in various metric-affine gravity theories
\cite{hehl95,harko21,iosifidis22,khyllep23,koussour23} the definitions and
coefficient formulas for respective $R_{\ \beta\gamma\delta}^{\alpha},
T_{\ \beta\lambda}^{\alpha},$ and $Q_{\alpha\beta\gamma}$ are
provided/computed for arbitrary frame/ coordinate decompositions for an affine
connection $\Gamma_{\ \beta\lambda}^{\alpha}$ not considering N-adapted
geometric constructions.

In geometric flow and gravity theories (this work and in \cite{kazakh1,vmon05}%
, there are involved nontrivial nonmetricity d-tensors), there are used also
another important geometric d-objects:
\begin{align}
\mathbf{R}ic  &  = \{\mathbf{R}_{\ \beta\gamma}:=\mathbf{R}_{\ \beta
\gamma\alpha}^{\alpha}\},\mbox{ the Ricci d-tensor};\label{dricci}\\
\mathbf{R}sc  &  =\mathbf{g}^{\alpha\beta}\mathbf{R}_{\ \alpha\beta
},\mbox{ the scalar curvature}, \label{dscal}%
\end{align}
where $\mathbf{g}^{\alpha\beta}$ are the coefficients of the inverse d-tensor
of a d-metric (\ref{dm}).

The N-adapted coefficient formulas involving the coefficients (\ref{dm}) and
(\ref{hvdcon}) are provided in \cite{vmon05,vacaru18,vreview23}.

\subsection{Canonical metric-affine d-structures, LC-connection, and
nonmetricity}

For a nonholonomic metric-affine space $(\mathbf{V,N,g,D}),$ we can construct
different geometric and physical models defined by the same metric structure
but involving different linear connection structures. Additionally to
$\mathbf{D}$ (not determined by the metric), there are two other important
linear connection structures determined by a d-metric $\mathbf{g}$ (\ref{dm})
following such definitions:
\begin{equation}
(\mathbf{g,N})\rightarrow\left\{
\begin{array}
[c]{cc}%
\mathbf{\nabla:} & \mathbf{\nabla g}=0;\ _{\nabla}\mathcal{T}=0,\ \mbox{\
LC--connection };\\
\widehat{\mathbf{D}}: & \widehat{\mathbf{Q}}=0;\ h\widehat{\mathcal{T}%
}=0,v\widehat{\mathcal{T}}=0,\ hv\widehat{\mathcal{T}}\neq
0,\mbox{ the canonical
d-connection}.
\end{array}
\right.  \label{twocon}%
\end{equation}
Such an auxiliary d-connection defines a canonical distortion
relation\footnote{In our works, "hat" labels are used typically for geometric
d-objects defined by $\widehat{\mathbf{D}}$ (\ref{twocon}), when N-adapted
coefficients are computed \cite{vmon05,vacaru18,vreview23}:
$\widehat{\mathbf{D}}=\{\widehat{\mathbf{\Gamma}}_{\ \alpha\beta}^{\gamma
}=(\widehat{L}_{jk}^{i},\widehat{L}_{bk}^{a},\widehat{C}_{jc}^{i}%
,\widehat{C}_{bc}^{a})\},$ for $\widehat{L}_{jk}^{i} = \frac{1}{2}%
g^{ir}(\mathbf{e}_{k}g_{jr}+\mathbf{e}_{j}g_{kr}-\mathbf{e}_{r}g_{jk}),$
\par
$\widehat{L}_{bk}^{a}=e_{b}(N_{k}^{a})+\frac{1}{2}g^{ac}(\mathbf{e}_{k}%
g_{bc}-g_{dc}\ e_{b}N_{k}^{d}-g_{db}\ e_{c}N_{k}^{d}), \widehat{C}_{jc}^{i} =
\frac{1}{2}g^{ik}e_{c}g_{jk},\ \widehat{C}_{bc}^{a}=\frac{1}{2}g^{ad}%
(e_{c}g_{bd}+e_{b}g_{cd}-e_{d}g_{bc})$.},
\begin{equation}
\widehat{\mathbf{D}}[\mathbf{g}]=\nabla\lbrack\mathbf{g}]+\widehat{\mathbf{Z}%
}[\mathbf{g}], \label{cdist}%
\end{equation}
when the canonical distortion d-tensor, $\widehat{\mathbf{Z}}[\mathbf{g}]$,
and $\nabla\lbrack\mathbf{g}]$ are determined by the same metric structure
$\mathbf{g}$. In our works, we prefer to work with $\widehat{\mathbf{D}}$
which allow to decouple in some general off-diagonal forms (\ref{offd})
physically important systems of nonlinear PDEs. The LC-connection does not
have such a property (excepting some special diagonalizable ansatz), see
details in \cite{vmon05,vacaru18,vreview23}.

A general d-connection $\mathbf{D}$ (\ref{hvdcon}) can be characterized by
\begin{align}
\mathbf{D}  &  =\nabla+\mathbf{L}=\widehat{\mathbf{D}}+\widehat{\mathbf{L}%
},\label{disf}\\
&  \qquad\mbox{ where } \widehat{\mathbf{L}} =\mathbf{L}-\widehat{\mathbf{Z}%
}\nonumber
\end{align}
for the disformation d-tensor $\mathbf{L}=\{\mathbf{L}_{\ \beta\lambda
}^{\alpha}=\frac{1}{2}(\mathbf{Q}_{\ \beta\lambda}^{\alpha}-\mathbf{Q}%
_{\ \beta\ \lambda}^{\ \alpha}-\mathbf{Q}_{\ \lambda\ \beta}^{\ \alpha})\}$
with $\mathbf{Q}_{\alpha\beta\lambda}:=\mathbf{D}_{\alpha}\mathbf{g}%
_{\beta\lambda},$ for $\widehat{\mathbf{Q}}_{\alpha\beta\lambda}%
=\widehat{\mathbf{D}}_{\alpha}\mathbf{g}_{\beta\lambda}=0.$ Here, we use
boldface symbols $\mathbf{Q}$ and $\mathbf{L}$ because any tensor can be
transformed into a d-tensor if a N-connection structure is prescribed. This
holds true even $\nabla$ is not a d-connection (nevertheless, all
constructions can be performed with respect to N-elongated frames).

The two linear connection structure (\ref{hvdcon}) and distortion relations
(\ref{cdist}) and (\ref{disf}) allow us to define and compute respective
distortion relations of such geometric d-objects and, respective, objects:

\begin{itemize}
\item Nonmetricity d-vectors and vectors:%
\begin{align}
\mathbf{Q}_{\alpha}  &  =\mathbf{g}^{\beta\lambda}\mathbf{Q}_{\alpha
\beta\lambda}=\mathbf{Q}_{\alpha\ \lambda}^{\ \lambda},\ ^{\intercal
}\mathbf{Q}_{\beta}=\mathbf{g}^{\alpha\lambda}\mathbf{Q}_{\alpha\beta\lambda
}=\mathbf{Q}_{\alpha\beta}^{\quad\alpha}%
;\mbox{ and, correspondingly, }\nonumber\\
Q_{\alpha}  &  =g^{\beta\lambda}Q_{\alpha\beta\lambda}=Q_{\alpha\ \lambda
}^{\ \lambda},\ ^{\intercal}Q_{\beta}=g^{\beta\lambda}Q_{\alpha\beta}%
^{\quad\alpha}=Q_{\alpha\beta}^{\quad\alpha}, \label{nmdv}%
\end{align}
where there are used N-adapted frames (\ref{nader}) and (\ref{nadif}) and,
correspondingly, $e_{\alpha}$ and $e^{\beta}.$

\item Nonmetricity conjugate d-tensor and tensor:%
\begin{align}
\widehat{\mathbf{P}}_{\ \ \alpha\beta}^{\gamma}  &  =\frac{1}{4}%
(-2\widehat{\mathbf{L}}_{\ \alpha\beta}^{\gamma}+\mathbf{Q}^{\gamma}%
\mathbf{g}_{\alpha\beta}-\ ^{\intercal}\mathbf{Q}^{\gamma}\mathbf{g}%
_{\alpha\beta}-\frac{1}{2}\delta_{\alpha}^{\gamma}\mathbf{Q}_{\beta}-\frac
{1}{2}\delta_{\beta}^{\gamma}\mathbf{Q}_{\alpha})\nonumber\\
\mathbf{P}_{\ \ \alpha\beta}^{\gamma}  &  =\frac{1}{4}(-2\mathbf{L}%
_{\ \alpha\beta}^{\gamma}+\mathbf{Q}^{\gamma}\mathbf{g}_{\alpha\beta
}-\ ^{\intercal}\mathbf{Q}^{\gamma}\mathbf{g}_{\alpha\beta}-\frac{1}{2}%
\delta_{\alpha}^{\gamma}\mathbf{Q}_{\beta}-\frac{1}{2}\delta_{\beta}^{\gamma
}\mathbf{Q}_{\alpha})\mbox{ and }\nonumber\\
P_{\ \ \alpha\beta}^{\gamma}  &  =\frac{1}{4}(-2L_{\ \alpha\beta}^{\gamma
}+Q^{\gamma}g_{\alpha\beta}-\ ^{\intercal}Q^{\gamma}g_{\alpha\beta}-\frac
{1}{2}\delta_{\alpha}^{\gamma}Q_{\beta}-\frac{1}{2}\delta_{\beta}^{\gamma
}Q_{\alpha}), \label{nmcjdt}%
\end{align}
where the formulas from the 3-d line for a general $D=\{\Gamma_{\ \beta\gamma
}^{\alpha}(u)\}$ (which can be not a d-connection).

\item Nonmetricity scalar for respective d-connection and LC-connection:%
\begin{equation}
\widehat{\mathbf{Q}}=-Q_{\alpha\beta\lambda}\widehat{\mathbf{P}}^{\alpha
\beta\lambda}, Q=-Q_{\alpha\beta\lambda}\mathbf{P}^{\alpha\beta\lambda
}\mbox{ and } Q =-Q_{\alpha\beta\lambda}P^{\alpha\beta\lambda}. \label{nmsc}%
\end{equation}
Here we note that, in general, the scalar $\widehat{\mathbf{Q}}\neq0$ even
$\widehat{\mathbf{Q}}_{\alpha\beta\lambda}=0$ by definition
($\widehat{\mathbf{Q}}$ is not constructing only by contracting components of
$\widehat{\mathbf{Q}}_{\alpha\beta\lambda}$).
\end{itemize}

Above geometric d-objects and objects can be used for elaborating and study
models of nonmetric geometric flows and modified gravity theories. The
motivation and priority of "hat" variables is that we can decouple and
integrate in certain general off-diagonal forme corresponding physically
important systems of nonlinear PDEs as we shall prove in next section and
Appendix \ref{appendixa}.

\subsection{Curvature, torsion \& nonmetricity tensors for nonholonomic
metric-affine spaces}

A nonholonomic metric-affine space $\mathbf{V}$ with a prescribed metric and d-metric structure, $\mathbf{g}=\{\underline{g}_{\alpha\beta}\}=(hg,vg),$ see formulas (\ref{dm}) and (\ref{offd}), is characterized by a corresponding multi-connection structure of type $(\nabla,D,\widehat{\mathbf{D}}%
,\mathbf{D})$, when the first two linear connections and the next two
d-connections are correspondingly non-adapted and adapted to a prescribed
N-connection structure $\mathbf{N.}$ In result, we can define different types
of curvature, Ricci, torsion and nonmetricity tensors and related scalars.
This is important for formulating and investigating geometric and physical
properties of nonmetric geometric flow and nonmetric gravitational theories.
The goal of this subsection is to introduce necessary conventions and define
in necessary abstract and N-adapted coefficient forms the above mentioned
geometric objects and d-objects.

The fundamental geometric d-objects (\ref{fundgeom}) and, with respective
N-adapted coefficients (\ref{fundgeomncoef}), the Ricci d-tensor and curvature
scalars (\ref{dricci}) and (\ref{dscal}) can be defined and computed in
standard forms for any nonholonomic metric-affine space $(\mathbf{V,N,g,D}%
=\nabla+\mathbf{L}=\widehat{\mathbf{D}}+\widehat{\mathbf{L}})\,$ as we
considered in \cite{kazakh1,vmon05}. Such d-objects are important for
formulating nonmetric geometric theories and nonmetric gravitational
equations. The last ones can be defined as a subclass of nonmetric
nonholonomic Ricci soliton equations and written in alternative forms with
(\ref{nmdv}), (\ref{nmcjdt}) and (\ref{nmsc}) encoded into certain effective
sources and nonholonomic constraints.

The N-adapted and, respective, general frame/coordinate coefficients of the
Riemannian d-curvatures and curvature can be labeled and computed as
\begin{equation}
\mathcal{R}=\{\mathbf{R}_{\ \beta\gamma\delta}^{\alpha}%
\},\ \widehat{\mathcal{R}}=\{\widehat{\mathbf{R}}_{\ \beta\gamma\delta
}^{\alpha}\} \mbox{ and } \mathcal{\breve{R}}=\{\breve{R}_{\ \beta\gamma
\delta}^{\alpha}\}. \label{dcurvatures}%
\end{equation}
Such values involve the same d-metric/metric structure $\mathbf{g}%
=(hg,vg)=\{\underline{g}_{\alpha\beta}\},$ see parameterizations (\ref{dm})
and (\ref{offd}), but different linear connections. We omit explicit
cumbersome formulas with $h$- and $v$-indices presented in \cite{vmon05}. It
should be noted the coefficients (\ref{dcurvatures}) are subjected to certain
distortion relations
\begin{equation}
\mathbf{R}_{\ \beta\gamma\delta}^{\alpha}=\breve{R}_{\ \beta\gamma\delta
}^{\alpha}+\mathbf{\breve{Z}}_{\ \beta\gamma\delta}^{\alpha}%
=\widehat{\mathbf{R}}_{\ \beta\gamma\delta}^{\alpha}+\widehat{\mathbf{Z}%
}_{\ \beta\gamma\delta}^{\alpha}, \label{curvatdistor}%
\end{equation}
where the coefficients of the distortion d-tensors $\mathbf{\breve{Z}%
}_{\ \beta\gamma\delta}^{\alpha}$ and $\widehat{\mathbf{Z}}_{\ \beta
\gamma\delta}^{\alpha}$ can be computed using respective distortion relations
(\ref{cdist}) and (\ref{disf}).

Contracting indices in (\ref{dcurvatures}) and (\ref{curvatdistor}), we define
and compute the coefficients of respective Ricci d-tensors and Ricci tensor,
see formulas (\ref{dricci}),
\begin{align}
\mathbf{R}ic  &  =\breve{R}ic+\breve{Z}ic=\widehat{\mathbf{R}}%
ic+\widehat{\mathbf{Z}}%
ic,\mbox{ for respective coefficients }\label{driccidist}\\
&  \mathbf{R}ic=\{\mathbf{R}_{\beta\gamma}=\mathbf{R}_{\ \beta\gamma\alpha
}^{\alpha}\};\breve{R}ic=\{\breve{R}_{\beta\gamma}=\breve{R}_{\ \beta
\gamma\alpha}^{\alpha}\},\breve{Z}ic=\{\breve{Z}_{\beta\gamma}=\breve
{Z}_{\ \beta\gamma\alpha}^{\alpha}\};\nonumber\\
&  \widehat{\mathbf{R}}ic=\{\widehat{\mathbf{R}}_{\beta\gamma}%
=\widehat{\mathbf{R}}_{\ \beta\gamma\alpha}^{\alpha}\},\widehat{\mathbf{Z}%
}ic=\{\widehat{\mathbf{Z}}_{\beta\gamma}:=\widehat{\mathbf{Z}}_{\ \beta
\gamma\alpha}^{\alpha}\}.\nonumber
\end{align}
Then, contracting respectively these formulas with $\mathbf{g}^{\alpha\beta},$
we compute the corresponding scalar curvatures (\ref{dscal}),
\begin{align*}
\mathbf{R}sc  &  =\mathbf{g}^{\alpha\beta}\mathbf{R}_{\ \alpha\beta}=\breve
{R}sc+\breve{Z}sc=\widehat{\mathbf{R}}sc+\widehat{\mathbf{Z}}sc,\mbox{
where }\\
&  \breve{R}sc=g^{\beta\gamma}\breve{R}_{\beta\gamma},\breve{Z}sc=g^{\beta
\gamma}\breve{Z}_{\beta\gamma};\widehat{\mathbf{R}}sc=\mathbf{g}^{\alpha\beta
}\widehat{\mathbf{R}}_{\alpha\beta},\widehat{\mathbf{Z}}sc=\mathbf{g}%
^{\alpha\beta}\widehat{\mathbf{Z}}_{\alpha\beta}.
\end{align*}

The nonmetric fundamental d-objects and objects (\ref{nmdv}), (\ref{nmcjdt}),
(\ref{nmsc}) \ and (\ref{dcurvatures}), (\ref{curvatdistor}) encode certain
nontrivial d-torsion and torison fields. Let us begin with the formulas for a
general affine connection $D=\{\Gamma_{\ \alpha\beta}^{\gamma}\}$ with
coefficients stated with respect to arbitrary frame, or coordinate, bases,
$\partial_{\alpha}$ or $e_{\alpha},$ and their dual. The fundamental geometric
objects (\ref{fundgeom}) are defined by $D$ with coefficients formulas
(\ref{fundgeomncoef})
\[
\mathcal{R}=\mathbf{\{}R_{\ \beta\gamma\delta}^{\alpha}\},\mathcal{T}%
=\{T_{\ \alpha\beta}^{\gamma}\},\mathcal{Q}=\mathbf{\{}Q_{\alpha\beta\gamma
}:=D_{\alpha}g_{\beta\gamma}\}.
\]
Such coefficient formulas are provided in \cite{hehl95,vmon05}.

The distortions formulas (\ref{cdist}) with distinguishing of coefficients for
$\nabla=\{\breve{\Gamma}_{\ \alpha\beta}^{\gamma}\}$ can be parameterized:
\begin{align}
\Gamma_{\ \alpha\beta}^{\gamma}  &  =\breve{\Gamma}_{\ \alpha\beta}^{\gamma
}+K_{\ \alpha\beta}^{\gamma}+\ ^{q}Z_{\ \alpha\beta}^{\gamma}%
,\mbox{ where }\label{cdist1}\\
K_{\alpha\beta\gamma}  &  =\frac{1}{2}(T_{\alpha\beta\gamma}+T_{\beta
\gamma\alpha}-T_{\gamma\alpha\beta}),\ S_{\gamma\ }^{\ \alpha\beta}=\frac
{1}{2}(K_{\ \ \ \gamma}^{\alpha\beta}+\delta_{\gamma}^{\alpha}T_{\ \ \ \tau
}^{\tau\beta}-\delta_{\gamma}^{\beta}T_{\ \ \ \tau}^{\tau\alpha}),\nonumber\\
\ ^{q}Z_{\ \alpha\beta\gamma}  &  =\frac{1}{2}(Q_{\alpha\beta\gamma}%
-Q_{\beta\gamma\alpha}-Q_{\gamma\beta\alpha}).\nonumber
\end{align}
Such geometric objects can be used for defining (as in formulas (\ref{dricci})
and (\ref{dscal})) the Ricci tenor and three scalar values for $D$ considered
in the Weyl--Cartan geometry:%
\begin{align*}
Ric[D]  &  =\{R_{\ \beta\gamma}:=R_{\ \beta\gamma\alpha}^{\alpha}\},\\
Rsc[D]  &  =R=g^{\beta\gamma}\ R_{\beta\gamma},\ ^{s}T=\ S_{\gamma
\ }^{\ \alpha\beta}T_{\ \alpha\beta}^{\gamma},\ Q=\ ^{q}Z_{\ \beta\alpha
}^{\alpha}\ ^{q}Z_{\quad\mu}^{\beta\mu}-\ ^{q}Z_{\ \beta\mu}^{\alpha}%
\ ^{q}Z_{\quad\alpha}^{\beta\mu}.
\end{align*}
Here we note that for the LC-connection as $\nabla=\{\breve{\Gamma}%
_{\ \alpha\beta}^{\gamma}\},$ when the corresponding tensor and scalar
geometric objects are defined satisfy such properties:
\begin{align*}
Ric[\nabla]  &  =\{\breve{R}_{\beta\gamma}:=\breve{R}_{\ \beta\gamma\alpha
}^{\alpha}\},T_{\ \alpha\beta}^{\gamma}\equiv0\\
Rsc[\nabla]  &  =\breve{R}=g^{\beta\gamma}\breve{R}_{\beta\gamma}%
,\ ^{s}T=\ S_{\gamma\ }^{\ \alpha\beta}T_{\ \alpha\beta}^{\gamma}%
\equiv0,\ Q=\ ^{q}Z_{\ \beta\alpha}^{\alpha}\ ^{q}Z_{\quad\mu}^{\beta\mu
}-\ ^{q}Z_{\ \beta\mu}^{\alpha}\ ^{q}Z_{\quad\alpha}^{\beta\mu}\equiv0.
\end{align*}

A d-connection $\mathbf{D=\{\mathbf{\Gamma}}_{\ \alpha\beta}^{\gamma}\}$ with
nontrivial nonmetric d-tensor $\mathbf{Q}_{\gamma\alpha\beta}$ allows to
compute the N-adapted coefficients of distortion d-tensors with respect to
$\widehat{\mathbf{D}}=\{\widehat{\mathbf{\mathbf{\Gamma}}}_{\ \alpha\beta
}^{\gamma}\},$%
\begin{align}
\mathbf{\mathbf{\Gamma}}_{\ \alpha\beta}^{\gamma}  &
=\widehat{\mathbf{\mathbf{\Gamma}}}_{\ \alpha\beta}^{\gamma}+\mathbf{K}%
_{\ \alpha\beta}^{\gamma}+\ ^{q}\widehat{\mathbf{Z}}_{\ \alpha\beta}^{\gamma
},\mbox{
where }\label{cdist2}\\
\mathbf{K}_{\alpha\beta\gamma}  &  =\frac{1}{2}(\mathbf{T}_{\alpha\beta\gamma
}+\mathbf{T}_{\beta\gamma\alpha}-\mathbf{T}_{\gamma\alpha\beta}),\ \mathbf{S}%
_{\gamma\ }^{\ \alpha\beta}=\frac{1}{2}(\mathbf{K}_{\ \ \ \gamma}^{\alpha
\beta}+\delta_{\gamma}^{\alpha}\mathbf{T}_{\ \ \ \tau}^{\tau\beta}%
-\delta_{\gamma}^{\beta}\mathbf{T}_{\ \ \ \tau}^{\tau\alpha}),\nonumber\\
\ ^{q}\mathbf{Z}_{\ \alpha\beta\gamma}  &  =\frac{1}{2}(\mathbf{Q}%
_{\alpha\beta\gamma}-\mathbf{Q}_{\beta\gamma\alpha}-\mathbf{Q}_{\gamma
\beta\alpha}).\nonumber
\end{align}
In above formulas, we can compute the canonical distortion $\mathbf{T}%
_{\ \alpha\beta}^{\gamma}= \widehat{\mathbf{T}}_{\ \alpha\beta}^{\gamma
}+\widehat{\mathbf{Z}}_{\ \alpha\beta}^{\gamma}$ and $\ ^{q}\mathbf{Z}%
_{\ \alpha\beta}^{\gamma}=\ ^{q}\widehat{\mathbf{Z}}_{\ \alpha\beta}^{\gamma}$
involving the disfunction (\ref{disf}) computed for the canonical distortion
(\ref{cdist}) from $\widehat{\mathbf{D}}=\{\widehat{\Gamma}_{\ \alpha\beta
}^{\gamma}\}.$ This defines a nonholonomic Weyl-Cartan geometry when
corresponding Ricci scalar, d-torsion scalar and $Q$-scalars are
\[
\mathbf{R}sc[\mathbf{D}]=\ ^{s}R=\mathbf{g}^{\beta\gamma}\ \mathbf{R}%
_{\beta\gamma},\ ^{s}\mathbf{T}=\ \mathbf{S}_{\gamma\ }^{\ \alpha\beta
}\mathbf{T}_{\ \alpha\beta}^{\gamma},\ ^{q}\mathbf{Q}=\ ^{q}\mathbf{Z}%
_{\ \beta\alpha}^{\alpha}\ ^{q}\mathbf{Z}_{\quad\mu}^{\beta\mu}-\ ^{q}%
\mathbf{Z}_{\ \beta\mu}^{\alpha}\ ^{q}\mathbf{Z}_{\quad\alpha}^{\beta\mu}.
\]
In such formulas, we can always separate certain canonical "hat" components,
which can be determined by some classes of exact/parametric off-diagonal
solutions for physically important nonlinear PDEs when the $Q$-distortions are
encoded in effective sources.

\subsection{ $f(Q)$ gravity in non-adapted or canonical dyadic variables}

In this subsection, we analyze two possibilities to formulate models of
nonmetric gravity theory, when 1) the gravitational Lagrange density
$\ ^{g}\mathcal{L\approx}f(Q)$ is a functional of the nonmetricity scalar $Q,$
or 2) it is a functional $\ ^{g}\widehat{\mathcal{L}}\mathcal{\approx
}\widehat{f}(\widehat{\mathbf{Q}})$ of $\widehat{\mathbf{Q}},$ see formulas
(\ref{nmsc}). The first formulation is considered in the bulk of nonmetricity
theories \cite{hehl95,harko21,iosifidis22,khyllep23,koussour23}, when physical
solutions are constructed for certain diagonal metrics depending on a radial
coordinate (quasi-stationary solutions) or on a time like coordinate
(cosmological solutions). For the second formulation, it is possible to apply
the AFCDM and construct generic off-diagonal solutions as we proved in
\cite{vmon05,bubuianu17,vacaru18,vreview23}. Such a modified geometric method
can be applied also for nonmetric geometric flow theories
\cite{kazakh1,vmon05} (in this work, we extend it for cosmological
$\widehat{\mathbf{Q}}$--deformed solutions).

\vskip5pt
It should be noted that we can formulate theories with gravitational Lagrange
density $\ ^{g}\mathcal{L\approx}f(\mathbf{Q})$, where $\mathbf{Q}$ is defined
by a general d-connection $\mathbf{D.}$ Using nonholonomic distributions and
distortions of connections, we can re-define the geometric constructions for
some effective theories with $\ ^{g}\mathcal{L}\approx\ ^{g}%
\widehat{\mathcal{L}}+\ ^{e}\widehat{\mathcal{L}},$ where $\ ^{e}%
\widehat{\mathcal{L}}(\widehat{\mathbf{Q}},...)$ is a functional of
$\widehat{\mathbf{Q}}$ and certain canonical scalars of d-torsion, distortions
and disfunctions (parameterized in different forms (\ref{cdist}),
(\ref{disf}),(\ref{cdist1}), or (\ref{cdist2})). In this work, we study for
simplicity models with $\ ^{g}\widehat{\mathcal{L}} \approx\widehat{f}%
(\widehat{\mathbf{Q}})$ when physically important systems of nonlinear PDEs
can be decoupled and integrated in certain general forms for generic
off-diagonal cosmological metrics and nontrivial nonmetricity. Such classes of
locally anisotropic cosmological solutions can be restricted and modified for
$f(Q)$ theories by imposing additional nonholonomic constraints; or distorted
for more general configurations for $f(\mathbf{Q}).$

\subsubsection{Nonmetric $f(Q)$ gravity in non-adapted variables \newline and
distortions of LC-connections}

Using the Lagrange densities $\ ^{g}\mathcal{L}=\frac{1}{2\kappa}f(Q),$ for
gravitational constant $\kappa,$ and the matter Lagrangian $\ ^{m}%
\mathcal{L}(g_{\alpha\beta},\phi),$ with matter fields $\phi$ and
pseudo-Riemannian measure $\sqrt{|g|}d^{4}u=\sqrt{|g_{\alpha\beta}|}d^{4}u,$
we can construct a nonmetric gravity theory for the action%
\begin{equation}
\mathcal{S}=\int\sqrt{|g|}d^{4}u(\ ^{g}\mathcal{L}+\ ^{m}\mathcal{L}).
\label{anmg}%
\end{equation}
By varying $\mathcal{S}$ on $g^{\alpha\beta}$ and $\Gamma_{\ \alpha\beta
}^{\gamma}$ for $f_{Q}:=\partial f/\partial Q,$ we obtain respective nonmetric
gravitational field equations (see \cite{jim18,khyllep23,koussour23}):%
\begin{align}
\frac{2}{\sqrt{|g|}}\nabla_{\gamma}(\sqrt{|g|}f_{Q}P_{\ \ \alpha\beta}%
^{\gamma})+\frac{1}{2}fg_{\alpha\beta}+f_{Q}(P_{\beta\mu\nu}Q_{\alpha}%
^{\ \mu\nu}-2P_{\alpha\mu\nu}Q_{\quad\beta}^{\mu\nu})  &  =\kappa
\ ^{m}T_{\alpha\beta}\label{gfeq1a}\\
\mbox{ and }\nabla_{\alpha}\nabla_{\beta}(\sqrt{|g|}f_{Q}P_{\ \ \gamma
}^{\alpha\beta})  &  =0. \label{gfeq1b}%
\end{align}
In these formulas, $P_{\ \ \alpha\beta}^{\gamma}$ and $Q_{\ \ \alpha\beta
}^{\gamma}$ are computed as in formulas (\ref{nmdv}), (\ref{nmcjdt}) and
(\ref{nmsc}). The matter energy-momentum in (\ref{gfeq1a}) is defined
variationally as
\begin{equation}
\ ^{m}T_{\alpha\beta}:=-\frac{2}{\sqrt{|g|}}\frac{\delta(\sqrt{|g|}%
\ ^{m}\mathcal{L})}{\delta g^{\alpha\beta}}=\ ^{m}\mathcal{L}g_{\alpha\beta
}+2\frac{\delta(\ ^{m}\mathcal{L})}{\delta g^{\alpha\beta}}, \label{emt}%
\end{equation}
where the second formula holds if $\ ^{m}\mathcal{L}$ does not depend in
explicit form on $\Gamma_{\ \alpha\beta}^{\gamma}$. In this work, we shall
study only such models of gravitational and matter field interactions. For
cosmological applications, we can use a barotropic perfect fluid approximation
when
\begin{equation}
\ ^{m}T_{\alpha\beta}=(\rho+p)v_{\alpha}v_{\beta}+pg_{\alpha\beta},
\label{emtc}%
\end{equation}
with isotropic pressure $\rho,$ energy density $\rho$ and the 4-velocity
vector $v_{\alpha}.$ We can generalize such formulas for certain locally
anisotropic models with (\ref{emtc}) describing locally anisotropic fluid
matter, when with respect to N-adapted frames the data $(\rho,p,v_{\alpha})$
may depend on a temperature/ geometric flow parameter $\tau$ and on spacetime
coordinates $u=\{u^{\alpha}\},$ parameterized as $\left(  \rho(\tau
,u),p(\tau,u),v_{\alpha}(\tau,u)\right)  .$

The covariant representation of nonmetric gravitational equations
(\ref{gfeq1a}) was formulated in \cite{jhao22} in a form using explicitly the
Einstein tensor, $\breve{E}:=\breve{R}ic-\frac{1}{2}g\breve{R}sc,$ for
$\nabla.$ There were studied also possible physical implications of some
equivalent effective Einstein equations with effective energy-momentum tensor
which may describe dark energy \cite{jhao22,koussour23}.\footnote{We note that
in this work we follow a different system of notations and chose an opposite
sign before $\ ^{m}T_{\alpha\beta}$ in (\ref{gfeq1a}) and (\ref{emt}).} Such
effective gravitational field equations can be written in the form%
\begin{align}
\breve{E}_{\alpha\beta}  &  =\frac{\kappa}{f_{Q}}\ ^{m}T_{\alpha\beta}%
+\ ^{DE}T_{\alpha\beta},\mbox{ where }\label{gfeq2a}\\
\ ^{DE}T_{\alpha\beta}  &  =\frac{1}{2}g_{\alpha\beta}(\frac{f}{f_{Q}%
}-Q)+2\frac{f_{QQ}}{f_{Q}}\nabla_{\gamma}(QP_{\ \ \alpha\beta}^{\gamma}).
\label{deemt}%
\end{align}
The nonmetric modifications of GR are encoded into the effective
energy-momentum tensor $\ ^{DE}T_{\alpha\beta}$ (\ref{deemt}). We note that
the left label DE is used because it is considered that nonmetricity can be
used to describe dark energy effects on modern cosmology (there are hundred of
such works, see discussion and references in
\cite{jhao22,hehl95,harko21,iosifidis22,khyllep23,koussour23}). In our
approach, we argue that additionally various DE and DM effects can be modelled
also by generic off-diagonal terms of metrics for exact/parametric solutions
in GR or MGTs and this holds true for theories of nonmetric geometric flows
and various metric-affine gravity theories \cite{kazakh1,vmon05}. Another
important remark is that we have to assume $f_{Q}<0$ if we consider modified
gravitational field equations (\ref{gfeq1a}) or (\ref{gfeq2a}) as
$Q$-deformations of the standard Einstein equations in GR.

Let us discuss the properties of the connection field equations (\ref{gfeq2a})
studied in \cite{jhao22,koussour23}:\ Such systems of nonlinear PDEs are
trivially satisfied in a model-independent manner for various classes of
geometric constructions and solutions \cite{de22,koussour23}. For instance, we
can consider the conditions $R_{\ \beta\gamma\delta}^{\alpha}=0,$ but
$\breve{R}_{\ \beta\gamma\delta}^{\alpha}\neq0$. This consists a class of
restrictions for developing the $f(Q)$-theory. The term "coincident gauge"
refers to such circumstances when there are certain coordinate frames for
which $\Gamma_{\ \alpha\beta}^{\gamma}=0,$ but $\breve{\Gamma}_{\ \alpha\beta
}^{\gamma}\neq0,$ and we can concentrate our sole attention to find physically
important solutions, for instance, of the $Q$-modified Einstein equations
(\ref{gfeq2a}). To study more general classes of metric-affine gravity and
nonmetric geometric flow theories \cite{kazakh1} is important to "relax" the
conditions $R_{\ \beta\gamma\delta}^{\alpha}=0,$ considering non-zero values.
In next subsection, we analyze similar conditions in terms of the canonical
d-connection structure $\widehat{\mathbf{D}}$ with canonical distortions
(\ref{cdist2}).

\subsubsection{Nonmetric $f(\protect\widehat{\mathbf{Q}})$ gravity in
canonical nonholonomic variables}

The systems of nonlinear PDEs (\ref{gfeq1a}) and (\ref{gfeq2a}) can not
decoupled and integrated in some general forms for generic off-diagonal
metrics formulated in arbitrary frames and coordinates. To generate exact and
parametric solutions for such nonlinear systems of equations is possible if
the geometric constructions are performed of nonholonomic dyadic variables
introduced in section \ref{ss21} and using distortions from
$\widehat{\mathbf{D}}$ of necessary type (non) linear connection structures.
In such a N-adapted approach, we write the Lagrange density as $\ ^{g}%
\widehat{\mathcal{L}}=\frac{1}{2\kappa}\widehat{f}(Q)$ and use the measure
$\sqrt{|\mathbf{g}_{\alpha\beta}|}\delta^{4}u,$ where $\delta^{4}u=
du^{1}du^{2}\delta u^{3}\delta u^{4}$ for $\delta u^{a}=\mathbf{e}^{a}$ being
N-elongated differentials of type (\ref{nadif}). We construct nonmetric
nonholonomic gravity theories in canonical N-adapted variables for the action%
\begin{equation}
\widehat{\mathcal{S}}=\int\sqrt{|\mathbf{g}_{\alpha\beta}|}\delta^{4}%
u(\ ^{g}\widehat{\mathcal{L}}+\ ^{m}\widehat{\mathcal{L}}), \label{actnmc}%
\end{equation}
where hats state that all Lagrange densities and geometric objects are written
in nonholonomic dyadic form, with boldface indices and using
$\widehat{\mathbf{D}}$. We can perform for $\widehat{\mathcal{S}}$ the same
variational procedure as in previous subsection but in N-adapted form, on
$\mathbf{g}^{\alpha\beta},$ and on $\mathbf{D}= \{\mathbf{\Gamma}%
_{\ \alpha\beta}^{\gamma}\}$ with distortions from $\widehat{\mathbf{\Gamma}%
}_{\ \alpha\beta}^{\gamma}$ as in (\ref{cdist2}), for $\widehat{f_{Q}%
}:=\partial\widehat{f}/\partial\widehat{Q}.$ If we chose $\widehat{\mathcal{S}%
}=\mathcal{S}$ with a $\mathcal{S}$ of type (\ref{anmg}), we obtain equivalent
classes of nonmetric gravity theories but formulated for different types of
nonholonomic or holonomic variables.

So, following a N-adapted variational procedure, or applying "pure" geometric
methods \cite{misner} with distoritions of geometric objects
\cite{kazakh1,vmon05,bubuianu17,vacaru18,vreview23}, we obtain such nonmetric
gravitational field equations defined in nonholonomic canonical diadic
variables
\begin{align}
\frac{2}{\sqrt{|\mathbf{g}|}}\widehat{\mathbf{D}}_{\gamma}(\sqrt
{|g|}\widehat{f_{Q}}\widehat{\mathbf{P}}_{\ \ \alpha\beta}^{\gamma})+\frac
{1}{2}\widehat{f}\mathbf{g}_{\alpha\beta}+\widehat{f_{Q}}(\widehat{\mathbf{P}%
}_{\beta\mu\nu}\mathbf{Q}_{\alpha}^{\ \mu\nu}-2\widehat{\mathbf{P}}_{\alpha
\mu\nu}\mathbf{Q}_{\quad\beta}^{\mu\nu})  &  =\kappa\widehat{\mathbf{T}%
}_{\alpha\beta}\label{cfeq3a}\\
\mbox{ and }\widehat{\mathbf{D}}_{\alpha}\widehat{\mathbf{D}}_{\beta}%
(\sqrt{|\mathbf{g}|}\widehat{f_{Q}}\widehat{\mathbf{P}}_{\ \ \gamma}%
^{\alpha\beta})  &  =0. \label{cfeq3b}%
\end{align}
In these formulas, $\widehat{\mathbf{P}}_{\ \ \alpha\beta}^{\gamma}$ and
$\mathbf{Q}_{\ \ \alpha\beta}^{\gamma}$ are defined as in formulas
(\ref{nmdv}), (\ref{nmcjdt}) and (\ref{nmsc}), when $\widehat{\mathbf{T}%
}_{\alpha\beta}$ are defined by N-adapted variations as in (\ref{emt}), or in
the form (\ref{emtc}), but with respect to N-elongated frames (\ref{nader})
and (\ref{nadif}).

The covariant representation of (\ref{cfeq3a}) can be formulated by analogy to
the formulas $\breve{E}$ as in previous subsection but using explicitly the
Einstein tensor, $\widehat{\mathbf{E}}:= \widehat{\mathbf{R}}ic-\frac{1}%
{2}\mathbf{g}\widehat{\mathbf{R}}sc,$ for $\widehat{\mathbf{D}}.$ Here we
consider a form of nonmetric gravitational equations with $\widehat{\mathbf{R}%
}ic=\{\widehat{\mathbf{R}}_{\alpha\beta}\}$ in the left side and certain
effective sources $\widehat{\yen }ic=\{\widehat{\yen }_{\alpha\beta}\}$ in the
right side. Such a representation is convenient for applying the AFCDM for
constructing off-diagonal solutions, see details in Appendix \ref{appendixa}.
Distorting the equations (\ref{gfeq2a}), for $\nabla\rightarrow
\widehat{\mathbf{D}}=\nabla+\widehat{\mathbf{Z}}$ (\ref{cdist}) and
parameterizations (\ref{cdist2}), or redefining in N-adapted form the
variation calculus\footnote{certain physical implications of some equivalent
effective Einstein equations were studied \cite{jhao22,koussour23}}, we
formulate the effective source (encoding both nonmetric geometric distortions
and matter fields) in the form%
\begin{align}
\widehat{\mathbf{R}}_{\alpha\beta}  &  =\widehat{\yen }_{\alpha\beta
},\mbox{ for }\label{cfeq4a}\\
\widehat{\yen }_{\alpha\beta}  &  =\ ^{e}\widehat{\mathbf{Y}}_{\alpha\beta
}+\ ^{m}\widehat{\mathbf{Y}}_{\alpha\beta}. \label{ceemt}%
\end{align}
The source $\widehat{\yen }_{\alpha\beta}$ (\ref{ceemt}) is defined by two
d-tenors: 1) $\ ^{e}\widehat{\mathbf{Y}}_{\alpha\beta}=\breve{Z}%
ic_{\alpha\beta}-\widehat{\mathbf{Z}}ic_{\alpha\beta}$ is of geometric
distorting nature, which can be computed in explicit form following formulas
(\ref{cdist}), (\ref{cdist1}) and (\ref{cdist2}) for (\ref{driccidist}). 2)
The energy-momentum d-tensor $\ ^{m}\widehat{\mathbf{Y}}_{\alpha\beta}$ of the
matter fields in (\ref{ceemt}) encodes the nonmetricity scalar
$\widehat{\mathbf{Q}}$ and d-tensor $\widehat{\mathbf{P}}_{\ \ \alpha\beta
}^{\gamma}$ defined respectively by formulas (\ref{nmdv}), (\ref{nmcjdt}) and
(\ref{nmsc}).\footnote{To keep certain compatibility with the Einstein
equations in GR for zero distortions and zero nonmetricity, we consider
distortions of Ricci d-tensors and tensors related by formulas
(\ref{driccidist}) with $\mathbf{R}ic=\breve{R}ic+\breve{Z}%
ic=\widehat{\mathbf{R}}ic+ \widehat{\mathbf{Z}}ic.$ For such distortions, the
nonmetric modified Einstein equations (\ref{gfeq2a}) can be written in the
form%
\[
\breve{E}_{\alpha\beta}=\frac{\kappa}{f_{Q}}\ ^{m}T_{\alpha\beta}%
+\ ^{DE}T_{\alpha\beta}=\kappa\breve{T}_{\alpha\beta},\mbox{ or }\breve
{R}_{\alpha\beta}=\breve{\Upsilon}_{\alpha\beta},\mbox{ for }\breve{\Upsilon
}_{\alpha\beta}=\kappa(\breve{T}_{\alpha\beta}-\frac{1}{2}g_{\alpha\beta
}\breve{T}),\mbox{ for }\breve{T}=g^{\alpha\beta}\breve{T}_{\alpha\beta}.
\]
The corresponding distortions of the Ricci d-tensors and Ricci tensor can be
computed following formulas
\[
\breve{R}ic_{\alpha\beta}+\breve{Z}ic_{\alpha\beta}=\widehat{\mathbf{R}%
}ic_{\alpha\beta}+\widehat{\mathbf{Z}}ic_{\alpha\beta}=\breve{\Upsilon
}_{\alpha\beta}+\breve{Z}ic_{\alpha\beta};\widehat{\mathbf{R}}ic_{\alpha\beta
}+\widehat{\mathbf{Z}}ic_{\alpha\beta}=\breve{\Upsilon}_{\alpha\beta}%
+\breve{Z}ic_{\alpha\beta},\widehat{\mathbf{R}}ic_{\alpha\beta}=\breve
{Z}ic_{\alpha\beta}-\widehat{\mathbf{Z}}ic_{\alpha\beta}+\breve{\Upsilon
}_{\alpha\beta}.
\]
In canonical nonholonomic dyadic variables and for $\nabla\rightarrow
\widehat{\mathbf{D}},$ above formulas transform respectively into those for
(\ref{cfeq4a}) and (\ref{ceemt}).}

The properties of the d-connection field equations (\ref{cfeq3b}) with
$\widehat{\mathbf{D}}$ are different from the connection field equations
(\ref{gfeq2a}) with $\nabla.$ Our goal is to formulate some well-defined
conditions when such systems of nonlinear PDEs are trivially satisfied (for
certain classes of models and their off-diagonal solutions) nonmetric
nonholonomic geometric flow nonmetric gravitational field equations and
solutions generalizing the constructions from \cite{de22,koussour23}; see also
the end of previous subsection. In N-adapted form, we can consider the
conditions 1] $\mathbf{R}_{\ \beta\gamma\delta}^{\alpha}=0,$ but when
$\breve{R}_{\ \beta\gamma\delta}^{\alpha}\neq0$ and $\widehat{\mathbf{R}%
}_{\ \beta\gamma\delta}^{\alpha}\neq0.$ We can always impose such constraints
and construct different classes of $f(Q)$ or $\widehat{f}(\widehat{Q})$
gravity theories. The term canonical nonholonomic "coincident gauge" can be
used for certain circumstances when in N-adapted form there are certain
coordinate frames for which $\mathbf{D}=\{\Gamma_{\ \alpha\beta}^{\gamma
}=0\},$ but $\breve{\Gamma}_{\ \alpha\beta}^{\gamma}\neq0$ and/or
$\widehat{\Gamma}_{\ \alpha\beta}^{\gamma}\neq0.$ 2] In a more general
context, we can consider nonmetric phase space determined by conditions
\begin{equation}
\widehat{\mathbf{D}}_{\beta}(\sqrt{|\mathbf{g}|}\widehat{f_{Q}}%
\widehat{\mathbf{P}}_{\ \ \gamma}^{\alpha\beta})=const. \label{dconnectcontr}%
\end{equation}
This imposes some nonholonomic covariant and N-adapted constraints on the
class of affine d-connections $\mathbf{D}=\{\Gamma_{\ \alpha\beta}^{\gamma}\}$
and respective nonmetric d-tensors $\mathbf{Q}_{\alpha\beta\gamma}.$ This way
(considering assumptions of type 1] or 2]), we can study various types of
nonmetric gravity theories but constrained in some forms which allow to
construct general classes of exact/ parametric solutions when concentrating on
physically important solutions of nonmetric modified Einstein equations
(\ref{cfeq4a}).

Finally, we note that in section \ref{sec4} and Appendix \ref{appendixa}, we construct and study off-diagonal solutions of the system of nonlinear PDEs (\ref{cfeq4a}) considering that certain restrictions on the classes of integration functions and generating functions and sources imposed on (\ref{ceemt}) allow also to satisfy the canonical d-connection constraints
(\ref{dconnectcontr}). In general, such solutions with $\widehat{\mathbf{D}}$ do not consist of examples of solutions of the system (\ref{gfeq2a}) with $\nabla. $ But imposing additional nonholonomic constraints of type $\widehat{\mathbf{D}}_{|\mathcal{T}=0}=\nabla,$ when the canonical nonholonomic torsion induced by N-coefficients become zero, $\widehat{\mathbf{T}}_{\ \alpha\beta}^{\gamma},$ we can extract exact/parametric solutions for (\ref{gfeq2a}) or (\ref{cfeq3a}). Such conditions can be satisfied by restricting respectively the class of generating functions and generating effective sources as we proved in \cite{kazakh1,bubuianu17,vacaru18,vacaru20,vreview23}. They modify also the constructions in Appendix \ref{appendixa} for nontrivial nonmetric cosmological configurations. In this work, we concentrate on locally anisotropic cosmological solutions of (\ref{cfeq4a}), and their nonmetric geometric flow evolution in terms of $\widehat{\mathbf{D}}$ considering that we can reduce them in terms of LC-configurations $\nabla$ by imposing
additional nonholonomic constraints.

\section{Nonmetric geometric flows and $f(Q)$ gravity}

\label{sec3}One of the main purposes of this work (stated for \textbf{Obj 4} in the introduction section) is to construct generic off-diagonal cosmological solutions of nonmetric gravitational equations(\ref{cfeq4a}) and analyze possible implications of such solutions in modern cosmology and DM and DE physics. Various classes of physically important such solutions were considered for nonholonomic MGTs \cite{bubuianu17,vacaru18,vacaru20,vreview23} and, in nonmetric form,  for quasi-stationary configurations, \cite{kazakh1}. Unfortunately, to investigating physical properties of off-diagonal metrics and nonmetric deformations of connection is not possible if we apply only the concept of Bekenstein--Hawking thermodynamics \cite{bek1,bek2,haw1,haw2} which works effectively for certain classes of solutions involving some
hypersurface, duality conditions, or holographic configurations. We can apply advanced nonholonomic geometric methods involving relativistic and nonmetric generalizations \cite{kazakh1} of the theory of geometric flows \cite{perelman1,hamilton82} and applications in GR and MGTs \cite{bubuianu23}. In such cases, we are able to define and compute another type, and more general, thermodynamic variables using the concept of W-entropy \cite{perelman1} which can generalized for nonmetric theories and corresponding quasi-stationary solutions \cite{kazakh1}. In this work, we elaborate on geometric and gravitational models that are related to the nonmetric effective sources 
$\ ^{e}\widehat{\mathbf{Y}}_{\alpha\beta}$ (\ref{ceemt}) and the canonical Ricci d-tensor $\widehat{\mathbf{R}}ic$
(\ref{driccidist}) describing off-diagonal cosmological solutions with topological QC structure. In this section, a model of nonmetric flows is elaborated for canonical deformations of $f(Q)$-gravity.

\subsection{Nonmetric geometric flow equations in canonical dyadic variables}

\label{ss31}We elaborate on the theory of nonmetric geometric flows in
canonical dyadic variables and families of geometric and physical data
\begin{equation}
(\mathbf{V,N}(\tau),\mathbf{g}(\tau),\mathbf{D}(\tau)=\nabla(\tau
)+\mathbf{L}(\tau)=\widehat{\mathbf{D}}(\tau)+\widehat{\mathbf{L}}(\tau
)\ ^{g}\widehat{\mathcal{L}}(\tau)+\ ^{m}\widehat{\mathcal{L}}(\tau
))\label{canonicalmafdata}%
\end{equation}
parameterized by a real parameter $\tau,0\leq\tau\leq\tau_{1}.$ The geometric
and physical objects for such theories depend, in general, on spacetime
coordinates. To simplify our notations, we shall write (for instance)
$\mathbf{g}(\tau)$ instead of $\mathbf{g}(\tau,u^{\beta})$ if that will not
result in ambiguities. For any fixed value $\tau=\tau_{0},$ such data define a
nonmetric gravity theory when the geometric d-objects are subjected to the
conditions to define solutions of the nonlinear system of PDEs (\ref{cfeq4a})
and (\ref{dconnectcontr}) with effective sources (\ref{ceemt}) as we
considered in previous section. According to G. Perelman, $\tau$ can be
treated as a temperature like parameter \cite{perelman1}. Such an
interpretation and corresponding statistical thermodynamical formulations of
the geometric flow theories can be considered also for relativistic and
modified background manifolds.

\subsubsection{$Q$-distorted R. Hamilton and D. Friedan equations}

The nonmetric geometric flow evolution equations can be postulated in
canonical dyadic variables in the form:
\begin{align}
\partial_{\tau}g_{ij}(\tau)  &  =-2[\widehat{\mathbf{R}}_{ij}(\tau
)-\widehat{\yen }_{ij}(\tau)];\ \label{ricciflowr2}\\
\partial_{\tau}g_{ab}(\tau)  &  =-2[\widehat{\mathbf{R}}_{ab}(\tau
)-\widehat{\yen }_{ab}(\tau)];\nonumber\\
\widehat{\mathbf{R}}_{ia}(\tau)  &  =\widehat{\mathbf{R}}_{ai}(\tau
)=0;\widehat{\mathbf{R}}_{ij}(\tau)=\widehat{\mathbf{R}}_{ji}(\tau
);\widehat{\mathbf{R}}_{ab}(\tau)=\widehat{\mathbf{R}}_{ba}(\tau).\ \nonumber
\end{align}
The equations (\ref{ricciflowr2}) are written in N-adapted frames, where
$\widehat{\square}(\tau)=\widehat{\mathbf{D}}^{\alpha}(\tau
)\widehat{\mathbf{D}}_{\alpha}(\tau)$ is the canonical d'Alambert operator.
The conditions $\widehat{\mathbf{R}}_{ia}(\tau)=\widehat{\mathbf{R}}_{ai}%
(\tau)=0$ are imposed for $\widehat{R}ic[\widehat{\mathbf{D}}%
]=\{\widehat{\mathbf{R}}_{\alpha\beta}=[\widehat{R}_{ij},\widehat{R}%
_{ia},\widehat{R}_{ai},\widehat{R}_{ab}]\}$ if we elaborate on a theory with
symmetric d-metrics evolving under nonmetric nonholonomic Ricci flows. Such
constraints are not considered, for instance, in nonassociative geometric flow
theories with nonsymmetric metrics \cite{bubuianu23}. Systems of nonlinear
PDEs of similar type are studied in \cite{kazakh1} for models of nonmetric
geometric flows related to $f(R,T,Q,T_{m})$ theories of gravity. In this
paper, the $Q$-deformations, distorting relations (\ref{cdist2}) and sources
$\widehat{\yen }_{\alpha\beta}$ (\ref{ceemt}) are different from the sources
studied in that work.

The geometric flow equations (\ref{ricciflowr2}) consist of certain generalizations of the R. Hamilton equations \cite{hamilton82} postulated for $\nabla.$ Here we note that equivalent equations were considered a few years
before the mentioned mathematical works (by D. Friedan who was inspired by research on string theory and condensed matter physics \cite{friedan2,friedan3}). We can derive nonmetric variants of geometric flow equations considering an approach which is similar to the abstract geometric method from \cite{misner} when certain $\tau$-running fundamental geometric objects Ricci tensors and generalized sources are distorted to canonical nonholonomic data (\ref{canonicalmafdata}). In section \ref{sec4} and Appendix \ref{appendixa}, we prove that such systems of nonlinear PDEs are well--defined because they can be decoupled and integrated in certain general forms and certain classes of solutions describe physically important processes. Here we note that it is not possible to formulate and prove some general forms of
nonmetric Thurston-Poincar\'{e} conjectures, as it was considered in \cite{perelman1}, because there are an infinite number of non-Riemannian geometries when various topological and geometric analysis constructions are possible, being more sophisticate and not unique. Nevertheless, we can study certain physically important implications of nonmetric geometric flow and nonmetric gravity theories if we are able to find exact/parametric solutions of systems of nonlinear PDEs of type (\ref{ricciflowr2}). For the classes of solutions with topological QC structures as in Appendix \ref{appendixb}, the
approach involve nontrivial topological configurations.

\subsubsection{$\tau$-running Einstein equations, nonmetric geometric flows,
and $f(Q)$ Ricci solitons}

We can consider the term $\partial_{\tau}\mathbf{g}_{\mu^{\prime}\nu^{\prime}%
}(\tau)$ as an additional effective source defining $\tau$-running of
geometric flows of Ricci d-tensors. Using $\tau$-families of vierbein
transforms $\mathbf{e}_{\ \mu^{\prime}}^{\mu}(\tau)=\mathbf{e}_{\ \mu^{\prime
}}^{\mu}(\tau, u^{\gamma})$ and their dual transform $\mathbf{e}_{\nu}%
^{\ \nu^{\prime}}(\tau,u^{\gamma})$ with $\mathbf{e}_{\ }^{\mu}(\tau
)=\mathbf{e}_{\ \mu^{\prime}}^{\mu}(\tau)du^{\mu^{\prime}},$ we can introduce
N-adapted effective sources
\begin{equation}
\ \quad^{\intercal}\yen _{\ \nu}^{\mu}(\tau)=\mathbf{e}_{\ \mu}^{\mu^{\prime}%
}(\tau)\mathbf{e}_{\nu}^{\ \nu^{\prime}}(\tau)[~\yen _{\mu^{\prime}\nu
^{\prime}}(\tau)-\frac{1}{2}~\partial_{\tau}\mathbf{g}_{\mu^{\prime}%
\nu^{\prime}}(\tau)]=[\ _{h}\yen (\tau,{x}^{k})\delta_{j}^{i},~\ \yen (\tau
,x^{k},y^{a})\delta_{b}^{a}]. \label{dsourcparam}%
\end{equation}
The data $\ ^{\intercal}\yen =[\ _{h}\yen ,\ \yen ]$ can be fixed as some
generating functions for effective matter sources encoding contributions both
by geometric flows and $Q$-deformations. Prescribing explicit values of
$\ _{h}\yen $ and $\yen $, we impose certain nonholonomic constraints on the
noncommutative geometric flow scenarios. It is not possible to solve in
general form systems of PDEs of type $\partial_{\tau}\mathbf{g}_{\mu^{\prime
}\nu^{\prime}}= 2(\yen _{\mu^{\prime}\nu^{\prime}}- \ ^{\intercal}%
\yen _{\mu^{\prime}\nu^{\prime}})$, for prescribed generating sources in
(\ref{dsourcparam}) and general effective distortions and matter
energy-momentum. We can search for solutions with small parameters resulting
in recurrent formulas for powers on such parameters. In a different approach,
certain approximations allow to model relativistic models of geometric locally
anisotropic geometric diffusion or nonlinear wave evolution. For applications
to modern cosmology, we can consider that generating sources of type
$[\ _{h}\yen , \yen ]$ impose certain restrictions on the geometric evolution
and dynamics of sources modified by $Q$-deformations.

Using the generating sources$\ ^{\intercal}\yen _{\ \nu}^{\mu}(\tau)$
(\ref{dsourcparam}), we can write the nonmetric geometric flow equations
(\ref{ricciflowr2}) as $\tau$-running and $Q$-deformed Einstein equations for
$\widehat{\mathbf{D}}^{\alpha}(\tau),$
\begin{equation}
\widehat{\mathbf{R}}_{\ \ \beta}^{\alpha}(\tau)=\ \ ^{\intercal}%
\yen _{\ \ \beta}^{\alpha}(\tau). \label{cfeq4af}%
\end{equation}
Constraining nonholonomically this system for zero canonical d-connections, we
model nonmetric $\tau$-evolution scenarios in terms of LC-data:
\begin{align}
\widehat{\mathbf{T}}_{\ \alpha\beta}^{\gamma}(\tau)  &
=0,\mbox{ for }\widehat{\mathbf{D}}_{|\widehat{\mathcal{T}}=0}(\tau
)=\nabla(\tau),\mbox{ when }\label{lccondf}\\
\breve{R}_{\beta\gamma}(\tau)  &  =\ \ ^{\intercal}\yen _{\beta\gamma}(\tau).
\label{gfeq2af}%
\end{align}
Such systems of nonlinear PDEs define $\tau$-running generalizations of
$Q$-modified Einstein equations (\ref{gfeq2a}). Here we note that the torsion
of the related affine connections $D$ subjected to conditions (\ref{gfeq1b})
may be nontrivial and that $\ ^{\intercal}\yen _{\beta\gamma}(\tau)$ encodes
$Q$-deformations.

In \cite{kazakh1}, we defined nonholonomic and nonmetric Ricci soliton
configurations is a self-similar one for the corresponding nonmetric geometric
flow equations. Fixing $\tau=\tau_{0}$ in (\ref{ricciflowr2}), we obtain the
equations for the $\widehat{f}(\widehat{Q})$ Ricci solitons:
\begin{align}
\widehat{\mathbf{R}}_{ij}  &  =\ ^{\intercal}\yen _{ij},\ \widehat{\mathbf{R}%
}_{ab}=\ ^{\intercal}\yen ,\label{canriccisol}\\
\widehat{\mathbf{R}}_{ia}  &  =\widehat{\mathbf{R}}_{ai}=0;\widehat{\mathbf{R}%
}_{ij}=\widehat{\mathbf{R}}_{ji};\widehat{\mathbf{R}}_{ab}=\widehat{\mathbf{R}%
}_{ba}.\nonumber
\end{align}
The nonholonomic variables can be chosen in such forms that (\ref{canriccisol}%
) are equivalent to nonmetric modified Einstein equations (\ref{cfeq4a}) for
$\widehat{\mathbf{D}}^{\alpha}(\tau_{0})$.\footnote{In various MGTs and GR, as
in the theory of Ricci flows of Riemannian metrics, the Einstein spaces and
certain nonholonomic deformations consist examples of (nonholonomic) Ricci
solitons.} For additional nonholonomic constraints, such equations define
solutions of the nonmetric Einstein equations (\ref{gfeq2a}) for $\nabla
(\tau_{0}).$

Let us discus the issue of formulating conservation laws for $Q$-deformed
systems. Using the canonical d-connection, we can check that for systems of
type (\ref{canriccisol}) and related nonholonomic modified Einstein equations
there are satisfied the conditions
\[
\widehat{\mathbf{D}}^{\beta}\widehat{\mathbf{E}}_{\ \ \beta}^{\alpha
}=\widehat{\mathbf{D}}(\widehat{\mathbf{R}}_{\ \ \beta}^{\alpha}-\frac{1}%
{2}\ ^{s}\widehat{R}\delta_{\ \ \beta}^{\alpha})\neq
0\mbox{ and }\widehat{\mathbf{D}}^{\beta}\ ^{\intercal}\yen _{\ \ \beta
}^{\alpha}\neq0.
\]
This is typical for nonholonomic systems and the issue of formulating
conservation laws becomes more sophisticate because of nonmetricities. Similar
problems exist also in nonholonomic mechanics when the conservation laws are
formulated by solving nonholonomic constraints or introducing some Lagrange
multiples associated to certain classes of nonholonomic constraints. Solving
the constraint equations (they may depend on local coordinates, velocities or
momentum variables), we can re-define the variables and formulate conservation
laws in some explicit or non-explicit forms. Such nonholonomic variables allow
us to introduce new effective (mechanical) Lagrangians and Hamiltonians. This
allow us to define conservation laws in certain standard form if
$\mathbf{Q}_{\alpha\beta\gamma}=0.$ In some general forms and for nonmetric
gravity theories, such constructions can be performed only for some special
classes of solutions and corresponding geometric flow/ nonholonomic Ricci
soliton models. Using distortions of connections, we can rewrite systems of
type (\ref{cfeq4a}), (\ref{gfeq2a}), (\ref{canriccisol}) etc. in terms of
$\nabla, $ when $\nabla^{\beta}E_{\ \ \beta}^{\alpha}=\nabla^{\beta
}T_{\ \ \beta}^{\alpha}=0$ for $\mathbf{Q}_{\alpha\beta\gamma}\rightarrow0$ as
in GR. Here we note that we shall formulate in next subsection a statistical
thermodynamic energy type variable which is related to G. Perelman's paradigm
for Ricci flows, which may encode also nonmetric contributions.

In Appendix \ref{appendixa}, we show how using the AFCDM the equations
(\ref{cfeq4a}) and, if necessary, (\ref{gfeq2a}) with nonholonomic constraints
(\ref{gfeq2af}) can be decoupled and integrated in general locally anisotropic
cosmological forms for certain prescribed nonmetric effective sources
(\ref{dsourcparam}). Fixing the running parameter, such (in general,
off-diagonal) solutions with nonmetricity describe certain nonmetric
cosmological spacetimes described by systems of nonlinear PDEs (\ref{cfeq4a})
and (\ref{canriccisol}).

\subsection{Statistical geometric thermodynamics for $f(Q)$ geometric flows}

\label{ssperelth}G. Perelman \cite{perelman1} introduced the so-called F- and W-functionals as certain Lyapunov type functionals, $\breve{F}(\tau,g,\nabla,\breve{R}sc)$ and $\breve{W}(\tau,g,\nabla,\breve{R}sc)$ depending on a temperature like parameter $\tau$ and fundamental geometric objects when $\breve{W}$ have properties of "minus entropy". Using $\breve{F}$ or $\breve{W},$ he elaborated a variational proof for geometric flow equations of Riemannian metrics, which was applied to developing a strategy for proving the Poincar\'{e}--Thorston conjecture. Respective details are provided in
mathematical monographs \cite{monogrrf1,monogrrf2,monogrrf3}. It is not possible to formulate and prove in some general forms such conjectures for non-Riemannian geometries.

\vskip5pt
G. Perelman's geometric and analytic methods have a number of perspectives in modern particle physics, gravity and cosmology, see reviews of results in \cite{bubuianu23,vacaru18,vacaru20}. The main point is that using relativistic
generalizations and various modifications of $\breve{W}$ we can formulate statistical and geometric thermodynamic models which are more general and very different from the constructions performed in the framework of the Bekenstein-Hawking paradigm \cite{bek1,bek2,haw1,haw2}. The geometric flow constructions and related statistical thermodynamic approach can be generalized for various supersymmetric/ noncommutative / nonassociative geometries etc. In \cite{kazakh1}, such constructions were extended in nonholonomic form for nonmetric geometric flow theories related to $f(R,T,Q,T_{m})$ gravity and applied for investigating important physical properties of quasi-stationary solutions in such theories. To develop such geometric methods for the case of  $f(Q)$ gravity when the nonmetric gravitational equations of type (\ref{gfeq1a}) involve directly the geometric objects $P_{\ \alpha\beta}^{\gamma}$ and $Q_{\ \alpha\beta}^{\gamma}$ it is convenient because the geometric flows are usually defined as Ricci flows. It is more appropriate  to study models when the $f(Q)$ nonmetric gravitational equations are of type (\ref{gfeq2a}) and then to distort the constructions to include nonholonomic Ricci soliton equations (\ref{canriccisol}). So, the goal of the next subsections is to formulate a nonmetric geometric thermodynamic model for nonmetric geometric flows of canonical d-connections in nonholonomic dyadic variables for $\tau$-running modified Einstein equations (\ref{cfeq4a}).

\subsubsection{Nonholonomic F-/ W-functionals for $Q$-deformed geometric flows and gravity}

We postulate the modified Perelman's functionals following the same principles as in
\cite{perelman1,monogrrf1,monogrrf2,monogrrf3,bubuianu23,vacaru18,vacaru20,kazakh1}
but for nonmetric geometric flows involving canonical nonholonomic geometric
objects and $\widehat{f}(\widehat{Q})$ distortions:
\begin{align}
\widehat{\mathcal{F}}(\tau)  &  =\int_{t_{1}}^{t_{2}}\int_{\Xi_{t}}%
e^{-\zeta(\tau)}\sqrt{|\mathbf{g}(\tau)|}\delta^{4}u[\widehat{f}%
(\widehat{\mathbf{R}}sc(\tau))+\ ^{e}\mathcal{L}(\tau)+\ ^{m}%
\widehat{\mathcal{L}}(\tau)+|\widehat{\mathbf{D}}(\tau)\ \zeta(\tau
)|^{2}],\label{fperelm4t}\\
\widehat{\mathcal{W}}(\tau)  &  =\int_{t_{1}}^{t_{2}}\int_{\Xi_{t}}\left(
4\pi\tau\right)  ^{-2}e^{-\zeta(\tau)}\sqrt{|\mathbf{g}(\tau)|}\delta
^{4}u[\tau(\widehat{f}(\widehat{\mathbf{R}}sc(\tau))+ \ ^{e}\mathcal{L}(\tau)+
\ ^{m}\widehat{\mathcal{L}}(\tau)+|\widehat{\mathbf{D}}(\tau)\zeta(\tau
)|^{2}+\zeta(\tau)-4], \label{wfperelm4t}%
\end{align}
In these formulas, the condition $\int_{t_{1}}^{t_{2}}\int_{\Xi_{t}}\left(
4\pi\tau\right)  ^{-2}e^{-\zeta(\tau)}\sqrt{|\mathbf{g}|}d^{4}u=1$ is imposed
on the normalizing function $\zeta(\tau)=\zeta(\tau,u)$; and $\tau$-families
of Lagrange densities $\ ^{g}\widehat{\mathcal{L}}+\ ^{m}\widehat{\mathcal{L}%
}$ are considered as in (\ref{actnmc}).

Performing a N-adapted variational calculus for $\widehat{\mathbf{D}}$ (which
is similar to that presented in \cite{perelman1,monogrrf1,monogrrf2,monogrrf3}
but N-adapted to nonmetric geometric structures) and redefining respectively
the normalizing functions and nonholonomic distributions, we prove the
nonmetric geometric flow equations (\ref{ricciflowr2}) with an additional
constraint equation for $\zeta(\tau)$:
\begin{equation}
\partial_{\tau}\zeta(\tau)=-\widehat{\square}(\tau)[\zeta(\tau)]+\left\vert
\widehat{\mathbf{D}}\ ^{n} \zeta(\tau)\right\vert ^{2}-\widehat{f}%
(\widehat{\mathbf{R}}sc(\tau))-\ ^{e}\mathcal{L}(\tau)-\ ^{m}%
\widehat{\mathcal{L}}(\tau). \label{normcond}%
\end{equation}
This equation for $\zeta(\tau)$ involves nonlinear partial differential
operators of first and second order and relates $\tau$-families of canonical
Ricci scalars $\ \widehat{\mathbf{R}}sc(\tau)$ and nonmetric sources
$\widehat{\yen }_{\alpha\beta}$ (\ref{ceemt}). Such a nonlinear PDE can't be
solved in a general form which do not allow us to study in general forms
models of flow evolution of topological configurations determined by arbitrary
nonmetric structures and distributions of effective and real matter fields. We
can prescribe a topological structure for an off-diagonal metric constructed
as an exact/parametric solution of nonholonomic system of nonlinear PDEs
(\ref{ricciflowr2}). In such a case, we can chose a convenient $\zeta(\tau)$
(it can be prescribed to be a constant, or zero) which states certain
constrains on nonmetric geometric evolution. Alternatively, we can solve the
equation (\ref{normcond}) in certain parametric forms and then to re-define
the constructions for arbitrary systems of reference.

It should be noted here that the functionals $\widehat{\mathcal{F}}$ and
$\widehat{\mathcal{W}}$ were postulated in canonical nonholonomic variables in
a form which is similar to the original F- and W-functionals \cite{perelman1},
which were introduced for 3-d Riemannian $\tau$-flows $(g(\tau),\nabla
(\tau)).$ In this work, we extend the constructions in $\widehat{f}%
(\widehat{Q})$-deformed 4-d Lorentz manifolds endowed with additional
nonholonomic distributions determined by effective Lagrange densities. We can
compute the functionals (\ref{fperelm4t}) and (\ref{wfperelm4t}) for any 3+1
splitting with 3-d closed hypersurface fibrations $\widehat{\Xi}_{t}$ and
considering nonholonomic canonical d-connections and respective geometric
variables. Similar computations are provided in
\cite{bubuianu23,vacaru18,vacaru20} and references therein. The
$\widehat{\mathcal{W}}$-functional possess the properties of "minus" entropy
if we re-define the normalizing function to "absorb" the contributions from
$Q\,$-distortions and matter fields. This can be also modelled by choosing
corresponding nonholonomic configurations along some causal curves taking
values $\widehat{\mathcal{W}}(\tau)$ on $\widehat{\Xi}_{t}.$

The $\widehat{f}(\widehat{Q})$ modified G. Perelman's functionals can be
defined and computed for any solution of nonmetric geometric flow equations
(for instance, of type (\ref{cfeq4a})), or for nonmetric Ricci soliton
equations (\ref{canriccisol}). They can be used for elaborating thermodynamic
models both for nonmetric quasi-periodic configurations as in \cite{kazakh1}
and for locally anisotropic cosmological solutions encoding nonmetricity data
(we shall provide details and analyze explicit examples in next sections).

\subsubsection{Canonical thermodynamic variables for $f(Q)$ theories}

Let us redefine the normalization function $\zeta(\tau)\rightarrow
\widehat{\zeta}(\tau),$ for
\[
\partial_{\tau}\zeta(\tau)+\widehat{\square}(\tau)[\zeta(\tau)]-\left\vert
\widehat{\mathbf{D}}\zeta(\tau)\right\vert ^{2}-\ ^{e}\mathcal{L}(\tau
)-\ ^{m}\widehat{\mathcal{L}}(\tau)=\partial_{\tau}\ \widehat{\zeta}%
(\tau)+\widehat{\square}(\tau)[\widehat{\zeta}(\tau)]-\left\vert
\widehat{\mathbf{D}}\widehat{\zeta}(\tau)\right\vert ^{2},
\]
when (\ref{normcond}) transforms into
\begin{equation}
\partial_{\tau}\widehat{\zeta}(\tau)=-\widehat{\square}(\tau)[\widehat{\zeta
}(\tau)]+\left\vert \widehat{\mathbf{D}}\widehat{\zeta}(\tau)\right\vert
^{2}-\widehat{f}(\widehat{\mathbf{R}}sc(\tau)). \label{normcondc}%
\end{equation}
In terms of a corresponding integration measure, the W-functional
(\ref{wfperelm4t}) can be written "without" the effective matter
source\footnote{nevertheless the effective nonmetric and matter sources are
encoded into geometric data if we consider an explicit class of solutions of
(\ref{cfeq4af})}
\begin{equation}
\widehat{\mathcal{W}}(\tau)=\int_{t_{1}}^{t_{2}}\int_{\Xi_{t}}\left(  4\pi
\tau\right)  ^{-2}e^{-\widehat{\zeta}(\tau)}\sqrt{|\mathbf{g}(\tau)|}%
\delta^{4}u[\tau(\widehat{f}(\widehat{\mathbf{R}}sc(\tau
))+|\widehat{\mathbf{D}}(\tau)\widehat{\zeta}(\tau)|^{2}+\widehat{\zeta}%
(\tau)-4]. \label{wf1}%
\end{equation}

On a metric-affine space $\mathcal{M}$ endowed with canonical geometric
variables and a nonholonomic (3+1) splitting,\footnote{such a conventional
splitting is necessary for defining thermodynamic variables; in another turn,
a nonholonomic 2+2 decomposition is important for generating off-diagonal
solutions} we introduce the statistical partition function
\begin{equation}
\ ^{q}\widehat{Z}(\tau)=\exp[\int_{\widehat{\Xi}}[-\widehat{\zeta}+2]\ \left(
4\pi\tau\right)  ^{-2}e^{-\widehat{\zeta}}\ \delta\widehat{\mathcal{V}}%
(\tau)], \label{spf}%
\end{equation}
where the volume element is defined and computed as
\begin{equation}
\delta\widehat{V}(\tau)=\sqrt{|\mathbf{g}(\tau)|}\ dx^{1}dx^{2}\delta
y^{3}\delta y^{4}\ . \label{volume}%
\end{equation}
A left label $q$ for $\ ^{q}\widehat{Z}(\tau)$ is used in order to emphasize
that nonmetric $Q$-contributions are encoded in $\mathbf{g}(\tau)$ considered
as a solution of nonassociative geometric flow/ gravitational equations.

Using $\ ^{q}\widehat{Z}$ (\ref{spf}) and $\widehat{\mathcal{W}}(\tau)$
(\ref{wf1}) and performing the variational procedure in canonical N-adapted
variables, on a closed region of $\mathcal{M}$ (which is similar to that
provided in section 5 of \cite{perelman1}), we can define and compute
respective (canonical geometric and statistical) thermodynamic variables:
\begin{align}
\ ^{q}\widehat{\mathcal{E}}\ (\tau)  &  =-\tau^{2}\int_{\widehat{\Xi}%
}\ \left(  4\pi\tau\right)  ^{-2}\left(  \widehat{f}(\widehat{\mathbf{R}%
}sc)+|\ \widehat{\mathbf{D}}\ \widehat{\zeta}|^{2}-\frac{2}{\tau}\right)
e^{-\widehat{\zeta}}\ \delta\widehat{V}(\tau),\label{qthermvar}\\
\ \ ^{q}\widehat{S}(\tau)  &  =-\int_{\widehat{\Xi}}\left(  4\pi\tau\right)
^{-2}\left(  \tau((\widehat{\mathbf{R}}sc)+|\widehat{\mathbf{D}}%
\widehat{\zeta}|^{2})+\widehat{\zeta}-4\right)  e^{-\widehat{\zeta}}
\ \delta\widehat{V}(\tau),\nonumber\\
\ \ ^{q}\widehat{\sigma}(\tau)  &  =2\ \tau^{4}\int_{\widehat{\Xi}}\left(
4\pi\tau\right)  ^{-2}|\ \widehat{\mathbf{R}}_{\alpha\beta}%
+\widehat{\mathbf{D}}_{\alpha}\ \widehat{\mathbf{D}}_{\beta}\widehat{\zeta
}_{[1]}-\frac{1}{2\tau}\mathbf{g}_{\alpha\beta}|^{2}e^{-\widehat{\zeta}}
\ \delta\widehat{V}(\tau).\nonumber
\end{align}
Such nonmetric geometric thermodynamic variables were defined in
\cite{kazakh1} but for a different class of metric-affine theories. The
fluctuation variable$\ ^{q}\widehat{\sigma}(\tau)$ can be written as a
functional of $\ \widehat{\mathbf{R}}_{\alpha\beta}$ even $\ ^{q}%
\widehat{\mathcal{E}}\ (\tau)$ and $\ ^{q}\widehat{S}(\tau)$ are functionals
of $\widehat{f}(\widehat{\mathbf{R}}sc)$ if we correspondingly re-define the
normalizing functions $\widehat{\zeta}\rightarrow\widehat{\zeta}_{[1]}$. We
omit such details in this work because we shall not compute $\ ^{q}%
\widehat{\sigma}(\tau)$ for certain classes of solutions. Fixing the
temperature in (\ref{qthermvar}), we can compute thermodynamic variables
$\left[  \ ^{q}\widehat{\mathcal{E}}(\tau_{0}),\ ^{q}\widehat{\mathcal{S}%
}(\tau_{0}), \ ^{q}\widehat{\sigma}(\tau_{0})\right]  $ for nonmetric Ricci
solitons (\ref{canriccisol}). Certain classes of solutions can be not
well-defined in general form if, for instance, $\ ^{q}\widehat{\mathcal{S}%
}(\tau_{0})<0.$ We have to restrict some classes of nonholonomic
distributions/distortions in order to generate physically viable solutions. In
some spacetime regions, the nonmetric $Q$-deformations may result in
un-physical models, but be well-defined for another nonholonomic conditions
because of different sign contributions. We have to investigate this in
explicit form for corresponding classes of exact/parametric solutions of
physically important systems of nonlinear PDEs.

\section{Off-diagonal cosmological solutions encoding nonmetric geometric flows}

\label{sec4}Recent observational data provided by James Webb Space Telescope, JWST, indicate that the Standard Cosmological Model, SCM, could require correcting or even a substantial revision \cite{foroconi23,boylan23,biagetti23}. Such modifications can be modelled as different cosmological scenarios determined by (non) metric geometric flow
models and MGTs, or GR, with generic off-diagonal cosmological metrics. To be able to compare our nonholonomic approach with predictions of $\Lambda$CDM cosmological models involving nonmetricity fields we provide some basic
formulas for the SCM. Then, we generate and analyse more general classes of nonmetric geometric flow and cosmological solutions. The main goal of this section is to prove that by applying the AFCDM (see a summary of results and basic references in Appendix \ref{appendixa}) we can elaborate on more general classes of cosmological theories with nonmetric geometric flows and MGTs with off-diagonal interactions. This opens new possibilities for revising the theory of accelerating Universe and the DE and DM paradigms. As explicit examples, we construct and analyse basic properties of off-diagonal cosmological metrics with topological quasicrystal, QC, structure (main definitions are provided in Appendix \ref{appendixb}) determined in different cases by quasi-periodic nonmetric geometric flows, effective and real matter sources, and/or off-diagonal components of metrics.

\subsection{Diagonal metrics for $f(Q)$ cosmology}

\label{primeqcosm}We cite \cite{khyllep23,jim18,jhao22,de22} for reviews on
$f(Q)$ gravity and cosmology with modified Einstein equations (\ref{anmg}), or
(\ref{gfeq1a}), or (\ref{gfeq2a}). For such a nonmetric spacetime, a prime
off-diagonal metric $\mathbf{\mathring{g}=}[\mathring{g}_{\alpha}%
,\ \mathring{N}_{i}^{a}]$ (\ref{offdiagdefr}) is written using general curved
coordinates $u^{\alpha}(x,y,z,t)$ and considered as a conformal transform,
\begin{equation}
\mathbf{\mathring{g}}_{\alpha\beta}\simeq\mathring{a}^{-2}(t)\ ^{RW}%
\mathbf{g}_{\alpha\beta}, \label{confofdrw}%
\end{equation}
of the Friedmann-Lema\^{\i}tre-Robertson-Walker, FLRW, diagonal metric
\begin{equation}
d\mathring{s}^{2}=\ ^{RW}\mathbf{g}_{\alpha\beta}(u)\mathbf{e}^{\alpha
}\mathbf{e}^{\beta}\simeq\mathring{a}^{2}(t)(dx^{2}+dy^{2}+dz^{2})-dt^{2}.
\label{flrw}%
\end{equation}
In this formula, $t$ is the cosmic time, $\mathring{a}(t)$ is the scale
factor; and $x^{\grave{\imath}}=(x,y,z)$ are the Cartesian coordinates. A
metric (\ref{flrw}) defines a homogeneous, isotropic, and spatially flat
cosmological spacetime as a solution of the Einstein equations in GR.
$Q$-modifications of the energy-momentum tensor for matter are parameterized
in the form (\ref{emtc}) and generalized to effective tensors for DE
(determined as in the modified gravitational equations (\ref{gfeq2a}) and
(\ref{deemt})). Such metrics define $Q$-deformations to diagonal $f(Q)$
cosmological models determined by modified Friedman equations
\begin{align}
3\mathring{H}^{2}  &  =\mathring{\rho}+\frac{1}{2}[f(\mathring{Q}%
)-\mathring{Q}]-\mathring{Q}[f_{Q}(\mathring{Q})-1],\label{friedmeq}\\
\lbrack2\mathring{Q}(f_{QQ}(\mathring{Q})+f_{Q}(\mathring{Q}))]\mathring
{H}^{\diamond}  &  =\frac{1}{4}[f(\mathring{Q})-2\mathring{Q}+2\mathring
{Q}(1-f_{Q}(\mathring{Q}))]-\frac{1}{2}\mathring{p},\nonumber
\end{align}
for the Hubble function $\mathring{H}:=$ $\mathring{a}^{\diamond}/\mathring
{a}$ and $\mathring{Q}=6\mathring{H}^{2}$ for in the above equations, see the
conventions on partial derivatives for $Q$--modified coefficients of the Ricci
d-tensor (\ref{riccist2c}). The conservation law (i.e. the matter
equation-of-state) involving the prime energy density and pressure of matter
fluid (respectively, $\mathring{\rho}$ and $\mathring{p}$) is
\[
\mathring{\rho}^{\diamond}+3\mathring{H}(1+\mathring{w})\mathring{\rho}=0,
\]
for the parameter $\mathring{w}:=\mathring{p}/\mathring{\rho}.$ In this work,
we follow our system of notation which is different from those used in other
cosmological papers with diagonalizable metrics.

For applications in DE and DM physics, the $Q$-modified Friedman equation
(\ref{friedmeq}) are usually written in effective form
\[
\ ^{m}\mathring{\Omega}+\ ^{Q}\mathring{\Omega}=1,
\]
where the energy density parameters are introduced as
\[
\ ^{m}\mathring{\Omega} =\mathring{\rho}/3\mathring{H}^{2}\mbox{ and }
\ ^{Q}\mathring{\Omega} =[\frac{1}{2}(f(\mathring{Q})-\mathring{Q}%
)-\mathring{Q}(f_{Q}(\mathring{Q})-1)]/3\mathring{H}^{2}.
\]
Such parameters allow to introduce some (prime) effective energy density and
pressure, respectively, as%
\[
\ ^{eff}\mathring{\rho} =\mathring{\rho}+\frac{1}{2}[f(\mathring{Q}%
)-\mathring{Q}]-\mathring{Q}[f_{Q}(\mathring{Q})-1]\mbox{ and } \ ^{eff}%
\mathring{p} =\frac{\mathring{\rho}(1+\mathring{w})}{2\mathring{Q}%
f_{QQ}(\mathring{Q})+f_{Q}(\mathring{Q})}-\frac{\mathring{Q}}{2}.
\]
In such thermodynamic variables, the total equation of state is written
\[
\ ^{eff}\mathring{w}:=\frac{^{eff}\mathring{p}}{^{eff}\mathring{\rho}%
}=-1+\frac{\ ^{m}\mathring{\Omega}(1+\mathring{w})}{2\mathring{Q}%
f_{QQ}(\mathring{Q})+f_{Q}(\mathring{Q})},
\]
when $\ ^{eff}\mathring{w}<-1/3$ for an accelerated universe.

A dynamical system analysis using cosmological parameters $(\ ^{m}%
\mathring{\Omega},\ ^{Q}\mathring{\Omega},\ ^{eff}\mathring{w})$ for various
types of $f(\mathring{Q})$ was performed in section III of \cite{khyllep23}.
In this work, we elaborate on more general classes of nonmetric off-diagonal
cosmological solutions which only for very special assumptions can be
characterized by effective $(\ ^{m}\Omega,\ ^{Q}\Omega,\ ^{eff}w)$ determined
by nonmetric geometric flows. For generic off-diagonal cosmological models,
the concepts of above type thermodynamical models and related DM and DE
theories are not applicable. We shall provide in section \ref{sec5} new
examples which involve nonmetric generalizations of G. Perelman thermodynamics
as considered in section \ref{ssperelth}.

\subsection{Ansatz for $\tau$-running nonmetric cosmological spacetimes}

In this subsection, we study a class of locally anisotropic cosmological
solutions encoding in general form nonmetric $Q$-deformations and geometric
flows on a temperature like parameter $\tau.$ Such generic off-diagonal
metrics can be used for explaining recent observational data provided by JWST,
\cite{foroconi23,boylan23,biagetti23}, when the cosmological evolution and
structure formation of the Universe are modelled in very different forms
comparing to constructions in the frameworks of the SCM.

Nonmetric geometric flows of locally anisotropic cosmological solutions of the
system of nonlinear PDEs (\ref{cfeq4af}) can be constructed for generic an
off-diagonal ansatz\footnote{with respect to coordinate dual frames, such a
d-metric can be represented as $\mathbf{\hat{g}}=\underline{\widehat{g}%
}_{\alpha\beta}(\tau,u)du^{\alpha}\otimes du^{\beta},$ when the off-diagonal
metrics are parameterized in the form
\[
\underline{\widehat{g}}_{\alpha\beta}(\tau,u)=\left[
\begin{array}
[c]{cccc}%
e^{\psi}+(n_{1})^{2}h_{3}+(w_{1})^{2}h_{4} & n_{1}n_{2}h_{3}+w_{1}w_{2}h_{4} &
n_{1}h_{3} & w_{1}h_{4}\\
n_{1}n_{2}h_{3}+w_{1}w_{2}h_{4} & e^{\psi}+(n_{2})^{2}h_{3}+(w_{2})^{2}h_{4} &
n_{2}h_{3} & w_{2}h_{4}\\
n_{1}h_{3} & n_{2}h_{3} & h_{3} & 0\\
w_{1}h_{4} & w_{2}h_{4} & 0 & h_{4}%
\end{array}
\right]  ,
\]
where respective dependencies of coefficients on $(\tau,x^{k},y^{3})$ are
omitted. Such a metric is generic off-diagonal if the W-coefficients defined
in footnote \ref{fnwcoeff} are not trivial.}
\begin{align}
\mathbf{\hat{g}}(\tau)  &  =g_{i}(\tau,x^{k})dx^{i}\otimes dx^{i}+h_{3}%
(\tau,x^{k},t)\mathbf{e}^{3}(\tau)\otimes\mathbf{e}^{3}(\tau)+h_{4}(\tau
,x^{k},t)\mathbf{e}^{4}(\tau)\otimes\mathbf{e}^{4}(\tau),\nonumber\\
&  \mathbf{e}^{3}(\tau)=dy^{3}+n_{i}(\tau,x^{k},t)dx^{i},\ \mathbf{e}^{4}%
(\tau)=dy^{4}+w_{i}(\tau,x^{k},t)dx^{i}. \label{lacosmans}%
\end{align}
This is a d-metric of type (\ref{dm}) when the d-metric and N-connection
coefficients depend on h-coordinates $x^{i},$ do not depend on the space
coordinate $y^{3},$ but respective v-components generically depend as a
v-coordinate on $y^{4}=t$ and being parameterized in N-adapted form as
\begin{equation}
g_{i}(\tau)=e^{\psi{(\tau,x^{j})}},g_{a}(\tau)=h_{a}(\tau,x^{k},t),\ N_{i}%
^{3}=n_{i}(\tau,x^{k},t),\,\,\,\,N_{i}^{4}=w_{i}(\tau,x^{k},t).
\label{lacosmdmnc}%
\end{equation}
We can find the solutions in explicit parametric form if we consider effective
sources (\ref{dsourcparam}) determined by generating sources
\begin{equation}
\quad^{\intercal}\yen _{\ \nu}^{\mu}(\tau)=\ ^{c}\yen _{\ \nu}^{\mu}%
(\tau)=[\ _{h}\yen (\tau,{x}^{k})\delta_{j}^{i},~\ _{v}\ \yen (\tau
,x^{k},t)\delta_{b}^{a}], \label{dsourcosm}%
\end{equation}
where the left label "c" is used for cosmological configurations with
h-v-decomposition. We shall use such splitting of effective generating
v-sources:
\begin{equation}
~\ _{v}\ \yen (\tau,x^{k},t)=\ ^{m}\ \yen (\tau,x^{k},t)+\ ^{e}\ \yen (\tau
,x^{k},t)+\ ^{DE}\yen (\tau,x^{k},t), \label{emefparam}%
\end{equation}
associated to respective $\tau$-running matter terms $\ ^{m}T_{\alpha\beta}$
(\ref{emt}), effective sources $\ ^{e}\widehat{\mathbf{Y}}_{\alpha\beta}$
(\ref{ceemt}) determined by distortions of connections encoding additional
terms $\partial_{\tau}\mathbf{g}_{\mu\nu}$ as in (\ref{dsourcparam}); and
effective DE terms $\ ^{DE}T_{\alpha\beta}$ (\ref{deemt}) computed by
respective $Q$-deformations of gravitational Lagrangians. We shall not use
similar decompositions for $\ _{h}\yen (\tau,{x}^{k})$ because they contribute
only to the sources of some 2-d Poisson equations which are linear and can be
integrated in general form (see details in Appendix \ref{appendixa}). In
another turn, the terms (\ref{emefparam}) are determined in parametric form by
different physical constants which allows to compute different type physical
implications to nonlinear geometric evolution and dynamical interactions in
conventional v-subspaces.

\subsection{Parametrization of d-metrics and sources describing nonmetric QC
evolution}

There are four possibilities to generate topological QCs and elaborate models
of their nonmetric geometric evolution by cosmological d-metrics
(\ref{lacosmans}):

\begin{enumerate}
\item We can consider that the v-components a d-metric and respective
N-coefficients are generated as functionals of a quasi-periodic function
\begin{equation}
\ ^{\widehat{1}}\theta^{a}\simeq\theta^{I}(u)=\theta_{\lbrack0]}%
^{I}+\mathbf{K}^{I}\mathbf{x,} \label{qp1}%
\end{equation}
with $\theta_{\lbrack0]}^{I}$ and $I=b,$ for $b=3,4.$ Such values are taken
for a crystal like structure as we explain for topological QCs in Appendix
\ref{appendixb} (for formulas (\ref{condqc}). For instance, the d-metric
coefficients (\ref{lacosmdmnc}) can be defined by a functional
$\ ^{\widehat{1}}\Psi(\tau,x^{k},t)= \Psi\lbrack\ ^{\widehat{1}}\theta
^{a}[\tau,x^{k},t]]$ as for (\ref{nonmcosmla}), when the generating sources
$\ _{v}\ \yen (\tau,x^{k},t)$ (\ref{emefparam}) are arbitrary ones (i. e. the
components $\ ^{m}\ \yen ,\ ^{e}\ \yen ,$ and $\ ^{DE}\yen $ do not involve
topological QC configurations). Correspondingly, the formulas
(\ref{lacosmdmnc}) are considered as functionals
\begin{equation}
h_{a}[\ ^{\widehat{1}}\theta^{a}]=h_{a}[\ ^{\widehat{1}}\theta^{a}[\tau
,x^{k},t]],\ n_{i}[\ ^{\widehat{1}}\theta^{a}]=n_{i}[\ ^{\widehat{1}}%
\theta^{a}[\tau,x^{k},t]],\,w_{i}[\ ^{\widehat{1}}\theta^{a}]=w_{i}%
[\ \ ^{\widehat{1}}\theta^{a}[\tau,x^{k},t]], \label{genfqc}%
\end{equation}
where the left label $\widehat{1}$ will be used for emphasizing quasi-periodic
functional dependencies for generating functions of the coefficients of
d-metrics and N-coefficients.

\item Topological QC configurations can be generated also by effective sources
$\ ^{e}\widehat{\mathbf{Y}}_{\alpha\beta}$ (\ref{ceemt}) if we consider that
such N-adapted coefficients are related by frame/coordinate frame transforms
(and including in nonholonomic form effective terms of type $\partial_{\tau
}\mathbf{g}_{\mu\nu},$ see explanations for formulas (\ref{dsourcparam})) to
some
\[
\ ^{e}\widehat{\mathbf{T}}_{\alpha\beta}:=-\frac{2}{\sqrt{|\mathbf{g}|}}%
\frac{\delta(\sqrt{|\mathbf{g}|}\ ^{e}\widehat{\mathcal{L}})}{\delta
\mathbf{g}^{\alpha\beta}}=\ ^{e}\widehat{\mathcal{L}}\mathbf{g}_{\alpha\beta
}+2\frac{\delta(\ ^{e}\widehat{\mathcal{L}})}{\delta\mathbf{g}^{\alpha\beta}%
}.
\]
Such a d-tensor is computed in N-adapted variational form as (\ref{emt}) but
for an effective quasi-periodic Lagrangian $\ ^{e}\widehat{\mathcal{L}%
}=\ ^{\theta2}\widehat{\mathcal{L}}$ (\ref{topqc2}), or $\ ^{e}%
\widehat{\mathcal{L}}=\ ^{\Theta2}\widehat{\mathcal{L}}$ (\ref{topqc2a}). In
functional form, we parameterize the corresponding topological QC generating
sources as
\begin{equation}
\ ^{e}\widehat{\mathbf{Y}}_{\alpha\beta}[\ \ ^{\widehat{2}}\theta^{a}%
]=\ ^{e}\widehat{\mathbf{Y}}_{\alpha\beta}[\ \ ^{\widehat{2}}\theta^{a}%
[\tau,x^{k},t]]\leftrightarrow\ ^{e}\Lambda, \label{gensourcqc2}%
\end{equation}
where the label $\widehat{2}$ emphasize that we consider quasi-periodic
structures related to canonical nonholonomic distortions of d-connections.

\item Quasi-periodic structures can be generated by effective DE sources
$\ ^{DE}\yen (\tau,x^{k},t)$ (\ref{deemt}) determined by $Q$-deformations of
sources. In N-adapted form, we write $\ ^{DE}\widehat{\mathbf{T}}_{\alpha
\beta}$ for generating source functionals in (\ref{dsourcparam}) and consider
DE generating sources as functionals
\begin{equation}
\ ^{DE}\widehat{\mathbf{Y}}_{\alpha\beta}[\ \ ^{\widehat{3}}\theta
^{a}]=\ ^{DE}\widehat{\mathbf{Y}}_{\alpha\beta}[\ \ ^{\widehat{3}}\theta
^{a}[\tau,x^{k},t]]\leftrightarrow\ ^{DE}\Lambda. \label{gensourcqc3}%
\end{equation}
We use the label $\ \widehat{3}$ for distinguishing quasi-periodic structures
of nonmetric origin.
\end{enumerate}

Applying nonlinear transforms (\ref{ntransf1}) and (\ref{ntransf2}), the quasi-periodic generating sources (\ref{gensourcqc2}) and (\ref{gensourcqc3}) can be transformed into respective effective cosmological constants
$\ ^{e}\Lambda$ and $\ ^{DE}\Lambda$ as in (\ref{effcosmconst}). This allows us to compute and distinguish contributions of different type quasi-periodic structures, which can be of distortion of connections and (non) metric geometric flow origin and/or of $Q$-deformation type. Because of nonlinear off-diagonal geometric evolution and gravitational and (effective) matter field interactions, considering nonmetric geometric data for $(\ ^{m}\Lambda\ ^{e}\Lambda,\ ^{DE}\Lambda),$ the coefficients of d-metrics and N-connections (in general frames of reference) became functionals on all types
$\ ^{\widehat{q}}\theta^{a},$ for $q=1,2,3,$; when (\ref{genfqc}) are re-parameterized as
\begin{equation}
h_{a}[\ ^{\widehat{q}}\theta^{a}]=h_{a}[\ ^{\widehat{q}}\theta^{a}[\tau
,x^{k},t]],\ n_{i}[\ ^{\widehat{q}}\theta^{a}]=n_{i}[\ ^{\widehat{q}}%
\theta^{a}[\tau,x^{k},t]],\,w_{i}[\ ^{\widehat{q}}\theta^{a}]=w_{i}%
[\ \ ^{\widehat{q}}\theta^{a}[\tau,x^{k},t]]. \label{genfqcparam}%
\end{equation}
Even for such general assumptions on generating data, we can construct exact and parametric solutions of physically important systems of nonlinear PDEs (\ref{cfeq4a})--(\ref{canriccisol}) if we apply the AFCDM summarized in
Appendix \ref{appendixa}. Such generic off-diagonal metrics and generalized (non) linear connections describe the generation and nonmetric flow evolution of cosmological topological QC structure, which may be applied in modern accelerating cosmology and DE and DM physics. In this section, we provide explicit examples and discuss the main physical properties which may be important for explaining recent observational data provided by JWST, \cite{foroconi23,boylan23,biagetti23}.

\subsection{Generic non $\Lambda$CDM cosmological solutions with $\tau$-running QC generating functions}

Such generic off-diagonal cosmological solutions of (\ref{cfeq4a}) are of type
(\ref{genfqc}) and generated by a functional $\ ^{\widehat{1}}\Psi=\Psi
\lbrack\ ^{\widehat{1}}\theta^{a}[\tau,x^{k},t]]$ introduced in $\tau$-family
of quadratic elements (\ref{nonmcosmla}), when
\begin{align}
ds^{2}(\tau)  &  = e^{\psi(\tau,x^{k})}[(dx^{1})^{2}+(dx^{2})^{2}%
]+\{g_{3}^{[0]}-\int dt\frac{[\ ^{\widehat{1}}\Psi^{2}]^{\diamond}}%
{4(\ _{v}\ \yen )}\}\label{nonmcosmla1}\\
&  \{dy^{3}+[\ _{1}n_{k}+\ _{2}n_{k}\int dt\frac{[(\ ^{\widehat{1}}\Psi
)^{2}]^{\diamond}}{4(\ _{v}\ \yen )^{2}|g_{3}^{[0]}-\frac{\int
dt[\ ^{\widehat{1}}\Psi^{2}]^{\diamond}}{4(\ _{v}\ \yen )|^{5/2}}}]dx^{k}\}
+\frac{[\ ^{\widehat{1}}\Psi^{\diamond}]^{2}}{4(\ _{v}\ \yen )^{2}%
\{g_{3}^{[0]}-\frac{\int dt[\ ^{\widehat{1}}\Psi^{2}]^{\diamond}}%
{4(\ _{v}\ \yen )\}}}(dt+\frac{\partial_{i}\ ^{\widehat{1}}\Psi}%
{\ ^{\widehat{1}}\Psi^{\diamond}}dx^{i})^{2}.\nonumber
\end{align}
The nonmetric geometric cosmological evolution described by (\ref{nonmcosmla1}%
) posses a topological QC structure even the effective sources $\ _{v}%
\ \yen =\ ^{m}\ \yen +\ ^{e}\ \yen +\ ^{DE}\yen $ (\ref{emefparam}) can be
some general ones not involving a quasi-periodic structure.

The nonlinear symmetries (\ref{ntransf1}) allow us to transform the generating
data
\begin{align}
&  (\Psi\lbrack\ ^{\widehat{1}}\theta^{a}[\tau,x^{k},t]],_{v}\ \yen (\tau
))\rightarrow(\ ^{\widehat{1}}h_{3}=h_{3}[\ ^{\widehat{1}}\theta^{a}%
[\tau,x^{k},t]],\ _{v}\ \yen (\tau,x^{k},t)),\nonumber\\
& \mbox{when} \ ^{\widehat{1}}h_{3}^{\diamond}(\tau)=-[(\ ^{\widehat{1}}%
\Psi)^{2}]^{\diamond}/4\ _{v}\ \yen (\tau). \label{auxform1}%
\end{align}
The solutions (\ref{nonmcosmla1}) can be written in the form
(\ref{offdsolgenfgcosmc}), when $h_{3}\rightarrow\ ^{\widehat{1}}h_{3}$ and
$\ ^{\widehat{1}}h_{3}$ is a generating functional for topological QC
structure $\ ^{\widehat{1}}\theta^{a}[\tau,x^{k},t]$ determining the
quasi-periodicity of the $v$-components of the d-metric and N-connection coefficients.

Using formula (\ref{auxform1}) and
\[
(\ ^{\widehat{1}}\Phi\lbrack\ ^{\widehat{1}}\theta^{a}[\tau,x^{k}%
,t]])^{2}=-4\widehat{\Lambda}(\tau)\ ^{\widehat{1}}h_{3}[\ ^{\widehat{1}%
}\theta^{a}[\tau,x^{k},t]], \mbox{ for }\widehat{\Lambda}(\tau)=\ ^{m}%
\Lambda(\tau)+\ ^{e}\Lambda(\tau)+\ ^{DE}\Lambda(\tau)
\]
(derived for nonlinear symmetries (\ref{ntransf1}) and (\ref{ntransf2}) and
for splitting of the effective $\tau$-running cosmological constant
(\ref{effcosmconst}) for (\ref{offdsolgenfgcosmc})), we re-write the d-metric
(\ref{nonmcosmla1}) in the form (\ref{offdiagcosmcsh})), when
\begin{align}
ds^{2}(\tau) =  &  g_{\alpha\beta}(\tau,x^{k},t,\ ^{\widehat{1}}\Phi
(\tau),\widehat{\Lambda}(\tau))du^{\alpha}du^{\beta}\nonumber\\
=  &  e^{\psi(\tau,x^{k})}[(dx^{1})^{2}+(dx^{2})^{2}]-\{g_{3}^{[0]}%
(\tau)-\frac{(\ ^{\widehat{1}}\Phi)^{2}(\tau)}{4\ \widehat{\Lambda}(\tau
)}\}\ \label{offdiagcosmcsh2}\\
&  \{dy^{3}+[\ _{1}n_{k}(\tau)+\ _{2}n_{k}(\tau)\int dt\frac{(\ ^{\widehat{1}%
}\Phi)^{2}(\tau)[(\ ^{\widehat{1}}\Phi)^{\diamond}(\tau)]^{2}}%
{|\ \widehat{\Lambda}(\tau)\int dt\ \ _{v}\ \yen (\tau)\ [(\ ^{\widehat{1}%
}\Phi)^{2}(\tau)]^{\diamond}|[g_{3}^{[0]}(\tau)-\frac{(\ ^{\widehat{1}}%
\Phi)^{2}(\tau)}{4\ \widehat{\Lambda}(\tau)}]^{5/2}}]\}\nonumber
\end{align}%
\[
-\frac{(\ ^{\widehat{1}}\Phi)^{2}(\tau)[(\ ^{\widehat{1}}\Phi)^{\diamond}%
(\tau)]^{2}}{|\ \widehat{\Lambda}(\tau)\int dt\ \ _{v}\ \yen (\tau
)[(\ ^{\widehat{1}}\Phi)^{2}(\tau)]^{\diamond}\ |[g_{3}^{[0]}(\tau
)-\frac{(\ ^{\widehat{1}}\Phi)^{2}(\tau)}{4\ \widehat{\Lambda}(\tau)}%
]}\{dt+\frac{\partial_{i}\ \int dt\ \ \ _{v}\ \yen (\tau)\ [(\ ^{\widehat{1}%
}\Phi)^{2}(\tau)]^{\diamond}}{\ \ _{v}\ \yen (\tau)\ [(\ (\ ^{\widehat{1}}%
\Phi)(\tau))^{2}]^{\diamond}}dx^{i}\}^{2}.
\]
This $\tau$-family of cosmological solutions is determined by generating data
$(\ ^{\widehat{1}}\Phi\lbrack\ ^{\widehat{1}}\theta^{a}[\tau,x^{k}%
,t]];\ ^{m}\Lambda(\tau)+ \ ^{e}\Lambda(\tau)+\ ^{DE}\Lambda(\tau
),\ _{v}\ \yen (\tau))$ and allow to distinguish and compare models with
different types of $\tau$-running effective cosmological constants. Such
constants approximate the contributions of standard matter fields, effective
distortions and $Q$-deformations. In general, we are not able to eliminate the
effective generating source $\ _{v}\ \yen (\tau)$ but we can always chose
certain nonholonomic configurations when the terms with some $\ (\ ^{m}%
\ \yen (\tau),\ ^{e}\ \yen (\tau),\ ^{DE}\yen (\tau))$ are smaller than the
terms with corresponding $(\ ^{m}\Lambda(\tau),\ ^{e}\Lambda(\tau
),\ ^{DE}\Lambda(\tau)).$ Fixing $\tau=\tau_{0},$ we define generic
off-diagonal cosmological solutions in nonmetric MGTs.

The $\tau$-running d-metrics (\ref{nonmcosmla1}) allow us to model cosmological scenarios with topological QC structure induced by generating functions when the effective matter sources$\ _{v}\ \yen (\tau)$ are not obligatory quasi-periodic. The priority of parametrization of solutions written in the form (\ref{offdiagcosmcsh2}) involving effective cosmological
constants is that we can compute corresponding thermodynamic variables for generating topological QC configurations (see section \ref{sec5}). Using such variables, we can distinguish contributions of nonholonomic distortions and off-diagonal terms, nonmetricity fields and metric fields and search for observational data which allows to decide which theory is more realistic and viable. Corresponding nonmetric generic off-diagonal cosmological scenarios are very different from those elaborated in the framework of the $\Lambda$CDM cosmological paradigm.

\subsection{$\tau$-running cosmology with topological QC generating sources}

Generic off-diagonal cosmological metrics can be generated by nonmetric
topological QC sources of type (\ref{gensourcqc2}) and/or (\ref{gensourcqc3})
when the generating functions $\Psi(\tau),\Phi(\tau)$ and $h_{3}(\tau)$ may
encode, or not, quasi-periodic structures. In this subsection, we provide a
modification of (\ref{offdiagcosmcsh2}) with $v$-components of d-metric and
N-connection coefficients of type (\ref{genfqcparam}), when
\begin{align}
ds^{2}(\tau)=  &  g_{\alpha\beta}(\tau,x^{k},t,\ ^{\widehat{1}}\Phi
(\tau),\widehat{\Lambda}(\tau))du^{\alpha}du^{\beta}\nonumber\\
=  &  e^{\psi(\tau,x^{k})}[(dx^{1})^{2}+(dx^{2})^{2}]-\{g_{3}^{[0]}%
(\tau)-\frac{(\ ^{\widehat{q}}\Phi)^{2}(\tau)}{4\ \widehat{\Lambda}(\tau
)}\}\{dy^{3}+[\ _{1}n_{k}(\tau)+\ \label{offdiagcosmcsh3}\\
&  \ _{2}n_{k}(\tau)\int dt\frac{(\ ^{\widehat{q}}\Phi)^{2}(\tau
)[(\ ^{\widehat{q}}\Phi)^{\diamond}(\tau)]^{2}}{|\ \widehat{\Lambda}(\tau)\int
dt\ (\ ^{m}\ \yen (\tau)+\ ^{e}\ \yen [\ ^{\widehat{2}}\theta^{a}%
[\tau]]+\ ^{DE}\yen [\ ^{\widehat{3}}\theta^{a}[\tau]])[(\ ^{\widehat{q}}%
\Phi)^{2}(\tau)]^{\diamond}|[g_{3}^{[0]}(\tau)-\frac{(\ ^{\widehat{q}}%
\Phi)^{2}(\tau)}{4\ \widehat{\Lambda}(\tau)}]^{5/2}}]\}\nonumber
\end{align}%
\begin{align*}
&  -\frac{(\ ^{\widehat{q}}\Phi)^{2}(\tau)[(\ ^{\widehat{q}}\Phi)^{\diamond
}(\tau)]^{2}}{|\ \widehat{\Lambda}(\tau)\int dt\ \ (\ ^{m}\ \yen (\tau
)+\ ^{e}\ \yen [\ \ ^{\widehat{2}}\theta^{a}[\tau]]+\ ^{DE}%
\yen [\ ^{\widehat{3}}\theta^{a}[\tau]])[(\ ^{\widehat{q}}\Phi)^{2}%
(\tau)]^{\diamond}\ |[g_{3}^{[0]}(\tau)-\frac{(\ ^{\widehat{q}}\Phi)^{2}%
(\tau)}{4\ \widehat{\Lambda}(\tau)}]}\\
&  \{dt+\frac{\partial_{i}\ \int dt\ \ (\ ^{m}\ \yen (\tau)+\ ^{e}%
\ \yen [\ \ ^{\widehat{2}}\theta^{a}[\tau]]+\ ^{DE}\yen [\ ^{\widehat{3}%
}\theta^{a}[\tau]])\ [(\ ^{\widehat{q}}\Phi)^{2}(\tau)]^{\diamond}}%
{(\ ^{m}\ \yen (\tau)+\ ^{e}\ \yen [\ ^{\widehat{2}}\theta^{a}[\tau
]]+\ ^{DE}\yen [\ ^{\widehat{3}}\theta^{a}[\tau]])[\ (\ ^{\widehat{q}}%
\Phi)^{2}(\tau)]^{\diamond}}dx^{i}\}^{2}.
\end{align*}
The $\tau$-family of cosmological solutions (\ref{offdiagcosmcsh3}) is
determined by generating data
\begin{equation}
(\ ^{\widehat{q}}\Phi\lbrack\ ^{\widehat{q}}\theta^{a}[\tau,x^{k}%
,t]];\ \widehat{\Lambda}(\tau)=\ ^{m}\Lambda(\tau)+\ ^{e}\Lambda(\tau
)+\ ^{DE}\Lambda(\tau),\ _{v}\ \yen (\tau)), \label{gendata1}%
\end{equation}
which allows us to compute the corresponding G. Perelman's thermodynamic
variables (\ref{qthermvar}) (see examples in section \ref{sec5}). In general
forms, such nonmetric cosmological models can't be embedded into the framework
of the $\Lambda$CDM cosmological paradigm. Such a general behaviour holds true
even we consider additional constraints for extracting LC-configurations as in
appendix \ref{asslcconf} and, for further restrictions to locally anisotropic
cosmological models in GR (with vanishing $\ ^{DE}\yen (\tau)\,$ and
$\ ^{DE}\Lambda(\tau)$). Nevertheless, even in such special cases, topological
QCs cosmological configurations can be generated because of generic
off-diagonal terms of d-metrics and related N-connection coefficients.

\subsection{Topological QC deformations of FLRW metrics under nonmetric
geometric flows}

Using the AFCDM, we can study how nonmetric geometric flows may result in
deformation of certain prime off-diagonal metric $\mathbf{\mathring{g}%
=}[\mathring{g}_{\alpha},\ \mathring{N}_{i}^{a}]$ (\ref{offdiagdefr}) together
with the formation of topological QC structure. An important physical example
is to consider that $\mathbf{\mathring{g}}_{\alpha\beta}\simeq\mathring
{a}^{-2}(t)\ ^{RW}\mathbf{g}_{\alpha\beta}$ as in (\ref{confofdrw}) and then
to generate parametric solutions of (\ref{cfeq4af}) defined by gravitational
$\eta$-polarizations (\ref{offdiagdefr}). Considering a generating function of
type (\ref{etapolgen}) $\eta_{3}\ (\tau)\simeq\ ^{\widehat{q}}\eta_{3}%
(\tau)=\eta_{3}[\ ^{\widehat{q}}\theta^{a}]$ as in (\ref{genfqcparam}) (in
particular cases, we can consider a functional dependence of type
(\ref{auxform1})) and generating sources of topological QC type
(\ref{gensourcqc2}) and/or (\ref{gensourcqc3}), we generate $\tau$-families of
cosmological d-metrics (\ref{offdncelepsilon}), when
\begin{align}
d\widehat{s}^{2}(\tau)=  &  \widehat{g}_{\alpha\beta}(\tau,x^{k}%
,t;\mathring{g}_{\alpha};\psi(\tau),\ ^{\widehat{q}}\eta_{3}(\tau
);(\ ^{m}\ \yen (\tau)+\ ^{e}\ \yen [\ ^{\widehat{2}}\theta^{a}[\tau
]]+\ ^{DE}\yen [\ ^{\widehat{3}}\theta^{a}[\tau]]))du^{\alpha}du^{\beta
}\nonumber\\
=  &  e^{\psi(\tau)}[(dx^{1})^{2}+(dx^{2})^{2}]+\ ^{\widehat{q}}\eta_{3}%
(\tau)\mathring{g}_{3}\{dy^{3}+[\ _{1}n_{k}(\tau)+\label{offdiagcosmcshqc}\\
&  \ _{2}n_{k}(\tau)\int dt\frac{[(\ ^{\widehat{q}}\eta_{3}(\tau)\mathring
{g}_{3})^{\diamond}]^{2}}{|\int dt\ (\ ^{m}\ \yen (\tau)+\ ^{e}%
\ \yen [\ ^{\widehat{2}}\theta^{a}[\tau]]+\ ^{DE}\yen [\ ^{\widehat{3}}%
\theta^{a}[\tau]])(\ ^{\widehat{q}}\eta_{3}(\tau)\mathring{g}_{3})^{\diamond
}|\ (\ ^{\widehat{q}}\eta_{3}(\tau)\mathring{g}_{3})^{5/2}}]dx^{k}%
\}^{2}\nonumber\\
&  -\frac{[(\ ^{\widehat{q}}\eta_{3}(\tau)\ \mathring{g}_{3})^{\diamond}]^{2}%
}{|\int dt\ \ _{v}\ (\ ^{m}\ \yen (\tau)+\ ^{e}\ \yen [\ ^{\widehat{2}}%
\theta^{a}[\tau]]+\ ^{DE}\yen [\ ^{\widehat{3}}\theta^{a}[\tau
]])(\ ^{\widehat{q}}\eta_{3}(\tau)\ \mathring{g}_{3})^{\diamond}%
|\ \ ^{\widehat{q}}\eta_{3}(\tau)\mathring{g}_{3}}\nonumber\\
&  \{dt+\frac{\partial_{i}[\int dt\ \ _{v}\ (\ ^{m}\ \yen (\tau)+\ ^{e}%
\ \yen [\ ^{\widehat{2}}\theta^{a}[\tau]]+\ ^{DE}\yen [\ ^{\widehat{3}}%
\theta^{a}[\tau]])\ (\ ^{\widehat{q}}\eta_{3}(\tau)\mathring{g}_{3}%
)^{\diamond}]}{\ (\ ^{m}\ \yen (\tau)+\ ^{e}\ \yen [\ ^{\widehat{2}}\theta
^{a}[\tau]]+\ ^{DE}\yen [\ ^{\widehat{3}}\theta^{a}[\tau]])(\ ^{\widehat{q}%
}\eta_{3}(\tau)\mathring{g}_{3})^{\diamond}}dx^{i}\}^{2}.\nonumber
\end{align}
For $\ (\ ^{\widehat{q}}\Phi\lbrack\ ^{\widehat{q}}\theta^{a}[\tau
,x^{k},t]])^{2}=-4\ \ \widehat{\Lambda}\eta_{3}[\ ^{\widehat{q}}\theta
^{a}]\mathring{g}_{\alpha},$ we can transform (\ref{offdiagcosmcshqc}) in a
variant of (\ref{offdiagcosmcsh3}), or (\ref{offdiagpolfr}), with $\eta
$-polarizations determined by the generating data
\begin{equation}
\left(  \ ^{\widehat{q}}h_{3}(\tau)=\eta_{3}[\ ^{\widehat{q}}\theta
^{a}]\mathring{g}_{\alpha};\ ^{m}\ \yen (\tau)+\ ^{e}\ \yen [\ ^{\widehat{2}%
}\theta^{a}[\tau]]+\ ^{DE}\yen [\ ^{\widehat{3}}\theta^{a}[\tau
]];\ \widehat{\Lambda}\ (\tau)=\ ^{m}\Lambda(\tau)+\ ^{e}\Lambda(\tau
)+\ ^{DE}\Lambda(\tau)\right)  . \label{gendata2}%
\end{equation}
The primary cosmological data define a $\Lambda$CDM model with $Q$%
-deformations as we considered in subsection \ref{primeqcosm}. More general
classes of solutions (\ref{offdiagcosmcshqc}) deform such metrics in generic
off-diagonal and describe arising of topological QC structures encoding via
polarization functions nonmetric geometric flows and corresponding effective
QC-sources. The physical interpretation of target classes of solutions is
possible in terms of statistical/ geometric thermodynamic variables
(\ref{qthermvar}). Such $\tau$-evolving locally anisotropic cosmological
models are very different from the theories of $\Lambda$CDM type. Choosing
corresponding parameters for the generating quasi-periodic structures (in
general, they are not just of QC type but subjected to additional deformations
because of random off-diagonal interactions and nonmetric geometric
evolution), we may try to explain observational data provided by JWST
\cite{foroconi23,boylan23,biagetti23}.

\subsection{$\Lambda$CDM configurations and small parametric deformations
inducing QC structures}

We can construct generic off-diagonal solutions for nonmetric geometric flows
defining topological QC structures which can be analyzed in the framework of
the $\Lambda$CDM paradigm but with nontrivial G. Perelman thermodynamic
variables. This is possible if we apply the AFCDM for a small $\varepsilon
$-parameter and respective $\chi$-polarizations as defined by formulas
(\ref{dataepsion}), (\ref{nonlinsymrex}) and (\ref{epsilongenfdecomp})
$\chi_{3}(\tau,x^{k},t)\simeq\ ^{\widehat{q}}\chi_{3}(\tau)=\chi
_{3}[\ ^{\widehat{q}}\theta^{a}]$ as a generating function with quasi-periodic
structure (as in (\ref{genfqcparam}). For some examples, we can use a
functional dependence of type (\ref{auxform1})) and generating sources of
topological QC type (\ref{gensourcqc2}) and/or (\ref{gensourcqc3}). For such
$\varepsilon$-linear decompositions, we obtain a small parametric version of
type (\ref{offdncelepsilon}) for the d-metrics (\ref{offdiagcosmcshqc}),
\begin{align*}
d\ \widehat{s}^{2}(\tau)  &  =\widehat{g}_{\alpha\beta}(\tau,x^{k}%
,t;\mathring{g}_{\alpha};\psi(\tau),\ ^{\widehat{q}}\chi_{3}(\tau
);(\ \ ^{\widehat{q}}\yen (\tau)=\ ^{m}\ \yen (\tau)+\ ^{e}%
\ \yen [\ ^{\widehat{2}}\theta^{a}[\tau]]+\ ^{DE}\yen [\ ^{\widehat{3}}%
\theta^{a}[\tau]]))du^{\alpha}du^{\beta}\\
&  =e^{\psi_{0}}(1+\varepsilon\ ^{\psi}\chi(\tau))[(dx^{1})^{2}+(dx^{2}%
)^{2}]+\zeta_{3}(\tau)(1+\varepsilon\ \ ^{\widehat{q}}\chi_{3}(\tau
))\ \mathring{g}_{3}%
\end{align*}%
\begin{align}
&  \{dy^{3}+[(\mathring{N}_{k}^{3})^{-1}[\ _{1}n_{k}(\tau)+16\ _{2}n_{k}%
(\tau)[\int dt\frac{\left(  [(\zeta_{3}(\tau)\mathring{g}_{3})^{-1/4}%
]^{\diamond}\right)  ^{2}}{|\int dt[\ ^{\widehat{q}}\yen (\tau)(\zeta_{3}%
(\tau)\mathring{g}_{3})]^{\diamond}|}]\label{offdncelepsilonq}\\
&  +\varepsilon\frac{16\ _{2}n_{k}(\tau)\int dt\frac{\left(  [(\zeta_{3}%
(\tau)\mathring{g}_{3})^{-1/4}]^{\diamond}\right)  ^{2}}{|\int
dt[\ ^{\widehat{q}}\yen (\tau)(\zeta_{3}(\tau)\mathring{g}_{3})]^{\diamond}%
|}(\frac{[(\zeta_{3}(\tau)\mathring{g}_{3})^{-1/4}\chi_{3})]^{\diamond}%
}{2[(\zeta_{3}(\tau)\mathring{g}_{3})^{-1/4}]^{\diamond}}+\frac{\int
dt[\ ^{\widehat{q}}\yen (\tau)(\zeta_{3}(\tau)\ ^{\widehat{q}}\chi_{3}%
(\tau)\mathring{g}_{3})]^{\diamond}}{\int dt[\ ^{\widehat{q}}\yen (\tau
)(\zeta_{3}(\tau)\mathring{g}_{3})]^{\diamond}})}{\ _{1}n_{k}(\tau
)+16\ _{2}n_{k}(\tau)[\int dt\frac{\left(  [(\zeta_{3}(\tau)\mathring{g}%
_{3})^{-1/4}]^{\diamond}\right)  ^{2}}{|\int dt[\ ^{\widehat{q}}%
\yen (\tau)(\zeta_{3}(\tau)\mathring{g}_{3})]^{\diamond}|}]}]\mathring{N}%
_{k}^{3}dx^{k}\}^{2}.\nonumber
\end{align}%
\begin{align*}
&  -\{\frac{4[(|\zeta_{3}(\tau)\mathring{g}_{3}|^{1/2})^{\diamond}]^{2}%
}{\mathring{g}_{3}|\int dt\{\ ^{\widehat{q}}\yen (\tau)(\zeta_{3}%
(\tau)\mathring{g}_{3})^{\diamond}\}|}-\varepsilon\lbrack\frac
{(\ ^{\widehat{q}}\chi_{3}(\tau)|\zeta_{3}(\tau)\mathring{g}_{3}%
|^{1/2})^{\diamond}}{4(|\zeta_{3}(\tau)\mathring{g}_{3}|^{1/2})^{\diamond}%
}-\frac{\int dt\{\ ^{\widehat{q}}\yen [(\zeta_{3}(\tau)\mathring{g}%
_{3})\ ^{\widehat{q}}\chi_{3}(\tau)]^{\diamond}\}}{\int dt\{\ ^{\widehat{q}%
}\yen (\tau)(\zeta_{3}(\tau)\mathring{g}_{3})^{\diamond}\}}]\}\mathring{g}%
_{4}\\
&  \{dt+[\frac{\partial_{i}\ \int dt\ \ ^{\widehat{q}}\yen (\tau)\zeta
_{3}^{\diamond}(\tau)}{(\mathring{N}_{i}^{3})\ \ ^{\widehat{q}}\yen (\tau
)\zeta_{3}^{\diamond}(\tau)}+\varepsilon(\frac{\partial_{i}[\int
dt\ \ ^{\widehat{q}}\yen (\tau)\ (\zeta_{3}(\tau)\ ^{\widehat{q}}\chi_{3}%
(\tau))^{\diamond}]}{\partial_{i}\ [\int dt\ \ ^{\widehat{q}}\yen (\tau
)\zeta_{3}^{\diamond}(\tau)]}-\frac{(\zeta_{3}(\tau)\ ^{\widehat{q}}\chi
_{3}(\tau))^{\diamond}}{\zeta_{3}^{\diamond}(\tau)})]\mathring{N}_{i}%
^{4}dx^{i}\}^{2}.
\end{align*}
We can re-define such small parametric formulas in terms of generating data
$(\ ^{\widehat{q}}\Phi(\tau),\ \widehat{\Lambda}(\tau)),$ when
\begin{equation}
(\ ^{\widehat{q}}\Phi\lbrack\ ^{\widehat{q}}\theta^{a}[\tau,x^{k}%
,t]])^{2}=-4\ \ \widehat{\Lambda}\zeta_{3}(\tau,x^{k},t)(1+\varepsilon
\ ^{\widehat{q}}\chi_{3}(\tau,x^{k},t))[\ ^{\widehat{q}}\theta^{a}%
]\mathring{g}_{\alpha}, \label{gendata3}%
\end{equation}
for $\ \widehat{\Lambda}(\tau)=\ ^{m}\Lambda(\tau)+\ ^{e}\Lambda(\tau
)+\ ^{DE}\Lambda(\tau).$ This allows to compute the G. Perelman thermodynamic
variables as in next section.

The off-diagonal parametric solutions (\ref{offdncelepsilonq}) can be parameterized, for instance, to define ellipsoidal deformations of spherical symmetric cosmological metrics into similar ones with ellipsoid symmetry or in certain N-adapted effective diagonal forms. Quasi-stationary solutions with nonmetricity fields are provided in a partner work \cite{kazakh1} for generating functions of type $\chi_{3}(\tau,x^{k},\varphi),$ with $x^{k}$ and $\varphi$ being space like coordinates and $\varepsilon$ considered as an eccentricity parameter for some rotoid type deformations. Similar techniques can be applied if $\chi_{3}(\tau,x^{k},\varphi)\rightarrow\ ^{\widehat{q}}\chi_{3}(\tau,x^{k},t)$ in order to generate ellipsoidal cosmological solutions encoding $Q$-deformations and quasi-periodic structures (for non-topological QCs, see \cite{bubuianu17,sv18}. We do not consider examples of such incremental work re-defining the constructions for topological QC configurations in this paper.

\subsection{$\tau$-evolution of nonmetric effective $\Lambda$CDM models}

New $\tau$-families of generic off-diagonal cosmological solutions of type (\ref{nonmcosmla1}), (\ref{offdiagcosmcsh2}), (\ref{offdiagcosmcsh3}), (\ref{offdiagcosmcshqc}) and (\ref{offdncelepsilonq}) motivate a new paradigm for elaborating cosmological models which describe nonlinear quasi-periodic structure formation. This way, new models of DE and DM theories can be elaborated that may encode possible nonmetric geometric evolution and various types of MGTs. Nevertheless, such modified cosmological models my possess some properties which can be described in the framework of the $\Lambda$CDM paradigm. In this subsection, we analyze such conditions for a prime metric $\mathbf{\mathring{g}}_{\alpha\beta}$ (\ref{confofdrw}) (being conformal to a FLRW metric (\ref{flrw}), in nonlinear coordinates) and transformed via frame/ coordinate transforms into a target metric (\ref{offdiagcosmcshqc}) or
(\ref{offdncelepsilonq}) and parameterized in the form
\begin{equation}
ds^{2}=\ ^{\widehat{q}}\mathbf{g}_{\alpha\beta}(\tau,t)\mathbf{e}^{\alpha
}\mathbf{e}^{\beta}\simeq\ ^{\widehat{q}}a^{2}(\tau,t)[dx^{2}+dy^{2}%
+(dz+\ ^{\widehat{q}}w_{i}(\tau,t)dx^{i})^{2}]-[dt+\ ^{\widehat{q}}n_{i}%
(\tau,t)dx^{i}]^{2}. \label{flrwqc}%
\end{equation}
In (\ref{flrwqc}), $t$ is the cosmic time, $x^{\grave{\imath}}=(x,y,z)$ are the Cartesian coordinates; 
$\ ^{\widehat{q}}a(\tau,t)\simeq a([\ ^{\widehat{q}}\theta^{a}(\tau,x,y,t)])$ defines a topological type QC scale factor, when $\ ^{\widehat{q}}w_{i}(\tau,t)$ and $\ ^{\widehat{q}}n_{i}(\tau,t)$ are proportional to a $\varepsilon$-parameter and can be stated to zero by choosing certain integration functions $\ _{1}n_{k}(\tau)$ and $\ _{2}n_{k}(\tau)$ and the generating functions are parameterized in a form to get small $\ ^{\widehat{q}}w_{i}(\tau,t).$

We suppose that an effective diagonal metric (in the chosen system of
reference and coordinated) defines a model of $\tau$-evolution of effective
homogeneous, isotropic, and spatially flat cosmological spacetimes, which for
a fixed temperature like parameter $\tau=\tau_{0}$ are solutions of the
$Q$-modified Einstein equations (\ref{gfeq2a}) and (\ref{deemt})). For such
configurations, the topological QC structure arise and evolve in diagonal form
with effective density $\ ^{\widehat{q}}\rho=\rho(\tau,t)=\frac{1}{8\pi^{2}%
}C_{ab}\varepsilon^{ij}N_{i}^{a}(\tau,t)N_{j}^{a}(\tau,t)$ as we explain for
(\ref{avdens1}). Effectively, such diagonal quasi-periodic effective matter is
described by a $\tau$-parameter conservation law (i.e. a family matter
equations-of-state parameters) involving and effective energy density and
certain pressure of an effective matter fluid (respectively, $\rho(\tau,t)$
and $p$),
\[
\ ^{\widehat{q}}\rho^{\diamond}+3\ ^{\widehat{q}}H(1+w)p=0,
\]
for the parameters $w(\tau):=$ $p/p$ and effective Hubble functions
$\ ^{\widehat{q}}H(\tau,t):=\ ^{\widehat{q}}a^{\diamond}/\ ^{\widehat{q}}a$
and $\ ^{\widehat{q}}Q=6\ ^{\widehat{q}}H^{2}$ which can be comuted for the
$Q$--modified coefficients of the Ricci d-tensor (\ref{riccist2c}). We note
that the nonmetric vacuum gravitational structure of such cosmological $\tau
$-running spacetime is not trivial because it encode $\varepsilon
$-deformations of topological QC deformations characterised by sources and
conservation laws of type (\ref{conslawsourc}) - (\ref{topologmobconstr}).

Metrics of type (\ref{flrwqc}) define $f(\ ^{\widehat{q}}Q)$ cosmological
models determined by modified Friedman equations
\begin{align}
3\ ^{\widehat{q}}H^{2}  &  =\ ^{\widehat{q}}\rho+\frac{1}{2}[f(\ ^{\widehat{q}%
}Q)-\ ^{\widehat{q}}Q]-\ ^{\widehat{q}}Q[f_{Q}(\ ^{\widehat{q}}%
Q)-1],\label{friedmeqc}\\
\lbrack2\ ^{\widehat{q}}Q(f_{QQ}(\ ^{\widehat{q}}Q)+f_{Q}(\ ^{\widehat{q}%
}Q))]\ ^{\widehat{q}}H^{\diamond}  &  =\frac{1}{4}[f(\ ^{\widehat{q}%
}Q)-2\ ^{\widehat{q}}Q+2\ ^{\widehat{q}}Q(1-f_{Q}(\ ^{\widehat{q}}%
Q))]-\frac{1}{2}p.\nonumber
\end{align}
We can use such equations for modeling topological QC of DE and DM, when the quasi-periodic $Q$-modified Friedman equation (\ref{friedmeqc}) are written $\ ^{m}\Omega+\ ^{Q}\Omega=1$, where the energy density parameters are introduced as
\[
\ ^{m}\Omega=\ ^{\widehat{q}}\rho/3\ ^{\widehat{q}}H^{2}\mbox{ and }\ ^{Q}%
\Omega=[\frac{1}{2}(f(\ ^{\widehat{q}}Q)-\ ^{\widehat{q}}Q)-\ ^{\widehat{q}%
}Q(f_{Q}(\ ^{\widehat{q}}Q)-1)]/3\ ^{\widehat{q}}H^{2}.
\]

A dynamical systems analysis can be performed as in section III of
\cite{khyllep23} (see also section (\ref{primeqcosm})) but for effective
$(\ ^{m}\Omega,\ ^{Q}\Omega,\ ^{eff}w)$ determined by nonmetric geometric
flows inducing topological QC structures. In nonholonomic variables keeping
the diagonal character of such configurations, we can consider the target
effective equation of state
\begin{equation}
\ ^{eff}w:=\frac{^{eff}p}{^{eff}\rho}=-1+\frac{\ ^{m}\Omega(1+w)}%
{2\ ^{\widehat{q}}Qf_{QQ}(\ ^{\widehat{q}}Q)+f_{Q}(\ ^{\widehat{q}}Q)},
\label{tauefeq}%
\end{equation}
energy density and pressure, respectively, as%
$$\ ^{eff}\rho   =\ ^{\widehat{q}}\rho+\frac{1}{2}[f(\ ^{\widehat{q}%
}Q)-\ ^{\widehat{q}}Q]-\ ^{\widehat{q}}Q[f_{Q}(\ ^{\widehat{q}}%
Q)-1]\mbox{ and } \ ^{eff}p   =\frac{\ ^{\widehat{q}}\rho(1+w)}{2\ ^{\widehat{q}}%
Qf_{QQ}(\ ^{\widehat{q}}Q)+f_{Q}(\ ^{\widehat{q}}Q)}-\frac{\ ^{\widehat{q}}%
Q}{2}.$$
The condition $\ ^{eff}w<-1/3$ holds for an accelerated universe. The formulas
(\ref{friedmeqc}) \ and (\ref{tauefeq}) depend in parametric form on effective
temperature $\tau.$ This can be used for modelling diagonal nonmetric
evolution on a time like parameter.

The equation of state (\ref{tauefeq}) can be completed with G. Perelman thermodynamic variables (\ref{qthermvar}) computed for a corresponding solution (\ref{flrwqc}).

\section{G. Perelman thermodynamics for topological QCs and nonmetric DE and DM}
\label{sec5} Physical properties of general $\tau$-families of off-diagonal cosmological solutions encoding nonmetric and topological QC structures for DE and DM configurations can't be studied in the framework of the $\Lambda CDM$
paradigm. Nevertheless, we can always characterize such models by respective G. Perelman thermodynamic variables, which can be computed in explicit form for all classes of solutions in geometric flow and gravity theories. The computations simplify substantially if we consider nonlinear symmetries transforming the data for generating functions and generating sources into equivalent ones re-defined in terms of equivalent new generating functions and effective $\tau$-running cosmological constants. We considered such transforms in previous section for $\widehat{\Lambda}\ (\tau)= \ ^{m}\Lambda(\tau)+\ ^{e}\Lambda(\tau)+\ ^{DE}\Lambda(\tau)$, see formulas (\ref{effcosmconst}). The goal of this section is to compute in explicit form the variables $\widehat{Z}$ (\ref{spf}) and $\ ^{q}\widehat{\mathcal{E}}\ (\tau
),\ ^{q}\widehat{S}(\tau)$ from (\ref{qthermvar}) for such off-diagonal classes of cosmological solutions.\footnote{We omit more cumbersome calculations for $\ ^{q}\widehat{\sigma}(\tau).$ Here we also note that for quasi-stationary configurations a similar nonmetric geometric calculus was provided in section 4 of \cite{kazakh1} (for instance, for nonmetric wormhole
solutions and solitonic hierarchies). The computations for cosmological configurations are certain sense dual to those for quasi-stationary ones, see details in \cite{vreview23,bubuianu17}. In this work, we consider a different class of nonmetric geometric flow and gravity theories, which involve different types of effective sources, generating functions and
nonlinear symmetries corresponding to topological QC configurations.}

For simplicity, we consider the same behavior for horizontal and vertical
$\tau$-running cosmological constants when $\ ^{h}\Lambda\ (\tau
)=\ ^{v}\Lambda\ (\tau)=\widehat{\Lambda}\ (\tau)$ and the canonical Ricci
scalar (see explanations for formulas (\ref{driccidist})) is computed as
\[
\widehat{\mathbf{R}}sc=2[\ ^{m}\Lambda(\tau)+\ ^{e}\Lambda(\tau)+\ ^{DE}%
\Lambda(\tau)]
\]
for any generating data for cosmological solutions taken in a form
(\ref{gendata1}), (\ref{gendata2}) or (\ref{gendata3}). The nonholonomic
conditions for normalizing functions $\widehat{\zeta}(\tau)$ are stated as
$\widehat{\mathbf{D}}_{\alpha}\ \widehat{\zeta}=0$ and approximate
$\widehat{\zeta}\approx0$. To simplify computations we can fix a
frame/coordinate system when such conditions are satisfied and then redefine
the constructions for arbitrary bases and normalizing functions. We obtain:%
\begin{align}
\ ^{q}\widehat{Z}(\tau)  &  =\exp[\int_{\widehat{\Xi}}\frac{1}{8\left(
\pi\tau\right)  ^{2}}\ \delta\ ^{q}\mathcal{V}(\tau)],\label{thermvar1}\\
\ ^{q}\widehat{\mathcal{E}}\ (\tau)  &  =-\tau^{2}\int_{\widehat{\Xi}}%
\ \frac{1}{8\left(  \pi\tau\right)  ^{2}}[2[\ ^{m}\Lambda(\tau)+\ ^{e}%
\Lambda(\tau)+\ ^{DE}\Lambda(\tau)]-\frac{1}{\tau}]\ \delta\ ^{q}%
\mathcal{V}(\tau),\nonumber\\
\ \ ^{q}\widehat{S}(\tau)  &  =-\ ^{q}\widehat{W}(\tau)=-\int_{\widehat{\Xi}%
}\frac{1}{4\left(  \pi\tau\right)  ^{2}}[\tau\lbrack\ ^{m}\Lambda(\tau
)+\ ^{e}\Lambda(\tau)+\ ^{DE}\Lambda(\tau)]-1]\delta\ ^{q}\mathcal{V}%
(\tau),\nonumber
\end{align}
where the topological QC cosmological data are encoded into $\delta
\ ^{q}\mathcal{V}(\tau)$ determined by the determinant of corresponding
off-diagonal solutions.

The volume form $\delta\ ^{q}\mathcal{V}(\tau)$ (\ref{volume}) can be computed
for cosmological d-metrics (\ref{offdiagcosmcshqc}), with $\eta$--polarization
functions, or (\ref{offdncelepsilonq}), for $\chi$--polarization functions,
and including data for corresponding nonmetric generating sources.
Respectively, we obtain
\begin{align}
\ ^{\widehat{q}}\Phi(\tau)  &  =2\sqrt{|[\ ^{m}\Lambda(\tau)+\ ^{e}%
\Lambda(\tau)+\ ^{DE}\Lambda(\tau)]\ ^{\widehat{q}}h_{3}(\tau)|}   =\ 2\sqrt{|[\ ^{m}\Lambda(\tau)+\ ^{e}\Lambda(\tau)+\ ^{DE}\Lambda
(\tau)]\ ^{\widehat{q}}\eta_{3}(\tau)\ \mathring{g}_{3}(\tau)|}\nonumber\\
&  \simeq2\sqrt{|[\ ^{m}\Lambda(\tau)+\ ^{e}\Lambda(\tau)+\ ^{DE}\Lambda
(\tau)]\ \zeta_{3}(\tau)\ \ \mathring{g}_{3}|}[1-\frac{\varepsilon}%
{2}\ ^{\widehat{q}}\chi_{3}(\tau)]. \label{genf1}%
\end{align}
Considering nonholonomic and nonmetric evolution models with trivial
integration functions $\ _{1}n_{k}=0$ and $\ _{2}n_{k}=0$ and introducing
formulas (\ref{genf1}) in (\ref{volume}), then separating terms with shell
$\tau$-running cosmological constants, we compute:
\begin{align*}
\ \delta\ ^{q}\mathcal{V}  &  =\delta\mathcal{V}[\tau, \ ^{m}\Lambda
(\tau)+\ ^{e}\Lambda(\tau)+\ ^{DE}\Lambda(\tau);\ _{h}^{q}\Upsilon(\tau),
\ ^{m}\yen [\theta^{a}[\tau]],\ ^{e}\yen [\theta^{a}[\tau]],\ ^{DE}%
\yen [\theta^{a}[\tau]];\psi(\tau),\ ^{\widehat{q}}h_{3}(\tau)]\\
&  =\delta\mathcal{V}(\ \ _{h}^{q}\Upsilon(\tau), \ \ ^{m}
\yen \ ^{\widehat{2}}[\theta^{a}[\tau]],\ ^{e} \yen [\theta^{a}[\tau
]],\ ^{DE}\yen [\theta^{a}[\tau]]; \ ^{m}\Lambda(\tau)+\ ^{e}\Lambda
(\tau)+\ ^{DE}\Lambda(\tau),\ ^{\widehat{q}}\eta_{3}(\tau)\mathring{g}_{3})\\
&  =\frac{1}{\ ^{m}\Lambda(\tau)+\ ^{e}\Lambda(\tau)+\ ^{DE}\Lambda(\tau
)}\ \delta\ _{\eta}\mathcal{V},\mbox{ where }\ \delta\ _{\eta}\mathcal{V}%
=\ \delta\ _{\eta}^{1}\mathcal{V}\times\delta\ _{\eta}^{2}\mathcal{V}.
\end{align*}
Such volume forms encoding topological QC structures are parameterized by
products of two functionals:%
\begin{align}
\delta\ _{\eta}^{1}\mathcal{V}  &  =\delta\ _{\eta}^{1}\mathcal{V}%
[\ ^{m}\Lambda(\tau)+\ ^{e}\Lambda(\tau)+\ ^{DE}\Lambda(\tau),\eta_{1}%
(\tau)\ \mathring{g}_{1}]\label{volumfuncts}\\
&  =e^{\widetilde{\psi}(\tau)}dx^{1}dx^{2}=\sqrt{|\ ^{m}\Lambda(\tau
)+\ ^{e}\Lambda(\tau)+\ ^{DE}\Lambda(\tau)|}e^{\psi(\tau)}dx^{1}%
dx^{2},\mbox{ for }\psi(\tau)\mbox{ being a solution of  }(\ref{eq1}%
),\nonumber\\
\delta\ _{\eta}^{2}\mathcal{V}  &  =\delta\ _{\eta}^{2}\mathcal{V}%
[\ \ ^{m}\ \yen [\theta^{a}[\tau]],\ ^{e}\ \yen [\theta^{a}[\tau
]],\ ^{DE}\yen [\theta^{a}[\tau]],\ ^{\widehat{q}}\eta_{3}(\tau)\ \mathring
{g}_{3}]\nonumber\\
&  =\frac{\partial_{t}|\ \ ^{\widehat{q}}\eta_{3}(\tau)\ \mathring{g}%
_{3}|^{3/2}}{\ \sqrt{|\int dt\ [\ \ ^{m}\ \yen (\theta^{a}[\tau])+\ ^{e}%
\ \yen (\theta^{a}[\tau])+\ ^{DE}\yen (\theta^{a}[\tau])]\{\partial
_{t}|\ \ ^{\widehat{q}}\eta_{3}(\tau)\ \mathring{g}_{3}|\}^{2}|}}\nonumber\\
&  \lbrack dt+\frac{\partial_{i}\left(  \int dt\ [\ \ ^{m}\ \yen (\theta
^{a}[\tau])+\ ^{e}\ \yen (\theta^{a}[\tau])+\ ^{DE}\yen (\theta^{a}%
[\tau])]\partial_{t}|\ \ ^{\widehat{q}}\eta_{3}(\tau)\ \mathring{g}%
_{3}|\right)  dx^{i}}{\ [\ ^{m}\ \yen (\theta^{a}[\tau])+\ ^{e}\ \yen (\theta
^{a}[\tau])+\ ^{DE}\yen (\theta^{a}[\tau])]\partial_{t}|\ \ ^{\widehat{q}}%
\eta_{3}(\tau)\ \mathring{g}_{3}|}]dt.\nonumber
\end{align}
In these formulas, we distinguish effective sources and cosmological constants
with labels $m,e$ and $DE$ because such functionals induce, or not,
topological QC structures of different types. The functions $\widetilde{\psi
}(\tau)$ are defined as a $\tau$--family of solutions of 2-d Poisson equations
with effective source $\ ^{m}\Lambda(\tau)+\ ^{e}\Lambda(\tau)+\ ^{DE}%
\Lambda(\tau),$ or we can use $\psi(\tau)$ for a respective source
$\ _{h}\Upsilon(\tau).$ Integrating on a closed hypersurface $\widehat{\Xi}$
the products of $h$- and $v$-forms from (\ref{volumfuncts}), we obtain a
running cosmological phase space volume functional
\begin{equation}
\ _{\eta}^{\shortmid}\mathcal{\mathring{V}}(\tau)=\int_{\ \widehat{\Xi}}%
\delta\ _{\eta}\mathcal{V}(\ _{h}\Upsilon(\tau),[\ ^{m}\ \yen (\theta^{a}%
[\tau])+\ ^{e}\ \yen (\theta^{a}[\tau])+\ ^{DE}\yen (\theta^{a}[\tau
])],\ \mathring{g}_{\alpha}) \label{volumfpsp}%
\end{equation}
determined by prescribed classes of generating $\eta$-functions, effective
generating topological QC cosmological sources $\left[  _{h}\Upsilon
(\tau),\ ^{m}\ \yen (\theta^{a}[\tau])+\ ^{e}\ \yen (\theta^{a}[\tau
])+\ ^{DE}\yen (\theta^{a}[\tau])\right]  ,$ coefficients of a prime s-metric
$\ \mathring{g}_{\alpha}$ and nonholonomic distributions defining the
hyper-surface $\widehat{\Xi}.$ The explicit values of volume forms $\ _{\eta
}^{\shortmid}\mathcal{\mathring{V}}(\tau)$ depend on the data we prescribe for
$\widehat{\Xi}$ the type of topological QC $Q$-deformations (encoded into
$\eta$- or $\zeta$-polarizations) which are used for deforming a prime
cosmological d-metric into cosmological topological QC as we considered in
section \ref{sec4}. We consider that it is always possible to compute
$\ _{\eta}^{\shortmid}\mathcal{\mathring{V}}(\tau)$ for certain nonlinear
quasi-periodic generating data and general $Q$-deformations. The thermodynamic
variables depend explicitly on the $\tau$-running effective cosmological
constants when such dependencies can be chosen in some form explaining
observational cosmological data and prescribing the evolution of off-diagonal
DE and DM configurations.

Introducing functional (\ref{volumfuncts}) into the formulas (\ref{thermvar1}%
), we compute
\begin{align}
\ \ ^{q}\widehat{Z}(\tau)  &  =\exp[\frac{\ _{\eta}^{\shortmid}%
\mathcal{\mathring{V}}(\tau)}{8\left(  \pi\tau\right)  ^{2}}%
]\ ,\label{thermvar2}\\
^{q}\widehat{\mathcal{E}}\ (\tau)  &  =[\frac{1}{\tau}-2(\ ^{m}\Lambda
(\tau)+\ ^{e}\Lambda(\tau)+\ ^{DE}\Lambda(\tau))]\frac{\ _{\eta}^{\shortmid
}\mathcal{\mathring{V}}(\tau)}{8\pi^{2}}\ ,\nonumber\\
\ \ ^{q}\widehat{S}(\tau)  &  =-\ ^{q}\widehat{W}(\tau)=[1-\tau(\ ^{m}%
\Lambda(\tau)+\ ^{e}\Lambda(\tau)+\ ^{DE}\Lambda(\tau))]\frac{\ _{\eta
}^{\shortmid}\mathcal{\mathring{V}}(\tau)}{4\left(  \pi\tau\right)  ^{2}%
}.\nonumber
\end{align}

We can define the effective volume functionals (\ref{volumfuncts}) and
geometric thermodynamic variables (\ref{thermvar2}) for further parametric
decompositions with $\varepsilon$-linear approximations (\ref{genf1}) and
$\varepsilon$-polarizations and find parametric formulas for $\tau$-flows and
$Q$-deformations of prime metrics,
\begin{align}
\delta\mathcal{V}  &  =\delta\mathcal{V}_{0}\mathcal{[}\tau,\ ^{m}\Lambda
(\tau)+\ ^{e}\Lambda(\tau)+\ ^{DE}\Lambda(\tau);\ _{h}\Upsilon(\tau
),[\ ^{m}\ \yen (\theta^{a}[\tau])+\ ^{e}\ \yen (\theta^{a}[\tau
])+\ ^{DE}\yen (\theta^{a}[\tau])];\psi(\tau),\ \mathring{g}_{\alpha}%
;\zeta_{3}(\tau)]\nonumber\\
&  +\varepsilon\delta\mathcal{V}_{1}\mathcal{[}\tau,\ ^{m}\Lambda(\tau
)+\ ^{e}\Lambda(\tau)+\ ^{DE}\Lambda(\tau);\ _{h}\Upsilon(\tau),
\ [\ ^{m}\ \yen (\theta^{a}[\tau])+\ ^{e}\ \yen (\theta^{a}[\tau
])+\ ^{DE}\yen (\theta^{a}[\tau])];\nonumber\\
&  \psi(\tau),\ \mathring{g}_{\alpha} ;\zeta_{3}(\tau),~\ ^{\widehat{q}}%
\chi_{3}(\tau)].\label{volumfd}%
\end{align}
Computing such a (\ref{volumfd}) for a family of off-diagonal cosmological
solutions constructed in previous section, we can define corresponding
$\varepsilon$-decompositions of the thermodynamic variables (\ref{thermvar2}),
expressed as
\begin{equation}
\widehat{\mathcal{W}}(\tau)=\widehat{\mathcal{W}}_{0}+\varepsilon
\widehat{\mathcal{W}}(\tau),\widehat{\mathcal{Z}}(\tau)=\widehat{\mathcal{Z}%
}_{0}\ \widehat{\mathcal{Z}}_{1}(\tau),\widehat{\mathcal{E}}(\tau
)=\widehat{\mathcal{E}}_{0}+\varepsilon\widehat{\mathcal{E}}_{1}%
(\tau),\widehat{\mathcal{S}}(\tau)=\widehat{\mathcal{S}}_{0}+\varepsilon
\widehat{\mathcal{S}}_{1}(\tau). \label{klinthvar}%
\end{equation}
In our works on nonmetric geometric flows and $\tau$-running cosmological theories, we do not present details of such cumbersome and incremental computations with $\varepsilon$--linear decomposition for 
$\delta \mathcal{V}=\delta\mathcal{V}_{0}+\varepsilon\delta\mathcal{V}_{1}$ (\ref{volumfd}) and resulting (\ref{klinthvar}) determined by corresponding $\chi$-polarization functions. The final conclusion of this section is that we can always define and compute in general forms the corresponding generalized Perelman thermodynamical variables for different sources with labels $m,e$ and $DE,$ and elaborate on structure formation and statistical thermodynamic description
of cosmological models with $\tau$-running topological QC structures.

\section{Conclusions}
\label{sec6} This is the second paper in a series of partner works devoted to nonmetric deformations of the theory of geometric flows, modified gravity, and applications in modern astrophysics and cosmology. The first one \cite{kazakh1} was devoted to constructing and analyzing essential physical properties of quasi-stationary generic off-diagonal solutions in such
theories. As the  next steps (for this work), our main goals were to generalize our geometric and statistical thermodynamic methods for generating cosmological solutions and elaborate on possible applications in dark energy, DE, and dark matter, DM, physics. Such a multi- and interdisciplinary research program is motivated by recent results on nonmetric $f(Q)$-modified gravity and cosmology \cite{harko21,iosifidis22,khyllep23,koussour23,jim18,jhao22,de22}, when new classes of generic off-diagonal exact and parametric solutions for modified gravity theories, MGTs, with non-Riemannian connections  \cite{bubuianu17,vacaru18,vreview23}   seem to describe in more adequate forms various observational data for accelerating cosmology recently provided by James Webb space telescope, JWST, \cite{foroconi23,boylan23,biagetti23}.

\vskip5pt
To describe generic off-diagonal exact and parametric solutions for nonmetric MGTs in the framework of the Bekenstein--Hawking entropy paradigm is not possible because, in general, nonmetric and off-diagonal cosmological scenarios (for instance, the solutions with topological QC-like structure studied in this work) do not involve certain horizon/ duality / holographic configurations. We had to elaborate on a new statistical/ geometric thermodynamic paradigm introduced for  geometric flow theories due to G.  Perelman  \cite{perelman1}; and developed in nonholonomic and relativistic forms for various MGTs and nonholonomic/ nonassociative/ nonmetric and other types modifications  \cite{vacaru20,bubuianu23,bubuianu23a}. In this work, in addition to gaining a more complete understanding of cosmological effects of $f(Q)$ MGTs, we also studied new classes of $\tau$-parametric generic off-diagonal and locally anisotropic cosmological solutions \cite{bubuianu17,vacaru18,vreview23}, elaborated on models with topological
quasi-crystal, QC structure (see also other models with space and time structure \cite{bubuianu17,sv18,else21}), and provided examples of how to compute G. Perelman thermodynamic variables for generic off-diagonal ansatz of cosmological metrics with topological QC structure.

\vskip5pt
Let us summarize and discuss the main new results and methods of this paper following the key ideas for solutions of objectives, \textbf{Obj 1- Obj 5}, formulated in the Introduction section:
\begin{enumerate}
\item In section \ref{sec2}, we formulated the $f(Q)$ gravity theories in nonholonomic canonical (2+2) variables which allowed us to prove general decoupling and integration properties of corresponding nonmetric deformed Einstein equations. Necessary technical results were provided in Appendix \ref{appendixa}  as a summary of the anholonomic frame and connection deformation method, AFCDM, for constructing exact and parametric solutions for physically important systems of nonlinear PDEs  in $f(Q)$ deformed nonmetric geometric flow and gravity theories. Such methods are not incremental but involve new ideas give new results for time dual nonholonomic transforms and nonmetric deformations of constructions from \cite{kazakh1,bubuianu17,vacaru18}. They  can be summarized as the solution of \textbf{Ob1} for elaborating a general geometric formalism for constructing off-diagonal cosmological solutions in nonmetric geometric flow and gravity theories. 

\item \textbf{Obj 2} was solved in section \ref{ss31} by formulating the $f(Q)$-distorted nonmetric geometric flow equations, derived as  $\tau$-running nonmetric Einstein equations and written in canonical nonholonomic variables as nonmetric Ricci solitons.  Such formulations allowed us to apply the AFCDM and find exact and parametric solutions of such systems of nonlinear PDEs in general off-diagonal forms.

\item  Then the statistical thermodynamics for nonholonomic $f(Q)$-distorted nonmetric geometric flows (i.e. the solution of  \textbf{Obj 3)} was formulated in section \ref{ssperelth}. Such models are different from those studied in \cite{kazakh1}  for  $f(R,T,Q,T_{m})$ nonmetric theories and their quasi-stationary solutions. Here we note that the nonmetric geometric flow $\tau$-parameter is a temperature one as in \cite{perelman1}, which can be exploited for describing new observational JWST data \cite{foroconi23,boylan23,biagetti23} when the cosmological evolution scenarios depends on a temperature like parameter  being formulated in some generic nonlinear forms encoding or generating  quasi-periodic structures, for instance, of topological QC type.

\item In section \ref{sec4} (providing solutions of   \textbf{Obj 4} to consider applications of nonmetric geometric flow methods in DE and DM physics), we constructed and analyzed the most important physical properties of $\tau$-families of generic off-diagonal cosmological solutions encoding nonmetricity and generating topological QC structures. Such solutions and defining toy topological models were formulated in nonholonomic forms (choosing  corresponding classes of polarization functions and quasi-periodic effective sources) which may describe the geometric evolution of DM structure and correspond to effective running cosmological constants modelling DE configurations. In general, such generic off-diagonal  cosmological scenarios are different from those defined for $\Lambda$CDM models and modifications.   

\item Physical properties of $\tau$-running families of nonmetric cosmological solutions, even for a fixed value of the geometric evolution parameter,  can't be described in general form using the Bekenstein-Hawking thermodynamic paradigm. This requests the formulation of a new geometric and statistical thermodynamic concept of G. Perelman entropy. In explicit form (solving the  \textbf{Obj 5}), we have shown how to compute Perelman's thermodynamic variables for nonmetric geometric flows and cosmological configurations with topological QC structure encoding $f(Q)$ nonmetric effects. We speculate how such variables may characterize the DE and DM physics determined by such a nonmetric cosmological scenarios. Necessary results on topological QC structures are provided in Appendix  \ref{appendixb} by generalizing the approach in canonical nonholonomic variables which are important for applying the AFCDM.
\end{enumerate}

The results listed and discussed in paragraphs 1-5 support the statements of the main {\bf Hypothesis} formulated in section \ref{sec1} in such senses: (1)  Nonmetric $f(Q)$  modifications of geometric flow and MGTs can be formulated in canonical nonholonomic variables which allow us to decouple and integrate in general forms physically important systems of nonlinear PDEs. (2) Such way constructed new classes of exact and parametric solutions are defined by generic off-diagonal metrics, nonmetric compatible affine connections and effective sources encoding nonmetric geometric flow deformations and possible quasi-periodic spacetime structures. (3) The cosmological solutions encoding various topological QC phases characterized by nonmetric versions of G. Perelman thermodynamic variables determine new features of the DE and DM physics and provide new geometric and thermodynamic methods for describing new observational JWST data.

\vskip5pt
 Nevertheless, there are a number of  important open questions and unsolved problems for nonmetricity physics which were laid  out in \cite{vmon05,vacaru18} and  (in questions on nonmetric geometric and information flow theories and gravity) QNGIFG1-4 from \cite{kazakh1}. Perhaps, the next in the order to be solved in future works is to formulate a mathematically self-consistent and physically viable version of $f(Q)$--modified Einstein-Dirac theory and generalizing the AFCDM for constructing exact and parametric solutions describing nonmetric gravitational and spinor systems. In a more general context, a more fundamental and difficult task is to investigate possibilities of how to connect theories of nonmetric geometric flows to modern string theory. Such constructions should involve certain models of nonmetric twist products and nonassociative star deformations related to the research program on nonassociative geometric and information flows \cite{bubuianu23,bubuianu23a}. In future works, we plan to report on progress in solving such problems.

\vskip6pt \textbf{Acknowledgments:} This research on nonmetric geometric flows and applications extends SV and EN visit programs (respectively supported by the Fulbright USA-Romania and the Ministry of Education and Science of the Republic of Kazakhstan) at the physics department at California State University at Fresno, USA. The program on AFCDM for theories with effective sources encoding nonmetricity consists also a background for some objectives of a SV visiting fellowship at CAS LMU in Munich, Germany. SV and EN are grateful to the host professors D. L\"{u}st and D. Singleton for kind support and collaboration.

\appendix
\setcounter{equation}{0} \renewcommand{\theequation}
{A.\arabic{equation}} \setcounter{subsection}{0}
\renewcommand{\thesubsection}
{A.\arabic{subsection}}

\section{Decoupling and integrability of nonmetric Ricci flows and cosmological equations}

\label{appendixa}

We review the anholonomic frame and connection deformation method, AFCDM),
extended for constructing generic off-diagonal cosmological solutions of
nonmetric geometric flow equations (\ref{cfeq4af}). Such solutions are "time"
dual to certain quasi-stationary solutions, which can be generated by similar
methods for different geometric variables; see details in sections 3.1, 3.2
and 3.5 of \cite{vreview23} and, for nonmetric geometric flows, see the
partner work \cite{kazakh1}. We apply abstract and symbolic geometric methods
(for GR, see \cite{misner}) and analytic methods when similar proofs are
considered in \cite{bubuianu17,vacaru18,vreview23} and references therein.

\subsection{Decoupling of nonlinear PDEs for nonmetric geometric cosmological flows}

We consider the off-diagonal ansatz (\ref{lacosmans}) with N-adapted
coefficients of the canonical Ricci d-tensor and effective sources
(\ref{dsourcosm}) for the system of nonlinear PDs (\ref{cfeq4af}) written in
the form:
\begin{align}
\widehat{R}_{1}^{1}(\tau)  &  =\widehat{R}_{2}^{2}(\tau)=\frac{1}{2g_{1}g_{2}%
}[\frac{g_{1}^{\bullet}g_{2}^{\bullet}}{2g_{1}}+\frac{(g_{2}^{\bullet})^{2}%
}{2g_{2}}-g_{2}^{\bullet\bullet}+\frac{g_{1}^{\prime}g_{2}^{\prime}}{2g_{2}%
}+\frac{\left(  g_{1}^{\prime}\right)  ^{2}}{2g_{1}}-g_{1}^{\prime\prime
}]=-\ _{h}\ \yen (\tau),\nonumber\\
\widehat{R}_{3}^{3}(\tau)  &  =\widehat{R}_{4}^{4}(\tau)=\frac{1}{2h_{3}h_{4}%
}[\frac{\left(  h_{3}^{\diamond}\right)  ^{2}}{2h_{3}}+\frac{h_{3}^{\diamond
}h_{4}^{\diamond}}{2h_{4}}-h_{3}^{\diamond\diamond}]=-\ \ _{v}\ \yen (\tau
),\label{riccist2c}\\
\widehat{R}_{3k}(\tau)  &  =\frac{h_{3}}{2h_{4}}n_{k}^{\diamond\diamond
}+\left(  \frac{3}{2}h_{3}^{\diamond}-\frac{h_{3}}{h_{4}}h_{4}^{\diamond
}\right)  \frac{\ n_{k}^{\diamond}}{2h_{4}}=0,\nonumber\\
\widehat{R}_{4k}(\tau)  &  =\frac{\ w_{k}}{2h_{3}}[h_{3}^{\diamond\diamond
}-\frac{\left(  h_{3}^{\diamond}\right)  ^{2}}{2h_{3}}-\frac{(h_{3}^{\diamond
})(h_{4}^{\diamond})}{2h_{4}}]+\frac{h_{3}^{\diamond}}{4h_{3}}(\frac
{\partial_{k}h_{3}}{h_{3}}+\frac{\partial_{k}h_{4}}{h_{4}})-\frac{\partial
_{k}(h_{4}^{\diamond})}{2h_{4}}=0.\nonumber
\end{align}
In our works, we also use brief notations of partial derivatives when
$\partial_{1}q(u^{\alpha}):=q^{\bullet},\partial_{2}q(u^{\alpha}):=q^{\prime
},\partial_{3}q(u^{\alpha}):=q^{\ast}$ and $\partial_{4}q(u^{\alpha
}):=q^{\diamond}$, for an arbitrary function $q(u^{\alpha}).$ The d-metric
ansatz (\ref{lacosmans}) and corresponding nonmetric geometric flow equations
(\ref{riccist2c}) posses a Killing symmetry on d-vector $\partial_{3}$, i.e.
the corresponding coefficients of the d-metric, N-connection (\ref{lacosmdmnc}%
) etc. do not depend on coordinate $u^{3}=y^{3}.$ To have at least one Killing
symmetry on a $v$-coordinate is important for generating exact/parametric
solutions of such off-diagonal equations in explicit form. In principle, we
can consider more general ansatz without Killing symmetries but such
constructions are technically more cumbersome and involve proofs on hundreds
of pages, see discussions and examples in \cite{bubuianu17,vacaru18,vreview23}%
.\footnote{Here we note that the "t-duality" of locally anisotropic
cosmological $\tau$-evolution systems (\ref{lacosmans}) and (\ref{riccist2c})
means that such equations and solutions can be transformed into similar ones
for quasi-stationary $\tau$-evolving configurations if $h_{3}(\tau
,x^{k},t)\rightarrow h_{4}(\tau,x^{k},y^{3}),h_{4}(\tau,x^{k},t)\rightarrow
h_{3}(\tau,x^{k},y^{3});N_{i}^{3}=n_{i}(\tau,x^{k},t)\rightarrow N_{i}%
^{4}=w_{i}(\tau,x^{k},y^{3}),\,\,\,\,N_{i}^{4}=w_{i}(\tau,x^{k},t)\rightarrow
N_{i}^{3}=n_{i}(\tau,x^{k},y^{3})$and $\partial_{4}q(u^{\alpha}):=q^{\diamond
}\rightarrow$ $\partial_{3}q(u^{\alpha}):=q^{\ast}$ etc. Nonmetric
quasi-stationary models and their solutions are studied in the partner work
\cite{kazakh1} for different metric-affine configurations and effective
sources.}

We can write the system of equations (\ref{riccist2c}) in a more simpler and
explicit decoupled form if we express the h-components of the d-metric as
$g_{i}(\tau)=e^{\psi(\tau,x^{k})}$; introduce the coefficients
\begin{equation}
\alpha_{i}(\tau):=h_{3}^{\diamond}\partial_{i}[\varpi(\tau)],\beta
(\tau):=h_{3}^{\diamond}(\tau)[\varpi(\tau)]^{\diamond},\gamma(\tau
):=(\ln|h_{3}(\tau)|^{3/2}/|h_{4}(\tau)|)^{\diamond}, \label{coeff}%
\end{equation}
and consider a generating function $\Psi(\tau)=\exp[\varpi(\tau)]$ for
$\varpi(\tau):=\ln|h_{3}^{\diamond}(\tau)/\sqrt{|h_{3}(\tau)h_{4}(\tau)}|$.
For simplicity, we do not write an explicit dependence of such coefficients on
$(\tau,x^{k})$, or $(\tau,x^{k},y^{3}),$ but express:
\begin{align}
\psi^{\bullet\bullet}+\psi^{\prime\prime}  &  =2\ \ _{h}\yen (\tau
),\label{eq1}\\
(\varpi)^{\diamond}h_{3}^{\diamond}  &  =2h_{3}h_{4}\ _{v}\yen (\tau
),\label{e2a}\\
\ n_{k}^{\diamond\diamond}+\gamma n_{k}^{\diamond}  &  =0,\label{e2b}\\
\beta w_{j}-\alpha_{j}  &  =0, \label{e2c}%
\end{align}
Let us explain the explicit decoupling property of this system of nonlinear
PDEs. The coefficients $g_{i}(\tau)$ are determined as solutions of a $\tau
$-family of 2-d Poisson equations (\ref{eq1}). The coefficients $h_{3}(\tau)$
and $h_{4}(\tau)$ are related as solutions of (\ref{e2a}) for any nontrivial
$\varpi(\tau)$ and $\ _{v}\ \yen (\tau).$ This allows us to prescribe a value
of $h_{3}(\tau)$ and find $h_{4}(\tau),$ or inversely. Having defined
$h_{a}(\tau),$ we can compute the coefficients $\beta(\tau)$ and $\alpha
_{i}(\tau)$ (using formulas (\ref{coeff})) and solve respective linear
equations for $w_{j}(\tau)$ from (\ref{e2c}). To find solutions for
$n_{k}(\tau)$ we have to integrate two times on $y^{4}=t$ in (\ref{e2b}), when
$\gamma(\tau)$ is determined by $h_{3}(\tau)$ and $h_{4}(\tau)$ as in
(\ref{coeff}).

Finally, we note that (\ref{eq1}) defines 2-d conformal flat h-components of a
d-metric in a form that allows a linear principle of superposition of
solutions. The equation (\ref{e2a}) is generic nonlinear and involve
additional nonlinearities determined by coefficients (\ref{coeff}). For a
linear decomposition of effective sources $\ _{v}\ \yen =\ ^{m}\ \yen +\ ^{e}%
\ \yen +\ ^{DE}\yen $ (\ref{emefparam}), the parametric contributions of
different types of sources encode nonlinear nonmetric effects. The
cosmological scenarios are generic off-diagonal even for small parametric
interactions, and result in nonlinear nonmetric geometric evolution for
nontrivial solutions for N-coefficients determined by (\ref{e2b}) and
(\ref{e2c}).

\subsection{Off-diagonal solutions for nonmetric cosmological configurations}

Applying the AFCDM \cite{bubuianu17,vacaru18,vreview23}, we can integrate
recurrently the system of nonlinear PDEs (\ref{eq1}) -- (\ref{e2c}). Any such
$\tau$-family of generic off-diagonal metrics (\ref{lacosmans}) is determined
by respective families of a generating function $\varpi(\tau)$ (equivalently,
$\Psi(\tau)$) and generating sources $\ _{h}\ \yen (\tau)$ and $\ _{v}%
\ \yen (\tau).$ In this subsection, we provide some classes and different
parameterizations of such solutions which are important for elaborating
cosmological models encoding nonmetric geometric evolution and generic
off-diagonal interactions. The explicit form of such solutions depends on the
type of parameterizations of generating functions and generating sources and
how such values are related to integration functions and physical constants.

\subsubsection{Nonlinear symmetries of nonmetric cosmological solutions}

Nonmetric evolution models of anisotropic cosmological solutions with Killing
symmetry and on $\partial_{3}$ and for effective sources (\ref{emefparam}) are
defined by such generic off-diagonal quasi-stationary $\tau$-families of
d-metrics:%
\begin{align}
ds^{2}(\tau)  &  =e^{\psi(\tau,x^{k})}[(dx^{1})^{2}+(dx^{2})^{2}%
]\label{nonmcosmla}\\
&  +\{g_{3}^{[0]}-\int dt\frac{[\Psi^{2}]^{\diamond}}{4(\ _{v}\ \yen )}%
\}\{dy^{3}+[\ _{1}n_{k}+\ _{2}n_{k}\int dt\frac{[(\Psi)^{2}]^{\diamond}%
}{4(\ _{v}\ \yen )^{2}|g_{3}^{[0]}-\int dt[\Psi^{2}]^{\diamond}/4(\ _{v}%
\ \yen )|^{5/2}}]dx^{k}\}\nonumber\\
&  +\frac{[\Psi^{\diamond}]^{2}}{4(\ _{v}\ \yen )^{2}\{g_{3}^{[0]}-\int
dt[\Psi^{2}]^{\diamond}/4(\ _{v}\ \yen )\}}(dt+\frac{\partial_{i}\Psi}%
{\Psi^{\diamond}}dx^{i})^{2}.\nonumber
\end{align}
Such an exact/ parametric solutions is determined by a generating function
$\Psi(\tau,x^{k},t),$ two generating effective sources $\ _{h}\ \yen (\tau
,x^{k})$ (encoded as a solution $\psi(\tau,x^{k})$ of 2-d Poisson equation
(\ref{eq1})) and $\ _{v}\ \yen (\tau,x^{k},y^{3})$, and integration functions
$\ _{1}n_{i}(\tau,x^{k}),\ _{2}n_{i}(\tau,x^{k})$ and $g_{3}^{[0]}(\tau
,x^{k}).$ If for such d-metrics there are considered parametric decompositions
as in (\ref{emefparam}), we generate recurrently compute $Q$-deformations and
other type parametric solutions.

By straightforward computations, we can check that a $\tau$-family of
solutions (\ref{nonmcosmla}) posses important nonlinear symmetries which allow
us to re-write such d-metric in terms of different types of generating
functions ($\Psi(\tau)$ or $\Phi(\tau)$) and transform effective sources into
$\tau$-running effective cosmological constants $\widehat{\Lambda}(\tau).$
Such nonlinear transforms, $(\Psi(\tau),\ _{v}\ \yen (\tau))\leftrightarrow
(\Phi(\tau),\ \widehat{\Lambda}(\tau)=const\neq0$ for any $\tau_{0}),$ are
defined by formulas%
\begin{align}
\frac{\lbrack\Psi^{2}]^{\diamond}}{\ \ _{v}\ \yen (\tau)}  &  =\frac{[\Phi
^{2}(\tau)]^{\diamond}}{\widehat{\Lambda}(\tau)},\mbox{ which can be
integrated as  }\label{ntransf1}\\
\Phi^{2}(\tau)  &  =\widehat{\Lambda}(\tau)\int dt(\ \ _{v}\ \yen )^{-1}%
[\Psi^{2}(\tau)]^{\diamond}\mbox{ and/or }\Psi^{2}(\tau)=(\widehat{\Lambda
}(\tau))^{-1}\int dt(\ \ _{v}\ \yen )[\Phi^{2}(\tau)]^{\diamond}.
\label{ntransf2}%
\end{align}
For linear decompositions of effective sources (\ref{emefparam}, we can
consider such nonlinear transforms into respective $\tau$-running effective
cosmological constants
\begin{equation}
\ _{v}\ \yen =\ ^{m}\ \yen +\ ^{e}\ \yen +\ ^{DE}\yen \leftrightarrow
\widehat{\Lambda}=\ ^{m}\Lambda+\ ^{e}\Lambda+\ ^{DE}\Lambda.
\label{effcosmconst}%
\end{equation}
We can prescribe nonholonomic dyadic structures when $\ ^{e}\ \yen =\ ^{e}%
\Lambda=0$ for metric compatible configurations and generating sources
$\ ^{e}\widehat{\mathbf{Y}}_{\alpha\beta}$ (\ref{ceemt}) or $\ ^{DE}%
\yen =\ ^{DE}\Lambda=0$ for trivial $Q$-deformations $\ ^{DE}T_{\alpha\beta}$
(\ref{deemt}). We can analyze cosmological scenarios with nonmetric geometric
flow evolution prescribing certain small values of $\ ^{e}\Lambda$ and/ or
$\ ^{DE}\Lambda$ even the terms $\ ^{m}\ \yen ,\ ^{e}\ \yen ,$ and$\ ^{DE}%
\yen $ may be not small and subjected to nonlinear symmetries (\ref{ntransf1})
and (\ref{ntransf2}).

Considering above stated nonlinear symmetries, we can write the quadratic
element (\ref{nonmcosmla}) in a form including effective cosmological
constants,%
\begin{align}
ds^{2}(\tau)  &  =g_{\alpha\beta}(\tau,x^{k},t,\Phi(\tau),\widehat{\Lambda
}(\tau))du^{\alpha}du^{\beta}=e^{\psi(\tau,x^{k})}[(dx^{1})^{2}+(dx^{2}%
)^{2}]-\{g_{3}^{[0]}(\tau)-\frac{\Phi^{2}(\tau)}{4\ \widehat{\Lambda}(\tau
)}\}\ \label{offdiagcosmcsh}\\
&  \{dy^{3}+[\ _{1}n_{k}(\tau)+\ _{2}n_{k}(\tau)\int dt\frac{\Phi^{2}%
(\tau)[\Phi^{\diamond}(\tau)]^{2}}{|\ \widehat{\Lambda}(\tau)\int
dt\ \ _{v}\ \yen (\tau)[\Phi^{2}(\tau)]^{\diamond}|[g_{3}^{[0]}(\tau)-\Phi
^{2}(\tau)/4\ \widehat{\Lambda}(\tau)]^{5/2}}]\}\nonumber\\
&  -\frac{\Phi^{2}(\tau)[\Phi^{\diamond}(\tau)]^{2}}{|\ \widehat{\Lambda}%
(\tau)\int dt\ \ _{v}\ \yen (\tau)[\Phi^{2}(\tau)]^{\diamond}\ |[g_{3}%
^{[0]}(\tau)-\Phi^{2}(\tau)/4\ \widehat{\Lambda}(\tau)]}\{dt+\frac
{\partial_{i}\ \int dt\ \ \ _{v}\ \yen (\tau)\ [\Phi^{2}(\tau)]^{\diamond}%
}{\ \ _{v}\ \yen (\tau)\ [\Phi^{2}(\tau)]^{\diamond}}dx^{i}\}^{2}.\nonumber
\end{align}
In these formulas, the indices run respectively $i,j,k,...=1,2;a,b,c,...=3,4;$
there are used: generating functions $\psi(\tau,x^{k})$ and $\Phi
(\tau,x^{k_{1}},t);$ generating sources $\ _{h}\ \yen (\tau,x^{k})$ and
$\ _{v}\ \yen (\tau,x^{k},t);$ effective cosmological constants
$\ \widehat{\Lambda}(\tau);$ and integration functions $\ _{1}n_{k}(\tau
,x^{j}),\ _{2}n_{k}(\tau,x^{j})$ and $g_{3}^{[0]}(\tau,x^{k}).$

Using \ (\ref{ntransf1}), we express $h_{3}^{\diamond}(\tau)=-[\Psi^{2}%
(\tau)]^{\diamond}/4\ _{v}\ \yen (\tau),$ which allows to compute (we can
chose necessary types of integration functions ) in a form computed with
$\Psi(\tau)$ from (\ref{nonmcosmla}).\footnote{We have to integrate on $t$ the
formula $[\Psi^{2}(\tau)]^{\diamond}=-4\int dt\ _{v}\ \yen (\tau
)h_{3}^{\diamond}(\tau)$ for any prescribed $h_{3}(\tau)$ and $\ _{v}%
\ \yen (\tau).$} This means that taking $(h_{3}(\tau),\ _{v}\ \yen (\tau))$ as
generating data, we can write the $\tau$-families of cosmological d-metrics
(\ref{nonmcosmla}) or (\ref{offdiagcosmcsh}) in another equivalent form:
\begin{align}
d\widehat{s}^{2}(\tau)  &  =\widehat{g}_{\alpha\beta}(\tau,x^{k},t;h_{4}%
(\tau),\ \ _{v}\ \yen (\tau))du^{\alpha}du^{\beta}=e^{\psi(\tau,x^{k}%
)}[(dx^{1})^{2}+(dx^{2})^{2}]\label{offdsolgenfgcosmc}\\
&  +h_{3}(\tau)\{dy^{3}+[\ _{1}n_{k}(\tau)+\ _{2}n_{k}(\tau)\int
dt\frac{[h_{3}^{\diamond}(\tau)]^{2}}{|\int dt[\ _{v}\ \yen (\tau)h_{3}%
(\tau)]^{\diamond}|\ [h_{3}(\tau)]^{5/2}}]dx^{k}\}\nonumber\\
&  -\frac{[h_{3}^{\diamond}(\tau)]^{2}}{|\int dt[\ _{v}\ \yen (\tau)h_{3}%
(\tau)]^{\diamond}|\ h_{3}(\tau)}\{dt+\frac{\partial_{i}[\int dt(\ _{v}%
\ \yen (\tau))\ h_{3}^{\diamond}(\tau)]}{\ \ _{v}\ \yen (\tau)\ h_{3}%
^{\diamond}(\tau)}dx^{i}\}^{2}\nonumber
\end{align}

We can express $\Phi^{2}(\tau)=-4\widehat{\Lambda}(\tau)h_{3}(\tau)$ (also
using the nonlinear symmetries (\ref{ntransf1}) and (\ref{ntransf2})) and
eliminate $\Phi(\tau)$ from the nonlinear quadratic element in
(\ref{offdiagcosmcsh}). In a form which is similar to (\ref{offdsolgenfgcosmc}%
), we obtain a $\tau$-family of solutions determined by generating data
$(h_{3}(\tau);\widehat{\Lambda}(\tau),\ _{v}\ \yen (\tau)).$

\subsubsection{$\tau$-evolution of $Q$-deformed Levi-Civita configurations}

\label{asslcconf}The nonmetric off-diagonal locally anisotropic cosmological
solutions considered in previous subsection were constructed for $\tau
$-families of canonical d--connections $\widehat{\mathbf{D}}(\tau).$ They
encode $Q$-deformations and posses nonholonomically induced d--torsion
coefficients $\widehat{\mathbf{T}}_{\ \alpha\beta}^{\gamma}(\tau),$ which are
completely defined by the N--connection and d--metric structures. We can
extract zero torsion LC-configurations for $q$-distortions of $\nabla(\tau)$
if we impose the conditions (\ref{lccondf}). By straightforward computations
for quasi-stationary configurations, we can verify that all canonical
d-torsion coefficients $\widehat{\mathbf{T}}_{\ \alpha\beta}^{\gamma}(\tau)$
vanish for ansatz (\ref{lacosmans}) if there are satisfied such conditions:
\begin{align}
\mathbf{e}_{i}(\tau)\ln\sqrt{|\ h_{3}(\tau)|}  &  =0,\ w_{i}^{\diamond}%
(\tau)=\mathbf{e}_{i}(\tau)\ln\sqrt{|\ h_{4}(\tau)|},\partial_{i}w_{j}%
(\tau)=\partial_{j}w_{i}(\tau)\mbox{ and }\nonumber\\
n_{i}^{\diamond}(\tau)  &  =0,n_{k}(\tau,x^{i})=0\mbox{ and }\partial_{i}%
n_{j}(\tau,x^{k})=\partial_{j}n_{i}(\tau,x^{k}). \label{zerot1}%
\end{align}
The solutions for necessary type of $w$- and $n$-functions depend on the class
of vacuum, non--vacuum, $Q$-deformed and other type cosmological metrics which
we attempt to generate.

We consider such a possibility to extract solutions which satisfy also the
conditions (\ref{zerot1}): If we prescribe a generating function $\Psi
(\tau)=\check{\Psi}(\tau,x^{i},t),$ for which $[\partial_{i}(\check{\Psi
})]^{\diamond}=\partial_{i}(\check{\Psi})^{\diamond},$ we solve the equations
for $w_{j}$ from (\ref{zerot1}) in explicit form if $\ _{v}\ \yen =const,$ or
if such an effective source can be expressed as a functional $\ _{v}%
\ \yen (\tau,x^{i},t)= \ _{v}\ \yen [\ \check{\Psi}(\tau)].$ Then, the
conditions $\partial_{i}w_{j}(\tau)=\partial_{j}w_{i}(\tau)$ can be solved by
any generating function $\check{A}=\check{A}(\tau,x^{k},t)$ for which
\[
w_{i}(\tau)=\check{w}_{i}(\tau)=\partial_{i}\ \check{\Psi}(\tau)/(\check{\Psi
}(\tau))^{\diamond}=\partial_{i}\check{A}(\tau).
\]
The equations for $n$-functions in (\ref{zerot1}) are solved for any
$n_{i}(\tau)=\partial_{i}[\ ^{2}n(\tau,x^{k})],$ i.e. is such N-connection
coefficients do not depend on time like coordinate.

Putting together all above formulas for respective classes of generating
functions with "inverse hats" and generating sources , we construct a
nonlinear quadratic elements for locally anisotropic cosmological solutions
(\ref{nonmcosmla}) with zero canonical d-torsion,
\begin{align}
d\check{s}^{2}(\tau)  &  =\check{g}_{\alpha\beta}(\tau,x^{k},t)du^{\alpha
}du^{\beta}=e^{\psi(\tau,x^{k})}[(dx^{1})^{2}+(dx^{2})^{2}]\label{lcqdef}\\
&  +\{h_{3}^{[0]}(\tau)-\int dt\frac{[\check{\Psi}^{2}(\tau)]^{\diamond}%
}{4(\ _{v}\ \yen (\tau)[\check{\Psi}(\tau)])}\}\{dy^{3}+\partial_{i}%
[\ ^{2}n(\tau,x^{k})]dx^{i}\}^{2}\nonumber\\
&  +\frac{[\check{\Psi}^{\diamond}(\tau)]^{2}}{4(\ _{v}\ \yen (\tau
)[\check{\Psi}(\tau)])^{2}\{h_{3}^{[0]}(\tau)-\int dt[\check{\Psi}%
(\tau)]^{\diamond}/4\ _{v}\ \yen (\tau)[\check{\Psi}(\tau)]\}}\{dt+[\partial
_{i}(\check{A}(\tau))]dx^{i}\}^{2}.\nonumber
\end{align}
Under nonmetric geometric flows, the d-metrics (\ref{lcqdef}) involve
LC-configurations for $\nabla(\tau)$ but encode also $Q$-deformations included
into $\ _{v}\ \yen (\tau).$ Such solutions can be generated for $\ _{v}
\yen =\ ^{m}\ \yen +\ ^{e}\ \yen +\ ^{DE}\yen $ (\ref{emefparam}), or for
$\ ^{e}\ \yen =0,$ or $\ ^{DE}\yen =0.$ We can compare and decide if ceratin
observational cosmological data are explained by any such solutions.

\subsubsection{Nonmetric geometric cosmological evolution with small
parametric polarizations}

We can re-define the solutions constructed in previous subsection in such a
form which allow to study $\tau$-families of $Q$-deformations of a
\textbf{prime} d-metric $\mathbf{\mathring{g}}$ (it can be an arbitrary one;
or a solution of some equations in GR or a MGT) into a $\tau$-family of
\textbf{target} d-metrics $\mathbf{g}(\tau)$ of type (\ref{lacosmans}),
\begin{equation}
\mathbf{\mathring{g}=}[\mathring{g}_{\alpha},\ \mathring{N}_{i}^{a}%
]\rightarrow\mathbf{g}(\tau)=[g_{\alpha}(\tau)=\eta_{\alpha}(\tau)\mathring
{g}_{\alpha},N_{i}^{a}(\tau)=\eta_{i}^{a}\ (\tau)\mathring{N}_{i}^{a}].
\label{offdiagdefr}%
\end{equation}
Such nonholonomic transforms encode nonmetric deformations which in our
approach are defined by some values $\eta_{\alpha}(\tau,x^{k},t)$ and
$\eta_{i}^{a}(\tau,x^{k},t)$ called $\tau$-running gravitational polarization
($\eta$-polarization) functions. The gravitational polarization functions are
determined by respective $\tau$-families of generating functions, generating
sources and effective cosmological constants,
\begin{align}
(\Psi(\tau),\ _{v}\ \yen (\tau))  &  \leftrightarrow(\mathbf{g}(\tau
),\ _{v}\ \yen (\tau))\leftrightarrow(\eta_{\alpha}(\tau)\ \mathring
{g}_{\alpha}\sim(\zeta_{\alpha}(\tau)(1+\varepsilon\chi_{\alpha}%
(\tau))\mathring{g}_{\alpha},\ _{v}\ \yen (\tau))\leftrightarrow\nonumber\\
(\Phi(\tau),\ \widehat{\Lambda}(\tau))  &  \leftrightarrow(\mathbf{g}%
(\tau),\ \widehat{\Lambda}(\tau))\leftrightarrow(\eta_{\alpha}(\tau
)\ \mathring{g}_{\alpha}\sim(\zeta_{\alpha}(\tau)(1+\varepsilon\chi_{\alpha
}(\tau))\mathring{g}_{\alpha},\ \widehat{\Lambda}(\tau)), \label{dataepsion}%
\end{align}
where $\varepsilon$ is a small parameter $0\leq\varepsilon<1,$ with some
$\zeta_{\alpha}(\tau,x^{k},t)$ and $\chi_{\alpha}(\tau,x^{k},t)$ (in brief, we
shall use the term $\chi$-polarization functions).

Using families of $\eta$- and/or $\chi$-polarizations, the nonlinear
symmetries (\ref{ntransf2}) can be written in the form:
\begin{align}
\lbrack\Psi^{2}(\tau)]^{\diamond}  &  =-\int dt\ \ _{v}\ \yen (\tau
)h_{3}^{\diamond}(\tau)\simeq-\int dt\ \ _{v}\ \yen (\tau)(\eta_{3}%
(\tau)\ \mathring{g}_{3})^{\diamond} \simeq-\int dt\ \ _{v}\ \yen (\tau
)[\zeta_{3}(\tau)(1+\varepsilon\ \chi_{3}(\tau))\ \mathring{g}_{3}]^{\diamond
},\nonumber\\
\Phi^{2}(\tau)  &  =-4\ \widehat{\Lambda}(\tau)h_{3}(\tau)\simeq
-4\ \ \widehat{\Lambda}(\tau)\eta_{3}(\tau)\mathring{g}_{3}\simeq
-4\ \ \widehat{\Lambda}(\tau)\ \zeta_{3}(\tau)(1+\varepsilon\chi_{3}%
(\tau))\ \mathring{g}_{3}. \label{nonlinsymrex}%
\end{align}
So, the nonlinear symmetries of $\tau$- and $Q$-deformations and generating
functions for families of off-diagonal $\eta$-transforms of type
(\ref{offdiagdefr}) can be parameterized for $\eta$-polarizations,
\begin{equation}
\psi(\tau)\simeq\psi(\tau,x^{k}),\eta_{3}\ (\tau)\simeq\eta_{3}(\tau,x^{k},t).
\label{etapolgen}%
\end{equation}

The $\eta$-polarization functions can be used for generating $\tau$-families
of locally anisotropic cosmological solutions of type (\ref{nonmcosmla}),
\begin{align}
d\widehat{s}^{2}(\tau)  &  =\widehat{g}_{\alpha\beta}(\tau,x^{k}%
,t;\mathring{g}_{\alpha};\psi(\tau),\eta_{3}(\tau);\ _{v}\ \yen (\tau
))du^{\alpha}du^{\beta}=e^{\psi(\tau)}[(dx^{1})^{2}+(dx^{2})^{2}%
]+\label{offdiagpolfr}\\
&  \eta_{3}(\tau)\mathring{g}_{3}\{dy^{3}+[\ _{1}n_{k}(\tau)+\ _{2}n_{k}%
(\tau)\int dt\frac{[(\eta_{3}(\tau)\mathring{g}_{3})^{\diamond}]^{2}}{|\int
dt\ \ _{v}\ \yen (\tau)(\eta_{3}(\tau)\mathring{g}_{3})^{\diamond}|\ (\eta
_{3}(\tau)\mathring{g}_{3})^{5/2}}]dx^{k}\}^{2}\nonumber\\
&  -\frac{[(\eta_{3}(\tau)\ \mathring{g}_{3})^{\diamond}]^{2}}{|\int
dt\ \ _{v}\ \yen (\tau)(\eta_{3}(\tau)\ \mathring{g}_{3})^{\diamond}%
|\ \eta_{3}(\tau)\mathring{g}_{3}}\{dt+\frac{\partial_{i}[\int dt\ \ _{v}%
\ \yen (\tau)\ (\eta_{3}(\tau)\mathring{g}_{3})^{\diamond}]}{\ \ _{v}%
\ \yen (\tau)(\eta_{3}(\tau)\mathring{g}_{3})^{\diamond}}dx^{i}\}^{2}%
.\nonumber
\end{align}
For $\Phi^{2}(\tau)=-4\ \ \widehat{\Lambda}h_{3}(\tau),$ we can transform
(\ref{offdiagcosmcsh}) in a variant of (\ref{offdiagpolfr}) with $\eta
$-polarizations determined by the generating data $(h_{3}(\tau
);\ \widehat{\Lambda}\ (\tau)).$ Here we note that it is difficult to
understand if such off-diagonal metrics with general $\eta$- and
$Q$-deformations may have, or not, certain physical importance even the
primary data possess certain important physical interpretation.

We can study $\tau$-flows and $Q$-deformations on a small parameter
$\varepsilon$ using $\varepsilon$-linear functions for generating data
(\ref{dataepsion}) and nonlinear symmetries (\ref{nonlinsymrex}). For such
approximations, the $\eta$-polarizations in (\ref{offdiagpolfr}) describe
nonholonomic deformations of a prime d-metric $\mathbf{\mathring{g}}$ into
so-called $\varepsilon$-parametric $\tau$-families of solutions with $\zeta
$-and $\chi$-coefficients, when
\begin{align}
\psi(\tau)  &  \simeq\psi(\tau,x^{k})\simeq\psi_{0}(\tau,x^{k})(1+\varepsilon
\ _{\psi}\chi(\tau,x^{k})),\mbox{ for }\ \label{epsilongenfdecomp}\\
\ \eta_{2}(\tau)  &  \simeq\eta_{2}(\tau,x^{k})\simeq\zeta_{2}(\tau
,x^{k})(1+\varepsilon\chi_{2}(\tau,x^{k})),\mbox{ we can consider }\ \eta
_{2}(\tau)=\ \eta_{1}(\tau);\nonumber\\
\eta_{3}(\tau)  &  \simeq\eta_{3}(\tau,x^{k},t)\simeq\zeta_{3}(\tau
,x^{k},t)(1+\varepsilon\chi_{3}(\tau,x^{k},t)).\nonumber
\end{align}
In such formulas, $\psi(\tau)$ and $\eta_{2}(\tau)=\ \eta_{1}(\tau)$ are such
way chosen to be determined by solutions of the 2-d Poisson equation
$\partial_{11}^{2}\psi(\tau)+\partial_{22}^{2}\psi(\tau)=2\ _{h}%
\ \yen (\tau,x^{k}),$ see (\ref{eq1}).

$\varepsilon$-parametric deformations (\ref{epsilongenfdecomp}) define such
$\tau$-families of locally anisotropic cosmological d-metrics with $\chi
$-generating functions,
\[
d\ \widehat{s}^{2}(\tau)=\widehat{g}_{\alpha\beta}(\tau,x^{k},t;\psi
(\tau),g_{3}(\tau);\ _{v}\ \yen (\tau))du^{\alpha}du^{\beta}=e^{\psi_{0}%
}(1+\varepsilon\ ^{\psi}\chi(\tau))[(dx^{1})^{2}+(dx^{2})^{2}]
\]%
\begin{align}
&  +\zeta_{3}(\tau)(1+\varepsilon\ \chi_{3}(\tau))\ \mathring{g}_{3}%
\{dy^{3}+[(\mathring{N}_{k}^{3})^{-1}[\ _{1}n_{k}(\tau)+16\ _{2}n_{k}%
(\tau)[\int dt\frac{\left(  [(\zeta_{3}(\tau)\mathring{g}_{3})^{-1/4}%
]^{\diamond}\right)  ^{2}}{|\int dt[\ _{v}\ \yen (\tau)(\zeta_{3}%
(\tau)\mathring{g}_{3})]^{\diamond}|}]\label{offdncelepsilon}\\
&  +\varepsilon\frac{16\ _{2}n_{k}(\tau)\int dt\frac{\left(  [(\zeta_{3}%
(\tau)\mathring{g}_{3})^{-1/4}]^{\diamond}\right)  ^{2}}{|\int dt[\ _{v}%
\ \yen (\tau)(\zeta_{3}(\tau)\mathring{g}_{3})]^{\diamond}|}(\frac{[(\zeta
_{3}(\tau)\mathring{g}_{3})^{-1/4}\chi_{3})]^{\diamond}}{2[(\zeta_{3}%
(\tau)\mathring{g}_{3})^{-1/4}]^{\diamond}}+\frac{\int dt[\ _{v}%
\ \yen (\tau)(\zeta_{3}(\tau)\chi_{3}(\tau)\mathring{g}_{3})]^{\diamond}}{\int
dt[\ _{v}\ \yen (\tau)(\zeta_{3}(\tau)\mathring{g}_{3})]^{\diamond}})}%
{\ _{1}n_{k}(\tau)+16\ _{2}n_{k}(\tau)[\int dt\frac{\left(  [(\zeta_{3}%
(\tau)\mathring{g}_{3})^{-1/4}]^{\diamond}\right)  ^{2}}{|\int dt[\ _{v}%
\ \yen (\tau)(\zeta_{3}(\tau)\mathring{g}_{3})]^{\diamond}|}]}]\mathring
{N}_{k}^{3}dx^{k}\}^{2}.\nonumber
\end{align}%
\begin{align*}
&  -\{\frac{4[(|\zeta_{3}(\tau)\mathring{g}_{3}|^{1/2})^{\diamond}]^{2}%
}{\mathring{g}_{3}|\int dt\{\ _{v}\ \yen (\tau)(\zeta_{3}(\tau)\mathring
{g}_{3})^{\diamond}\}|}-\varepsilon\lbrack\frac{(\chi_{3}(\tau)|\zeta_{3}%
(\tau)\mathring{g}_{3}|^{1/2})^{\diamond}}{4(|\zeta_{3}(\tau)\mathring{g}%
_{3}|^{1/2})^{\diamond}}-\frac{\int dt\{\ _{v}\ \yen [(\zeta_{3}%
(\tau)\mathring{g}_{3})\chi_{3}(\tau)]^{\diamond}\}}{\int dt\{\ _{v}%
\ \yen (\tau)(\zeta_{3}(\tau)\mathring{g}_{3})^{\diamond}\}}]\}\mathring
{g}_{4}\\
&  \{dt+[\frac{\partial_{i}\ \int dt\ _{v}\ \yen (\tau)\zeta_{3}^{\diamond
}(\tau)}{(\mathring{N}_{i}^{3})\ \ _{v}\ \yen (\tau)\zeta_{3}^{\diamond}%
(\tau)}+\varepsilon(\frac{\partial_{i}[\int dt\ _{v}\ \yen (\tau)\ (\zeta
_{3}(\tau)\chi_{3}(\tau))^{\diamond}]}{\partial_{i}\ [\int dt\ \ _{v}%
\ \yen (\tau)\zeta_{3}^{\diamond}(\tau)]}-\frac{(\zeta_{3}(\tau)\chi_{3}%
(\tau))^{\diamond}}{\zeta_{3}^{\diamond}(\tau)})]\mathring{N}_{i}^{4}%
dx^{i}\}^{2}%
\end{align*}
Such off-diagonal parametric solutions allow us to define, for instance,
ellipsoidal deformations of spherical symmetric cosmological metrics into
similar ones with ellipsoid symmetry. Locally anisotropic cosmological
d-metrics of type (\ref{offdncelepsilon}) can be generated for by certain
small parametric deformations and generating data $(\Phi(\tau
),\ \widehat{\Lambda}(\tau)).$

\setcounter{equation}{0} \renewcommand{\theequation}
{B.\arabic{equation}} \setcounter{subsection}{0}
\renewcommand{\thesubsection}
{B.\arabic{subsection}}

\section{A brief review on 2+2 spacetime topological quasicrystal structures}

\label{appendixb} We outline necessary results on nonholonomic
quasicrystalline structures which are necessary to elaborate on models of
interacting topological phase of (effective) matter and nonmetric
gravitational vacuum under geometric evolution, i.e. DE and DM phases.
Considered elasticity models develop the constructions with cosmological
quasicristals, QCs, \cite{bubuianu17,sv18} and do not consider nontrivial
quantized topological terms with far richer structure than their crystalline
counterparts studied in condensed matter physics \cite{else21}.

\subsection{Definition of spacetime QC and quasicrystalline topological
phases}

Let us state the definition of QC spacetime structure that we adopt in this
work. Similarly to \cite{else21}, we consider a countable set $\emph{Vec}%
^{\ast}$ with vectors $\mathbf{k}=\{k^{\alpha^{\prime}}\}\in\emph{Vec}^{\ast
}\subset T^{\ast}\mathbf{V}$ for a point $\mathbf{x}(u)=\{x_{\alpha^{\prime}%
}(u)\},$ for $u\in\mathbf{V}$ of a Lorentz manifold with $Q$-deformations. For
a local (relativistic) quantum mechanical (QM) model, the expectation value
$<\widehat{O}>$ of a local observable $\widehat{O}$ can be expressed as a
Fourier series%
\[
<\widehat{O}(\mathbf{x}(u))>=\sum\nolimits_{\mathbf{k}}a_{\mathbf{k}%
}e^{i\mathbf{kx}} \mbox{ or }<\widehat{\mathbf{O}}(\mathbf{x}(u))>=\sum
\nolimits_{\mathbf{k}}\mathbf{a}_{\mathbf{k}}e^{i\mathbf{kx}},
\]
where $\mathbf{kx}$ denotes the scalar product in $u$ and the coefficients
$a_{\mathbf{k}}= a_{\mathbf{k}}(\widehat{O})$ depend on the choice of operator
$\widehat{O}.$ We can consider boldface symbols for d-operators adapted to a
N-connection splitting (\ref{ncon}). A crystal structure can be generated by a
finite set $\emph{Vec}^{\ast}$ considered as a reciprocal lattice of a crystal
defined for $d^{\prime}$ spacial dimensions (called as primitive reciprocal
lattice vectors; for modelling spacetime and time like crystals with
nonholonomic structure, we can consider pseudo-Euclidean signatures
\cite{bubuianu17,sv18}). We say that such a system is a QC if $\emph{Vec}%
^{\ast}$ is defined in such a way but the smallest such set is of size greater
than $d^{\prime}.$

A quasicrystalline topological phase is defined by a family of Hamiltonians
such that the ground state is always gapped and when the above QC conditions
are satisfied. We can consider a QC with elasticity when $<\widehat{O}%
(\mathbf{x}(u))>^{\prime}=\sum\nolimits_{\mathbf{k}}a_{\mathbf{k}}%
e^{i(\phi_{\mathbf{k}}+\mathbf{kx)}}$ for an independent choice on
$\widehat{O}$ of phases subjected to the conditions $\phi_{\mathbf{{k}_{1}}%
}+\mathbf{{k}_{2}}= \phi\mathbf{_{{k}_{1}}+}\phi\mathbf{_{{k}_{2}}}[mod2\pi]$
for any $\mathbf{{k}_{1},{k}_{2}\in\emph{Vec}^{\ast}.}$ The $\phi$-filed can
be thought as a "order parameter" for some spontaneous symmetry-breaking
ground states, when the low-energy elastic deformation are considered as
long-wavelength fluctuations (for such an order parameter). For such models,
we replace $\phi_{\mathbf{k}}$ with a slow varying function $\phi_{\mathbf{k}%
}=\phi_{\mathbf{k}}(\mathbf{x,}t)=\phi_{\mathbf{k}}(u).$ Introducing phase
fields
\begin{align}
\theta_{\mathbf{k}}(u)  &  = \phi_{\mathbf{k}}(u)+\mathbf{kx,}%
\mbox{ for }\theta_{\mathbf{k}_{1}+\mathbf{k}_{2}}(u)=\theta_{\mathbf{k}_{1}%
}(u)+\theta_{\mathbf{k}_{2}}(u)\ [{mod}2\pi],\label{condqc}\\
&  \mbox{ when }<\widehat{O}(\mathbf{x}(u))>^{\prime}=\sum
\nolimits_{\mathbf{k}}a_{\mathbf{k}}e^{i\theta_{\mathbf{k}}(u)}.\nonumber
\end{align}
We approximate for a low-energy deformation $\nabla\theta_{\mathbf{k}%
}(u)\approx\mathbf{k}$, when for some vacuum state $\nabla\theta_{\mathbf{k}%
}(u)=\mathbf{k}$. For canonical N-adapted deformations, such formulas
transform into similar ones with $\nabla\theta_{\mathbf{k}}(u)\rightarrow
\widehat{\mathbf{D}}\theta_{\mathbf{k}}(u).$

For QC structures, the solutions of (\ref{condqc}) can be parameterized by a
set of reciprocal d-vectors $\mathbf{K}^{1},...,\mathbf{K}^{\widehat{d}}$ that
generate $\emph{Vec}^{\ast}$( $\widehat{d}>d^{\prime},$ for a crystal
structure, $\widehat{d}=d^{\prime}$). There are imposed such properties:

\begin{enumerate}
\item Every d-vector can be expressed as an integer linear combination of type
$\mathbf{k}=n_{I}\mathbf{K}^{I}\in\emph{Vec}^{\ast}$ (for
$I=1,2,...,\widehat{d}).$

\item The reciprocal d-vectors are linearly independent over integers, i.e. if
$n_{I}\mathbf{K}^{I}=0$ for some integers $n_{I},$ then all $n_{I}=0$ (we do
no discuss variants of definition of QCs when this condition is dropped, for
instance, when we have an over complete set of d-vectors).
\end{enumerate}

Using properties 1 and 2, we can introduce phase angle fields $\theta^{I}(u)$
and write the solutions of (\ref{condqc}) in the form $\theta_{\mathbf{k}%
}(u)=\theta^{I}(u)n_{I}(\mathbf{k})$, where $n_{I}(\mathbf{k})$ is a unique
integer vector and $\nabla\theta^{I}(u)\approx K^{I}$ (or $\widehat{\mathbf{D}%
}\theta^{I}(u)\approx\mathbf{K}^{I}$ for N-adapted canonical constructions).
Thus, we parameterize the elastic deformations in terms of $\widehat{d}$
fields $\theta^{I}(u).$ For a crystal type structure with $\widehat{d}%
=d^{\prime},$ this defines phonon modes; for a QC structure with
$\widehat{d}>d^{\prime}$ there are more elastic modes than in a crystal. In
this paper, we work directly with $\theta^{I}(u)$ even in condensed matter
physics one consider decompositions into certain phonon and phason modes, see
\cite{else21} and references therein.

Finally, we note that in the ground spacetime vacuum space, $\theta
^{I}(u)=\theta_{\lbrack0]}^{I}+ \mathbf{K}^{I}\mathbf{x,}$ with $\theta
_{\lbrack0]}^{I}$ taken for a crystal like structure. In such cases, the
expectation values of observable in a QC deformation are given by linear
mappings of a $d^{\prime}$ dimensional physical space (for modeling the
crystal structure) into a hyperplane slice through a $\widehat{d}$-dimensional
crystal "superspace".

\subsection{Topological terms for nonholonomic spacetime QCs and elasticity}

Let us consider a $\widehat{d}$-dimensional vector $C_{I},$ an antisymmetric
(matrix $\widehat{d}\times\widehat{d})$ $C_{IJ},$ an antisymmetric tensor
$C_{IJK},$ and a gauge field $\mathbf{A}_{\mu}(u)$ with local $U(1)$ symmetry.
Using such algebraic data, vector $A_{\mu}$, and angle space fields
$\theta^{I}(u),$ we can introduce such effective Lagrange densities for
topological QCs (see details in sections III and V of \cite{else21}):
\begin{align}
\ ^{\theta1}\widehat{\mathcal{L}}  &  =\ ^{\theta1}\widehat{\mathcal{L}}%
_{[0]}+\frac{1}{2\pi}C_{I}\varepsilon^{\mu\nu}A_{\mu}\mathbf{\partial}_{\nu
}(\theta^{I}),\mbox{ for }d^{\prime}=1;\nonumber\\
\ ^{\theta2}\widehat{\mathcal{L}}  &  = \ ^{\theta2}\widehat{\mathcal{L}%
}_{[0]}+\frac{1}{2\pi}C_{IJ}\varepsilon^{\mu\nu\gamma}A_{\mu}\mathbf{\partial
}_{\nu}(\theta^{I})\mathbf{\partial}_{\gamma}(\theta^{J}%
),\mbox{ for }d^{\prime}=2;\label{topqc2}\\
\ ^{\theta3}\widehat{\mathcal{L}}  &  = \ ^{\theta3}\widehat{\mathcal{L}%
}_{[0]}+\frac{1}{2\pi}C_{IJK}\varepsilon^{\mu\nu\gamma\sigma}A_{\mu
}\mathbf{\partial}_{\nu}(\theta^{I})\mathbf{\partial}_{\gamma}(\theta
^{J})\mathbf{\partial}_{\sigma}(\theta^{K}),\mbox{ for }d^{\prime}=3,\nonumber
\end{align}
where $\varepsilon^{...}$ are absolute antisymmetric tensors and terms with
label $[0]$ can be defined for a specific crystal like structure. In this
work, we shall elaborate on models of dimensions $d^{\prime}=2,3$. Such
configurations are invariant (modulo 2$\pi$) under large gauge transformations
of $A$ on nonholonomic spacetime manifold $\mathbf{V}$ and for linearized
equations each entry of $C$ can be quantized to be integer. This defines a
large family of symmetry-protected topological, SPT, phases which can be
partially classified as QCs with $U(1)$ symmetry by integral antisymmetric
rank-$d^{\prime}$ tensors of dimension $\widehat{d}.$ In not N-adapted form,
we can consider crystalline structures with $\widehat{d}=d^{\prime},$ when all
such tensors are some integer multiples of the Levi-Civita antisymmetric
tensors. For a given QC SPT phase with underlying reciprocal lattice
$\mathcal{L},$ the $C$-values depend on some choice of the generating
reciprocal vectors $K^{I}. $ Such phases are classified by the internal
cohomology $H^{d^{\prime}}(\mathcal{L}^{\ast},\mathbb{Z}),$ with integer
$\mathbb{Z},$ where $\mathcal{L}$ is the space of homomorphisms from
$\mathcal{L}^{\ast}$ to $U(1).$

For QC spacetime configurations, there are possible additional topological
terms which do not depend on $U(1)$ symmetry. If $\widehat{d}\geq d^{\prime
}+1,$ we can consider terms:
\begin{equation}
\ ^{\Theta1}\widehat{\mathcal{L}} = \frac{1}{2\pi}\Theta_{IJ}\varepsilon
^{\nu\gamma}\mathbf{\partial}_{\nu}(\theta^{I})\mathbf{\partial}_{\gamma
}(\theta^{J}),\mbox{ for }d^{\prime}=1;\ \ ^{\Theta2}\widehat{\mathcal{L}} =
\frac{1}{2\pi}\Theta_{IJK} \varepsilon^{\nu\gamma\sigma}\mathbf{\partial}%
_{\nu}(\theta^{I})\mathbf{\partial}_{\gamma}(\theta^{J})\mathbf{\partial
}_{\sigma}(\theta^{K})\mbox{ for }d^{\prime}=2; \label{topqc2a}%
\end{equation}
and so on (for higher dimensions). Such terms are classified by $H^{d^{\prime
}+1}(\mathcal{L}^{\ast},\mathbb{Z}).$

In this work, we study QC v-structures adapted to a N-splitting when
$\theta^{I}\simeq$ $\theta^{a}(x^{i},t),$ for $I\simeq a,$ and $\partial
_{j}\theta^{a}(x^{i},t)=K_{j}^{a}(x^{i},t).$ For generating off-diagonal
solutions, we can elaborate on models with $N_{j}^{a}(x^{i},t)\simeq K_{j}%
^{a}(x^{i},t)$ and define topological QC-configurations determined by certain
$\theta^{I}$ terms.

Finally, we note that the angle space fields $\theta^{I}(u)$ can be related to
another class of topological terms of Wess-Zumino type (see \cite{else21} and
references therein) for $\widehat{d}\geq d^{\prime}+2.$ There are not
canonically expressible local Lagrangians in $d^{\prime}+1$ spacetime
dimensions of an extension of the $\theta$-fields. Such terms are
characterized by $H^{d^{\prime}+1}(\mathcal{L}^{\ast},\mathbb{Z}).$ For
simplicity, in this work we work with QC structures of type $\ ^{\theta
2}\widehat{\mathcal{L}}$ (\ref{topqc2}) or $\ ^{\Theta2}\widehat{\mathcal{L}}$
(\ref{topqc2a}).

\subsection{ Mobility of dislocations and currents and charge density for
spacetime QCs}

In QCs, there are possible defects with dislocations which can move in
N-adapted form only in the directions of their Burgers d-vector. Here, we
provide necessary formulas for 2-d nonholonomic cosmological QC structures
when
\[
\frac{1}{2\pi}{\displaystyle\oint\nolimits_{C}}dx^{i}\ \mathbf{e}_{i}%
\theta^{a}(\tau,x^{k},t)=\mathbf{b}^{a}\in\mathbb{Z}^{d^{\prime}},d^{\prime
}=2,
\]
for a N-connection 2+2 splitting. In this formula, $C$ is a loop surrounding
the dislocations and $\mathbf{b}^{a}$ is the Burgers d-vectors. The
nonholonomic and topological mobility constraints of a dislocations can be
derived for a respective charge conservation law. We can consider a d-vector
$A_{\mu}$ in (\ref{topqc2}), or introduce certain effective values determined
by $\varepsilon^{\nu\gamma\sigma}\mathbf{e}_{\nu}(\theta^{a})$ in
(\ref{topqc2a}), define certain "charged" mobility of $\theta^{a}(\tau
,x^{k},t)$ for temperature $\tau.$ The corresponding current is defined as the
d-vector%
\[
\mathbf{J}^{\nu}=\frac{1}{8\pi^{2}}C_{ab}\varepsilon^{\nu\gamma\sigma
}\mathbf{e}_{\gamma}(\theta^{a})\mathbf{e}_{\sigma}(\theta^{b}).
\]
The holonomic conservation law%
\[
\mathbf{\partial}_{\mu}J^{\mu}=C_{ab}\varepsilon^{\mu\gamma\sigma}\left[
\mathbf{\partial}_{\mu}\mathbf{\partial}_{\gamma}(\theta^{a})-\mathbf{\partial
}_{\gamma}\mathbf{\partial}_{\mu}(\theta^{a})\right]  \mathbf{\partial
}_{\sigma}(\theta^{b})=0
\]
transforms into a nonholonomic relation%
\begin{equation}
\mathbf{e}_{\mu}\mathbf{J}^{\mu}=C_{ab}\varepsilon^{\mu\gamma\sigma}%
W_{\mu\gamma}^{\nu}\mathbf{e}_{\nu}(\theta^{a})\mathbf{e}_{\sigma}(\theta
^{b})=\ ^{J}\mathbf{Z}\neq0, \label{conslawsourc}%
\end{equation}
where the anholonomy coefficients are defined in footnote \ref{fnwcoeff} and
$\ ^{J}\mathbf{Z}$ is determined by N-coefficients. This reflects the fact
that for nonholonomic and/or nonmetric systems the conservation laws became
more sophisticate and may encode integration constants for corresponding
Lagrange densities and distortion terms.

We can introduce double nonholonomic 2+2 and 3+1 splitting for space indices
$i,j,...=1,2$ and $\acute{\imath},\acute{j},...=1,2,3,$ when $\partial
_{4}=\partial_{t}.$ For off-diagonal vacuum models close to the equilibrium
configurations, one approximates $\mathbf{\partial}_{t}(\theta^{a})\approx0,$
$\mathbf{\partial}_{i}(\theta^{a})\approx N_{i}^{a}(x^{k})$ and
$\mathbf{\partial}_{j^{\prime}}(\theta^{a})\approx K_{j^{\prime}}^{a}(x^{i}).$
In vicinity of such configurations, the nonholonmic distortions of the current
conservation law (\ref{conslawsourc}) is computed
\[
\mathbf{e}_{\mu}\mathbf{J}^{\mu}=C_{ab}\varepsilon^{ij}W_{4i}^{\nu}%
\mathbf{e}_{\nu}(\theta^{a})N_{j}^{b}=\ ^{J}\mathbf{Z,}%
\]
which in holonomic frames is equivalent to
\[
\mathbf{\partial}_{\mu}J^{\mu}=C_{ab}\varepsilon^{\acute{\imath}\acute{j}%
}\left[  \mathbf{\partial}_{4}\mathbf{\partial}_{\acute{j}}(\theta
^{a})-\mathbf{\partial}_{\acute{j}}\mathbf{\partial}_{4}(\theta^{a})\right]
K_{\acute{\imath}}^{a}(x^{i})=0.
\]
Such formulas allow us to describe dislocations for a Burgers vector $b^{a}$
at a position $x^{\acute{\imath}}(\zeta),$ with parameter $\zeta,$ moving at
velocity $v^{\acute{\imath}},$ when
\[
\left[  \mathbf{\partial}_{4}\mathbf{\partial}_{\acute{j}}-\mathbf{\partial
}_{\acute{j}}\mathbf{\partial}_{4}\right]  \theta^{a}=\varepsilon_{\acute
{j}\acute{k}}v^{\acute{k}}\delta\lbrack x^{\acute{\imath}}-x^{\acute{\imath}%
}(\zeta)],\mbox{ or }W_{4i}^{\nu}\mathbf{e}_{\nu}(\theta^{a})=\varepsilon
_{jk}v^{k}\delta\lbrack x^{i}-x^{i}(\zeta)].
\]
These local conservation law for (non) holonomic dislocations require that%
\begin{equation}
C_{ab}b^{a}K_{j^{\prime}}^{b}v^{j^{\prime}}=0,\mbox{ or }C_{ae}b^{a}N_{j}%
^{e}v^{j}=0. \label{topologmobconstr}%
\end{equation}
The last condition is not satisfied if we consider general nonholonomic
deformations. Equations (\ref{topologmobconstr}) define the topological and
nonholonomic mobility constraints of dislocations in a 2-d QC determined by
the SPT invariant $C_{ab}$ and the burgers vector $b^{a}.$ Thus, dislocations
in a spacetime QC are never completely immobilized and can move in some
directions because there is at least one direction $b^{a}$ which satisfies
(non) holonomic conditions of type (\ref{topologmobconstr}).

The effective gravitational vacuum fields $A_{\mu}$ in the action
(\ref{topqc2}) can be evaluated by an effective charge density $\rho=\delta
S/\delta A_{4}$ which for nonholonomic 2-d QC structures with
$\mathbf{\partial}_{j}(\theta^{a})\approx K_{j}^{a}(x^{i})\rightarrow
N_{j}^{a}(x^{i},t)$ and spacetime depending average density%
\begin{equation}
\rho(x^{i},t)=\frac{1}{8\pi^{2}}C_{ab}\varepsilon^{ij}N_{i}^{a}N_{j}^{a}.
\label{avdens1}%
\end{equation}
We assume that nearly gravitational vacuum equilibrium the non-topological
part of the action does not contribute to average of such effective charge
density but may result in nonholonomic modifications. Such an assumption is
reasonable for effective phonon and phason fields which for (non) metric/
holonomic gravitational interactions does not give any contribution to the
ground state charge density.

Finally, we note two simple tilling interpretations for QC structures with are
similar to those presented in Figures 1 and 2 of section VIII A. of
\cite{else21}. In the first case (called $k=2,$ for $\widehat{d}=4=d^{\prime
}+2$ and $d^{\prime}=2$), there are generated the so-called QC topological
phases with conventional 6 colours if no rotational symmetry is imposed.
Imposing eight-fold rotational symmetry, we obtain two independent variants
(for corresponding square and rhombus tiles). In the second case (conventional
$k=1$), four conventional colors can be generated and a 1-d SPT invariant is
assigned for no rotational symmetry imposed. One independent SPT invariant is
defined also if eightfold rotational symmetry is imposed. For equilibrium
gravitational vacuum configurations, such nonholonomic configurations with
nontrivial topology can be defined as quasi-stationary ones as in
\cite{kazakh1}, Off-diagonal (non) metric gravitational interactions with
average density $\rho(x^{i},t)$ (\ref{avdens1}) result in nonholonomic
deformations of QC structures which can be used for modeling DM and DE time
depending configurations.

\vskip5pt
{\bf Data Availability Statement:} No Data are necessary to be associated in the manuscript.
`

\end{document}